\documentclass[aps,prd,twocolumn,floatfix,superscriptaddress,balancelastpage,nofootinbib,amsmath,showpacs]{revtex4}
\usepackage{graphicx}
\usepackage{epstopdf}
\usepackage{epsfig}
\usepackage{bm}
\usepackage{dcolumn}
\usepackage{natbib}
\usepackage{amsmath}
\usepackage{amssymb}
\usepackage{color}
\usepackage{capt-of}

\newcommand{\be}{\begin{equation}}
\newcommand{\ee}{\end{equation}}
\newcommand{\bea}{\begin{eqnarray}}
\newcommand{\eea}{\end{eqnarray}}

\def\lsim{\mathrel{\raise.3ex\hbox{$<$\kern-.75em\lower1ex\hbox{$\sim$}}}}
\def\gsim{\mathrel{\raise.3ex\hbox{$>$\kern-.75em\lower1ex\hbox{$\sim$}}}}

\begin{document}









\title{The Characterization of the Gamma-Ray Signal from the Central Milky Way: \\ A Compelling Case for Annihilating Dark Matter}




\author{Tansu Daylan}
\affiliation{Department of Physics, Harvard University, Cambridge, MA}

\author{Douglas P.~Finkbeiner}
\affiliation{Department of Physics, Harvard University, Cambridge, MA}
\affiliation{Harvard-Smithsonian Center for Astrophysics, Cambridge, MA}

\author{Dan Hooper}
\affiliation{Fermi National Accelerator Laboratory, 
             Theoretical Astrophysics Group, Batavia, IL} 
\affiliation{University of Chicago, 
             Department of Astronomy and Astrophysics, Chicago, IL} 

\author{Tim Linden}
\affiliation{University of Chicago, 
             Kavli Institute for Cosmological Physics, Chicago, IL} 

\author{Stephen K.~N.~Portillo}
\affiliation{Harvard-Smithsonian Center for Astrophysics, Cambridge, MA}

\author{Nicholas L.~Rodd}
\affiliation{Center for Theoretical Physics, Massachusetts Institute of Technology, Boston, MA}

\author{Tracy R.~Slatyer}
\affiliation{Center for Theoretical Physics, Massachusetts Institute of Technology, Boston, MA} 
\affiliation{School of Natural Sciences, Institute for Advanced Study, Princeton, NJ}

\begin{abstract}

Past studies have identified a spatially extended excess of $\sim$1-3 GeV gamma rays from the region surrounding the Galactic Center, consistent with the emission expected from annihilating dark matter. We revisit and scrutinize this signal with the intention of further constraining its characteristics and origin. By applying cuts to the \textit{Fermi} event parameter CTBCORE, we suppress the tails of the point spread function and generate high resolution gamma-ray maps, enabling us to more easily separate the various gamma-ray components. Within these maps, we find the GeV excess to be robust and highly statistically significant, with a spectrum, angular distribution, and overall normalization that is in good agreement with that predicted by simple annihilating dark matter models. For example, the signal is very well fit by a 36-51 GeV dark matter particle annihilating to $b\bar{b}$ with an annihilation cross section of $\sigma v = (1-3)\times 10^{-26}$ cm$^3$/s (normalized to a local dark matter density of 0.4 GeV/cm$^3$). Furthermore, we confirm that the angular distribution of the excess is approximately spherically symmetric and centered around the dynamical center of the Milky Way (within $\sim$$0.05^{\circ}$ of Sgr A$^*$), showing no sign of elongation along the Galactic Plane. The signal is observed to extend to at least $\simeq10^{\circ}$ from the Galactic Center, disfavoring the possibility that this emission originates from millisecond pulsars.

\end{abstract}

\pacs{95.85.Pw, 98.70.Rz, 95.35.+d; FERMILAB-PUB-14-032-A, MIT-CTP 4533} 

\maketitle

\section{Introduction}
\label{intro}

Weakly interacting massive particles (WIMPs) are a leading class of candidates for the dark matter of our universe. If the dark matter consists of such particles, then their annihilations are predicted to produce potentially observable fluxes of energetic particles, including gamma rays, cosmic rays, and neutrinos. Of particular interest are gamma rays from the region of the Galactic Center which, due to its proximity and high dark matter density, is expected to be the brightest source of dark matter annihilation products on the sky, hundreds of times brighter than the most promising dwarf spheroidal galaxies. 

Over the past few years, several groups analyzing data from the \textit{Fermi} Gamma-Ray Space Telescope have reported the detection of a gamma-ray signal from the inner few degrees around the Galactic Center (corresponding to a region several hundred parsecs in radius), with a spectrum and angular distribution compatible with that anticipated from annihilating dark matter particles~\cite{Goodenough:2009gk,Hooper:2010mq,Boyarsky:2010dr,Hooper:2011ti,Abazajian:2012pn,Gordon:2013vta,Abazajian:2014fta}. More recently, this signal was shown to also be present throughout the larger Inner Galaxy region, extending kiloparsecs from the center of the Milky Way~\cite{Hooper:2013rwa,Huang:2013pda}. While the spectrum and morphology of the Galactic Center and Inner Galaxy signals have been shown to be compatible with that predicted from the annihilations of an approximately 30-40 GeV WIMP annihilating to quarks (or a $\sim$7-10 GeV WIMP annihilating significantly to tau leptons), other explanations have also been proposed. In particular, it has been argued that if our galaxy's central stellar cluster contains several thousand unresolved millisecond pulsars, they might be able to account for the emission observed from the Galactic Center~\cite{Hooper:2010mq,Abazajian:2010zy,Hooper:2011ti,Abazajian:2012pn,Gordon:2013vta,Abazajian:2014fta}. The realization that this signal extends well beyond the boundaries of the central stellar cluster~\cite{Hooper:2013rwa,Huang:2013pda} disfavors such interpretations, however.  In particular, pulsar population models capable of producing the observed emission from the Inner Galaxy invariably predict that \textit{Fermi} should have resolved a much greater number of such objects. Accounting for this constraint, Ref.~\cite{Hooper:2013nhl} concluded that no more than $\sim$5-10\% of the anomalous gamma-ray emission from the Inner Galaxy can originate from pulsars. Furthermore, while it has been suggested that the Galactic Center signal might result from cosmic-ray interactions with gas~\cite{Hooper:2010mq,Hooper:2011ti,Abazajian:2012pn,Gordon:2013vta}, the analyses of Refs.~\cite{Linden:2012iv} and~\cite{Macias:2013vya} find that measured distributions of gas provide a poor fit to the morphology of the observed signal.  It also appears implausible that such processes could account for the more spatially extended emission observed from throughout the Inner Galaxy.

In this study, we revisit the anomalous gamma-ray emission from the Galactic Center and the Inner Galaxy regions and scrutinize the \textit{Fermi} data in an effort to constrain and characterize this signal more definitively, with the ultimate goal being to confidently determine its origin. One way in which we expand upon previous work is by selecting photons based on the value of the \textit{Fermi} event parameter CTBCORE. Through the application of this cut, we select only those events with more reliable directional reconstruction, allowing us to better separate the various gamma-ray components, and to better limit the degree to which emission from the Galactic Disk leaks into the regions studied in our Inner Galaxy analysis.  We produce a new and robust determination of the spectrum and morphology of the Inner Galaxy and the Galactic Center signals. We go on to apply a number of tests to this data, and determine that the anomalous emission in question agrees well with that predicted from the annihilations of a 36-51 GeV WIMP annihilating mostly to $b$ quarks (or a somewhat lower mass WIMP if its annihilations proceed to first or second generation quarks). Our results now appear to disfavor the previously considered 7-10 GeV mass window in which the dark matter annihilates significantly to tau leptons~\cite{Hooper:2010mq,Hooper:2011ti,Gordon:2013vta,Hooper:2013rwa,Abazajian:2014fta} (the analysis of Ref.~\cite{Gordon:2013vta} also disfavored this scenario). The morphology of the signal is consistent with spherical symmetry, and strongly disfavors any significant elongation along the Galactic Plane. The emission decreases with the distance to the Galactic Center at a rate consistent with a dark matter halo profile which scales as $\rho \propto r^{-\gamma}$, with $\gamma \approx 1.1-1.3$. The signal can be identified out to angles of $\simeq 10^{\circ}$ from the Galactic Center, beyond which systematic uncertainties related to the Galactic diffuse model become significant. The annihilation cross section required to normalize the observed signal is $\sigma v \sim 10^{-26}$ cm$^3$/s, in good agreement with that predicted for dark matter in the form of a simple thermal relic.

The remainder of this article is structured as follows. In the following section, we review the calculation of the spectrum and angular distribution of gamma rays predicted from annihilating dark matter.  In Sec.~\ref{ctbcore}, we describe the event selection used in our analysis, including the application of cuts on the \textit{Fermi} event parameter CTBCORE. In Secs.~\ref{inner} and~\ref{center}, we describe our analyses of the Inner Galaxy and Galactic Center regions, respectively. In each of these analyses, we observe a significant gamma-ray excess, with a spectrum and morphology in good agreement with that predicted from annihilating dark matter. We further investigate the angular distribution of this emission in Sec.~\ref{morphology}, and discuss the dark matter interpretation of this signal in Sec.~\ref{darkmatter}. In Sec.~\ref{discussion} we discuss the implications of these observations, and offer predictions for other upcoming observations. Finally, we summarize our results and conclusions in Sec.~\ref{summary}. In the paper's appendices, we include supplemental material intended for those interested in further details of our analysis.

\bigskip

\begin{figure*}[t!]
\includegraphics[width=6.9in]{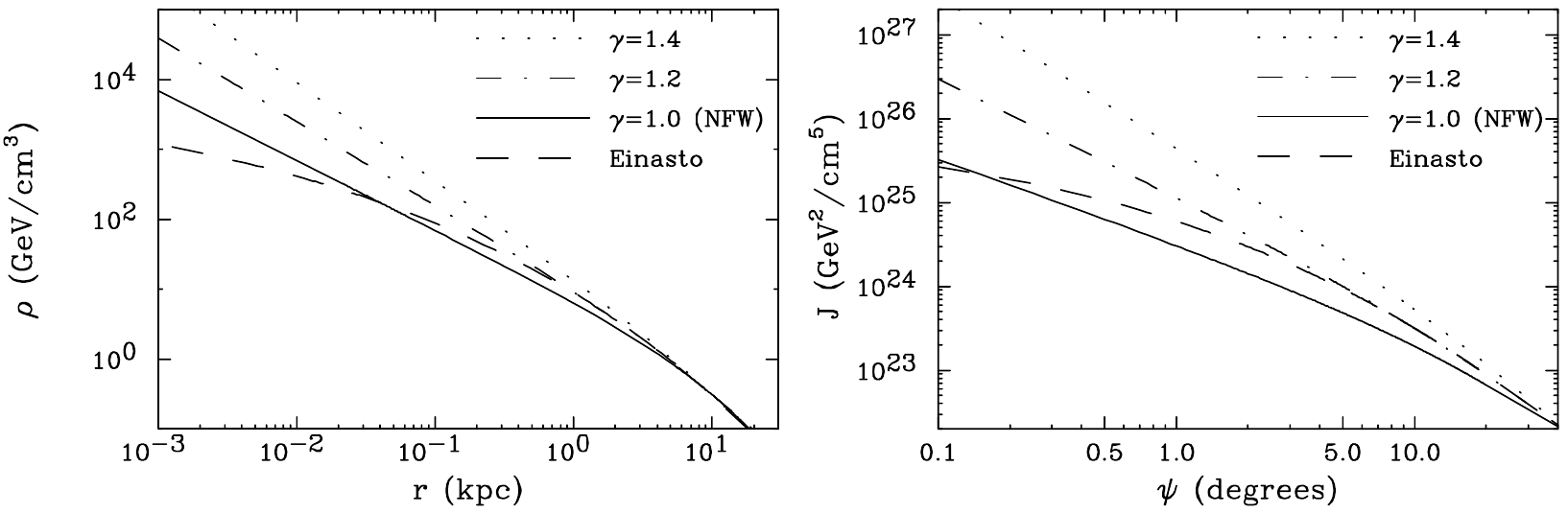}
\caption{Left frame: The dark matter density as a function of the distance to the Galactic Center, for several halo profiles, each normalized such that $\rho=0.4$ GeV/cm$^3$ at $r=8.5$ kpc. Right frame: The line-of-sight integral of the density squared, as defined in Eq.~\ref{J}, for the same set of halo profiles, as a function of the angular distance from the Galactic Center, $\psi$.}
\label{profiles}
\end{figure*}
\section{Gamma Rays From Dark Matter Annihilations in the Halo of the Milky Way}
\label{intro2}

Dark matter searches using gamma-ray telescopes have a number of advantages over other indirect detection strategies. Unlike signals associated with cosmic rays (electrons, positrons, antiprotons, etc), gamma rays are not deflected by magnetic fields. Furthermore, gamma-ray energy losses are negligible on galactic scales. As a result, gamma-ray telescopes can potentially acquire both spectral and spatial information, unmolested by astrophysical effects.

The flux of gamma rays generated by annihilating dark matter particles, as a function of the direction observed, $\psi$, is given by:
\begin{equation}
\Phi (E_{\gamma}, \psi) = \frac{\sigma v}{8 \pi  m_{X}^2}  \frac{\mathrm{d}N_{\gamma}}{\mathrm{d}E_\gamma}  \,  \int_{\text{los}} \rho^2(r ) \,\mathrm{d}l, 
\label{basic}
\end{equation}
where $m_X$ is the mass of the dark matter particle, $\sigma v$ is the annihilation cross section (times the relative velocity of the particles), $dN_{\gamma}/dE_{\gamma}$ is the gamma-ray spectrum produced per annihilation, and the integral of the density squared is performed over the line-of-sight (los). Although N-body simulations lead us to expect dark matter halos to exhibit some degree of triaxiality (see~\cite{Kuhlen:2007ku} and references therein), the Milky Way's dark matter distribution is generally assumed to be approximately spherically symmetric, allowing us to describe the density as a function of only the distance from the Galactic Center, $r$. Throughout this study, we will consider dark matter distributions described by a generalized Navarro-Frenk-White (NFW) halo profile~\cite{Navarro:1995iw,Navarro:1996gj}:
\begin{equation}
\rho( r)= \rho_0 \frac{(r/r_s)^{-\gamma}}{(1 + r/r_s)^{3-\gamma}}.
\label{gennfw}
\end{equation}  
Throughout this paper, we adopt a scale radius of $r_s =20$ kpc, and select $\rho_0$ such that the local dark matter density (at $8.5\,{\rm kpc}$ from the Galactic Center) is $0.4$ GeV/cm$^3$, consistent with dynamical constraints~\cite{Iocco:2011jz,Catena:2009mf}. Although dark matter-only simulations generally favor inner slopes near the canonical NFW value ($\gamma=1$)~\cite{Navarro:2008kc,Diemand:2008in}, baryonic effects are expected to have a non-negligible impact on the dark matter distribution within the inner $\sim$10 kiloparsecs of the Milky Way~\cite{1986ApJ...301...27B,1987ApJ...318...15R,Gnedin:2011uj,2004ApJ...616...16G,Governato:2012fa,Kuhlen:2012qw,Weinberg:2001gm,Weinberg:2006ps,Sellwood:2002vb,Valenzuela:2002np,Colin:2005rr}.  The magnitude and direction of such baryonic effects, however, are currently a topic of debate. With this in mind, we remain agnostic as to the value of the inner slope, and take $\gamma$ to be a free parameter.

\begin{figure*}[t!]
\includegraphics[width=3.35in]{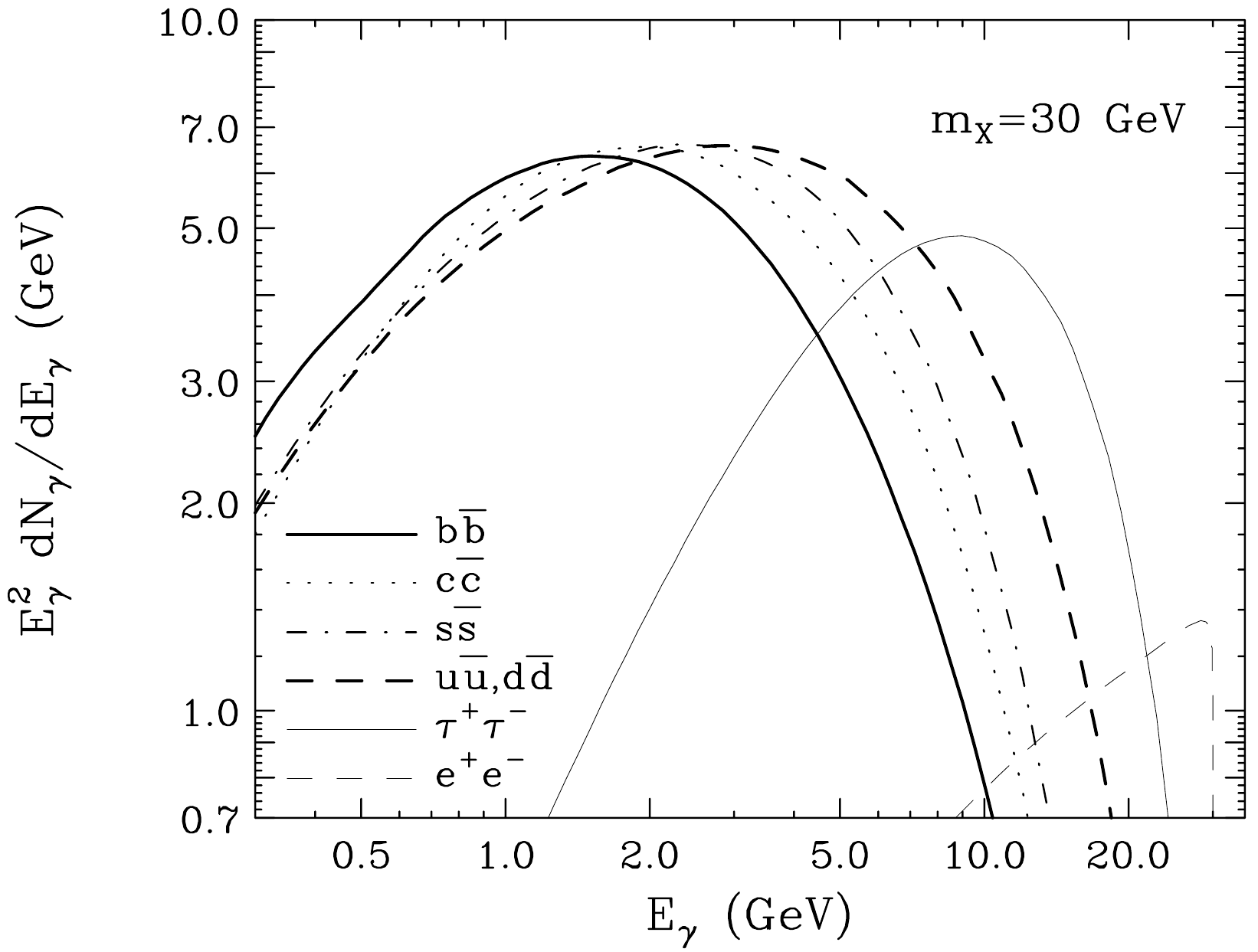}
\hspace{0.2in}
\includegraphics[width=3.35in]{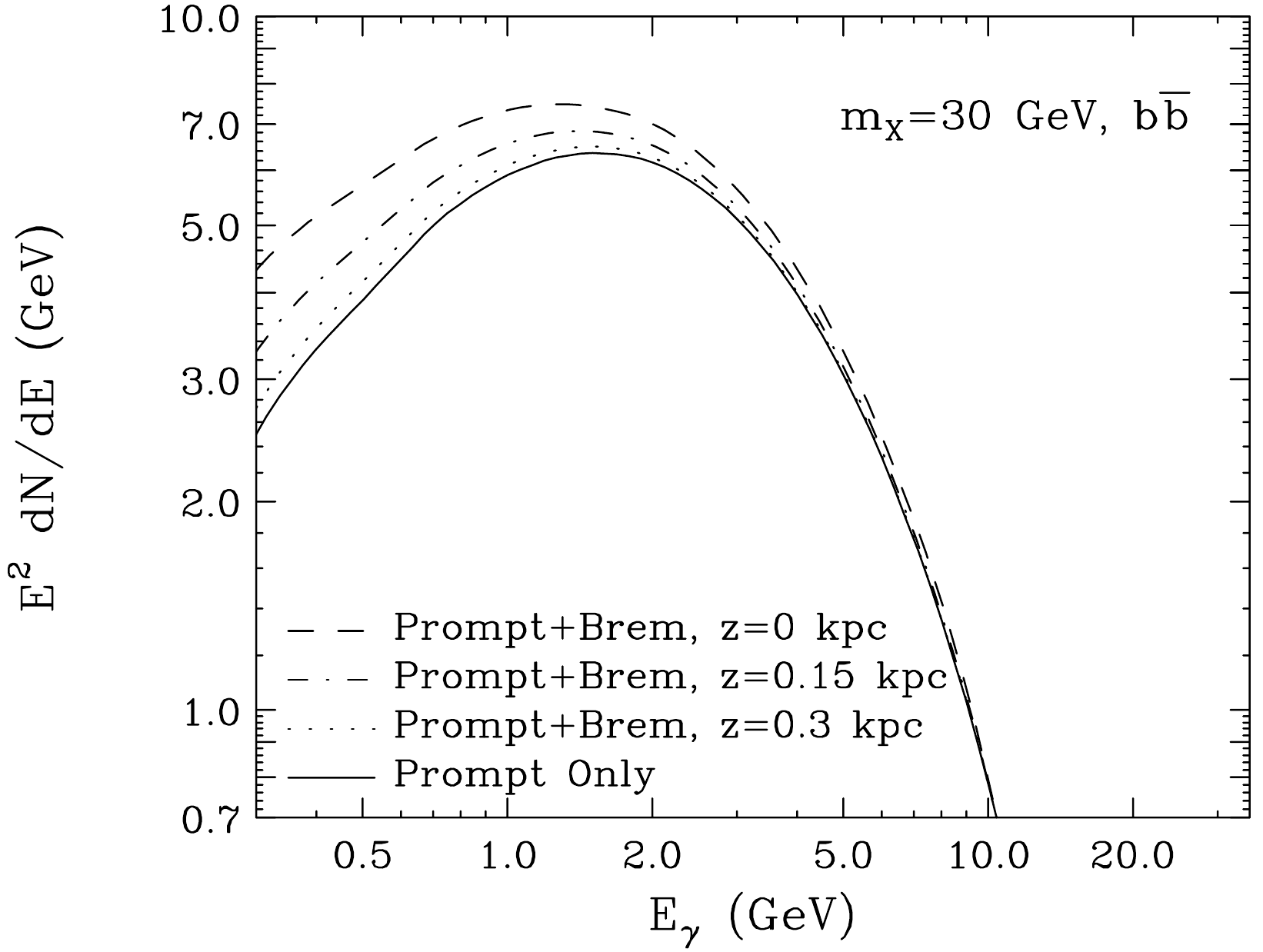}
\caption{Left frame: The spectrum of gamma rays produced per dark matter annihilation for a 30 GeV WIMP mass and a variety of annihilation channels. Right frame: An estimate for the bremsstrahlung emission from the electrons produced in dark matter annihilations taking place near the Galactic Center, for the case of a 30 GeV WIMP annihilating to $b\bar{b}$. At $|z| \lsim 0.3$ kpc ($|b| \lsim 2^{\circ}$) and at energies below $\sim$1-2 GeV, bremsstrahlung could potentially contribute non-negligibly. See text for details.}
\label{dnde}
\end{figure*}

In the left frame of Fig.~\ref{profiles}, we plot the density of dark matter as a function of $r$ for several choices of the halo profile. Along with generalized NFW profiles using three values of the inner slope ($\gamma$=1.0, 1.2, 1.4), we also show for comparison the results for an Einasto profile (with $\alpha=0.17$)~\cite{Springel:2008cc}. In the right frame, we plot the value of the integral in Eq.~\ref{basic} for the same halo profiles, denoted by the quantity, $J(\psi)$:
\begin{equation}
J(\psi)= \int_{\text{los}} \rho^2( r) \, dl,
\label{J}
\end{equation}
where $\psi$ is the angle observed away from the Galactic Center. In the NFW case (with $\gamma=1$), for example, the value of $J$ averaged over the inner degree around the Galactic Center exceeds that of the most promising dwarf spheroidal galaxies by a factor of $\sim$$50$~\cite{Ackermann:2013yva}. If the Milky Way's dark matter halo is contracted by baryons or is otherwise steeper than predicted by NFW, this ratio could easily be $\sim$$10^3$ or greater.

\begin{figure*}[t!]
\includegraphics[width=7.15in]{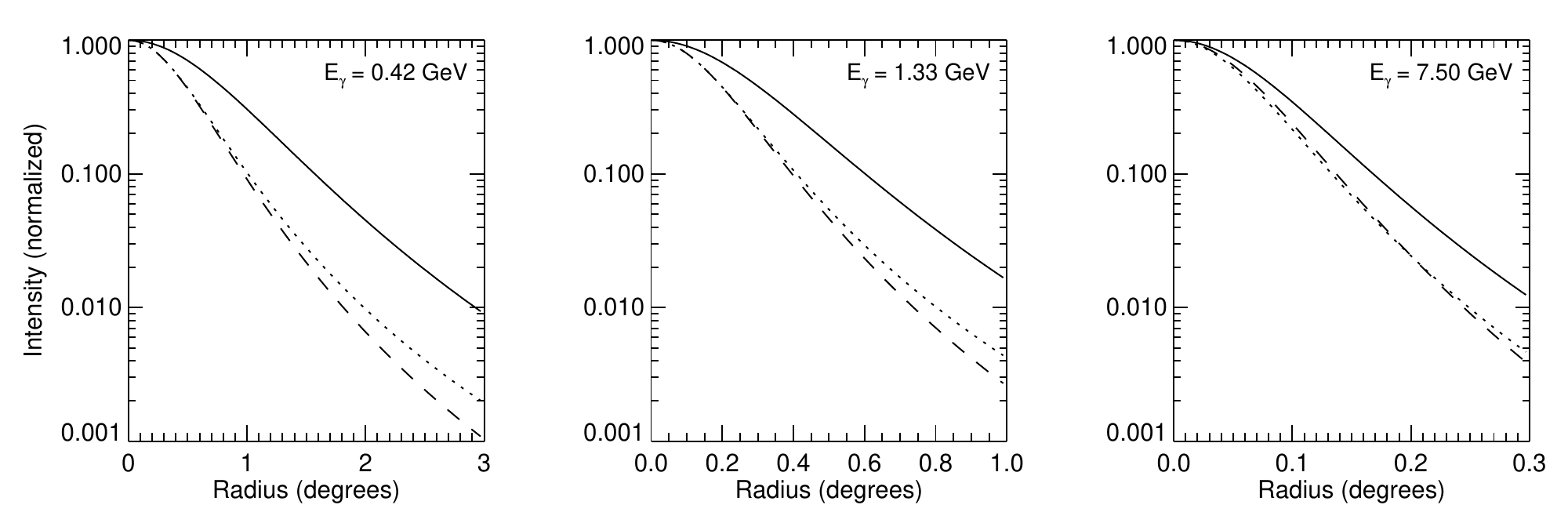}
\caption{The point spread function (PSF) of the \textit{Fermi} Gamma-Ray Space Telescope, for front-converting, Ultraclean class events. The solid lines represent the PSF for the full dataset, using the \textit{Fermi} Collaboration's default cuts on the parameter CTBCORE. The dotted and dashed lines, in contrast, denote the PSFs for the top two quartiles (Q2) and top quartile (Q1) of these events, respectively, as ranked by CTBCORE. See text for details.}
\label{psf}
\end{figure*}

The spectrum of gamma rays produced per dark matter annihilation, $dN_{\gamma}/dE_{\gamma}$, depends on the mass of the dark matter particle and on the types of particles produced in this process.  In the left frame of Fig.~\ref{dnde}, we plot $dN_{\gamma}/dE_{\gamma}$ for the case of a 30 GeV WIMP mass, and for a variety of annihilation channels (as calculated using PYTHIA~\cite{Sjostrand:2006za}, except for the $e^+ e^-$ case, for which the final state radiation was calculated analytically~\cite{Bergstrom:2004cy,Birkedal:2005ep}). In each case, a distinctive bump-like feature appears, although at different energies and with different widths, depending on the final state.

In addition to prompt gamma rays, dark matter annihilations can produce electrons and positrons which subsequently generate gamma rays via inverse Compton and bremsstrahlung processes. For dark matter annihilations taking place near the Galactic Plane, the low-energy gamma-ray spectrum can receive a non-negligible contribution from bremsstrahlung. In the right frame of Fig.~\ref{dnde}, we plot the gamma-ray spectrum from dark matter (per annihilation), including an estimate for the bremsstrahlung contribution. In estimating the contribution from bremsstrahlung, we neglect diffusion, but otherwise follow the calculation of Ref.~\cite{Cirelli:2013mqa}. In particular, we consider representative values of $\langle B \rangle =10 \, \mu$G for the magnetic field, and 10 eV$/$cm$^3$ for the radiation density throughout the region of the Galactic Center. For the distribution of gas, we adopt a density of 10 particles per cm$^3$ near the Galactic Plane ($z=0$), with a dependence on $z$ given by $\exp(-|z|/0.15 \, {\rm kpc})$. Within $\sim$$1^{\circ}$--\,$2^{\circ}$ of the Galactic Plane, we find that bremsstrahlung could potentially contribute non-negligibly to the low energy ($\lsim$\,1--2 GeV) gamma-ray spectrum from annihilating dark matter.


\section{Making Higher Resolution Gamma-Ray Maps with CTBCORE}
\label{ctbcore}

In most analyses of \textit{Fermi} data, one makes use of all of the events within a given class (Transient, Source, Clean, or Ultraclean). Each of these event classes reflects a different trade-off between the effective area and the efficiency of cosmic-ray rejection. Higher quality event classes also allow for somewhat greater angular resolution (as quantified by the point spread function, PSF). The optimal choice of event class for a given analysis depends on the nature of the signal and background in question.  The Ultraclean event class, for example, is well suited to the study of large angular regions, and to situations where the analysis is sensitive to spectral features that might be caused by cosmic ray backgrounds. The Transient event class, in contrast, is best suited for analyses of short duration events, with little background. Searches for dark matter annihilation products from the Milky Way's halo significantly benefit from the high background rejection and angular resolution of the Ultraclean class and thus can potentially fall into the former category.

As a part of event reconstruction, the \textit{Fermi} Collaboration estimates the accuracy of the reconstructed direction of each event. Inefficiencies and inactive regions within the detector reduce the quality of 
the information available for certain events. Factors such as whether an event is front-converting or back-converting, 
whether there are multiple tracks that can be combined into a vertex, and the amount of energy 
deposited into the calorimeter each impact the reliability of the reconstructed direction~\cite{Atwood:2009ez}. 

In their most recent public data releases, the \textit{Fermi} Collaboration has begun to include a greater body of information about each event, including a value for the parameter CTBCORE, which quantifies the reliability of the directional reconstruction. By selecting only events with a high value of CTBCORE, one can reduce the tails of the PSF, although at the expense of effective area~\cite{Atwood:2009ez}.

For this study, we have created a set of new event classes by increasing the CTBCORE cut from the default values used by the \textit{Fermi} Collaboration. To accomplish this, we divided all front-converting, Ultraclean events (Pass 7, Reprocessed) into quartiles, ranked by CTBCORE. Those events in the top quartile make up the event class Q1, while those in the top two quartiles make up Q2, etc. For each new event class, we calibrate the on-orbit PSF~\cite{Ackermann:2012kna,Ackermann:2013aaa} using the Geminga pulsar. Taking advantage of Geminga's pulsation, we remove the background by taking the difference between the on-phase and off-phase images. We fit the PSF in each energy bin by a single King function, and smooth the overall PSF with energy. We also rescale \textit{Fermi}'s effective area according to the fraction of events that are removed by the CTBCORE cut, as a function of energy and incidence angle.

\begin{figure*}[t!]
\includegraphics[width=3.5in]{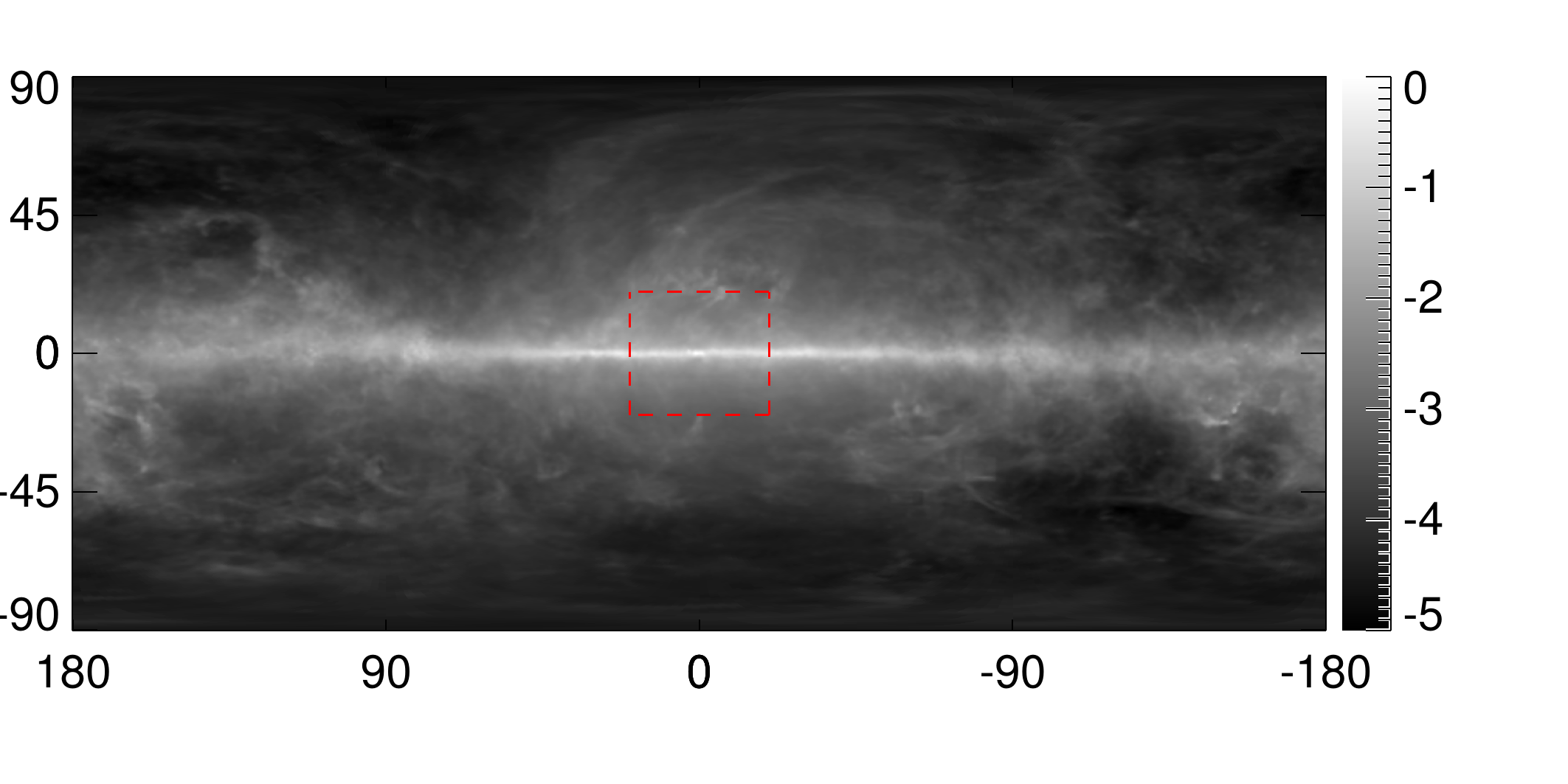}
\includegraphics[width=3.5in]{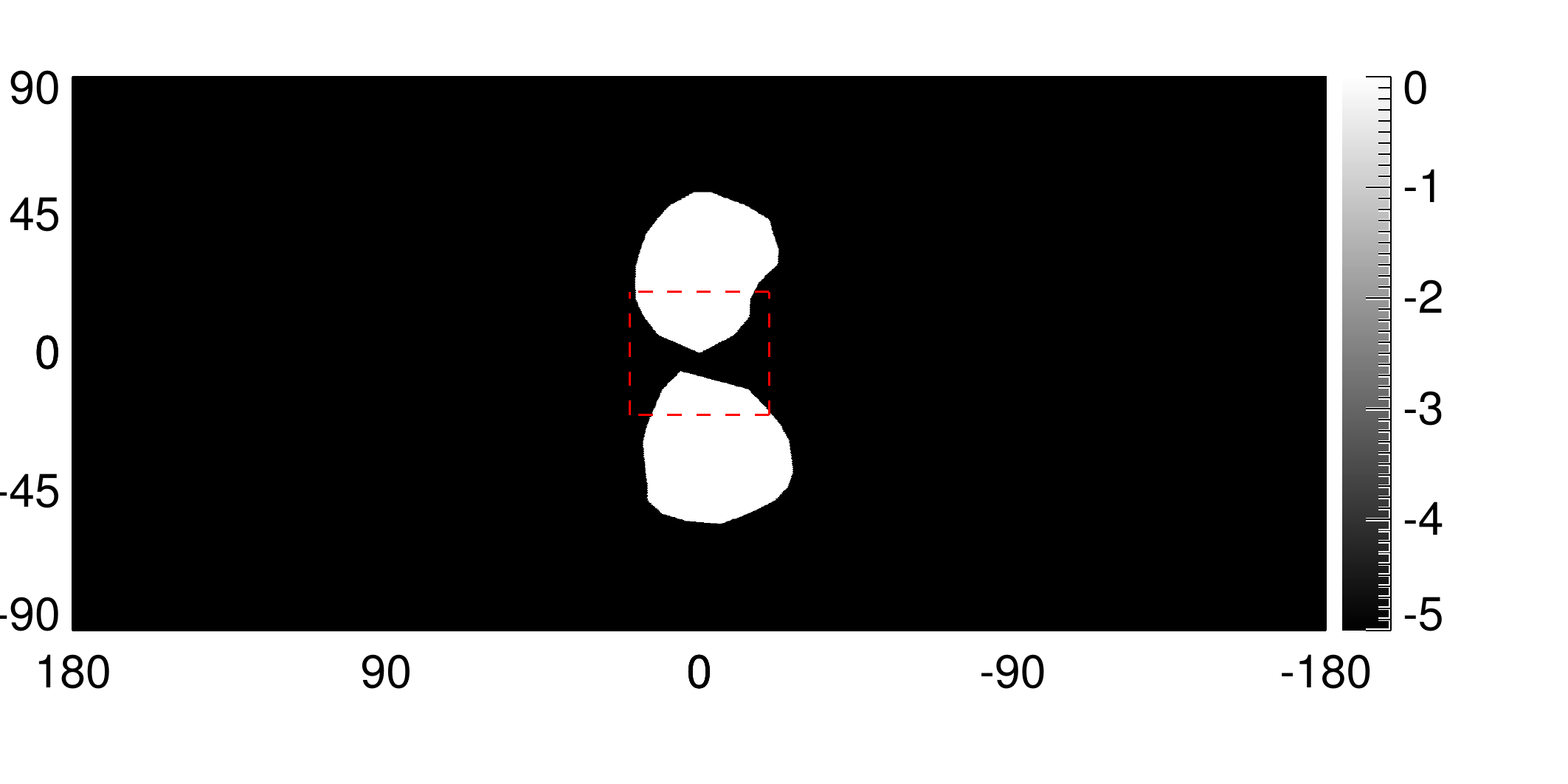}
\vspace{-0.1in}
\includegraphics[width=3.5in]{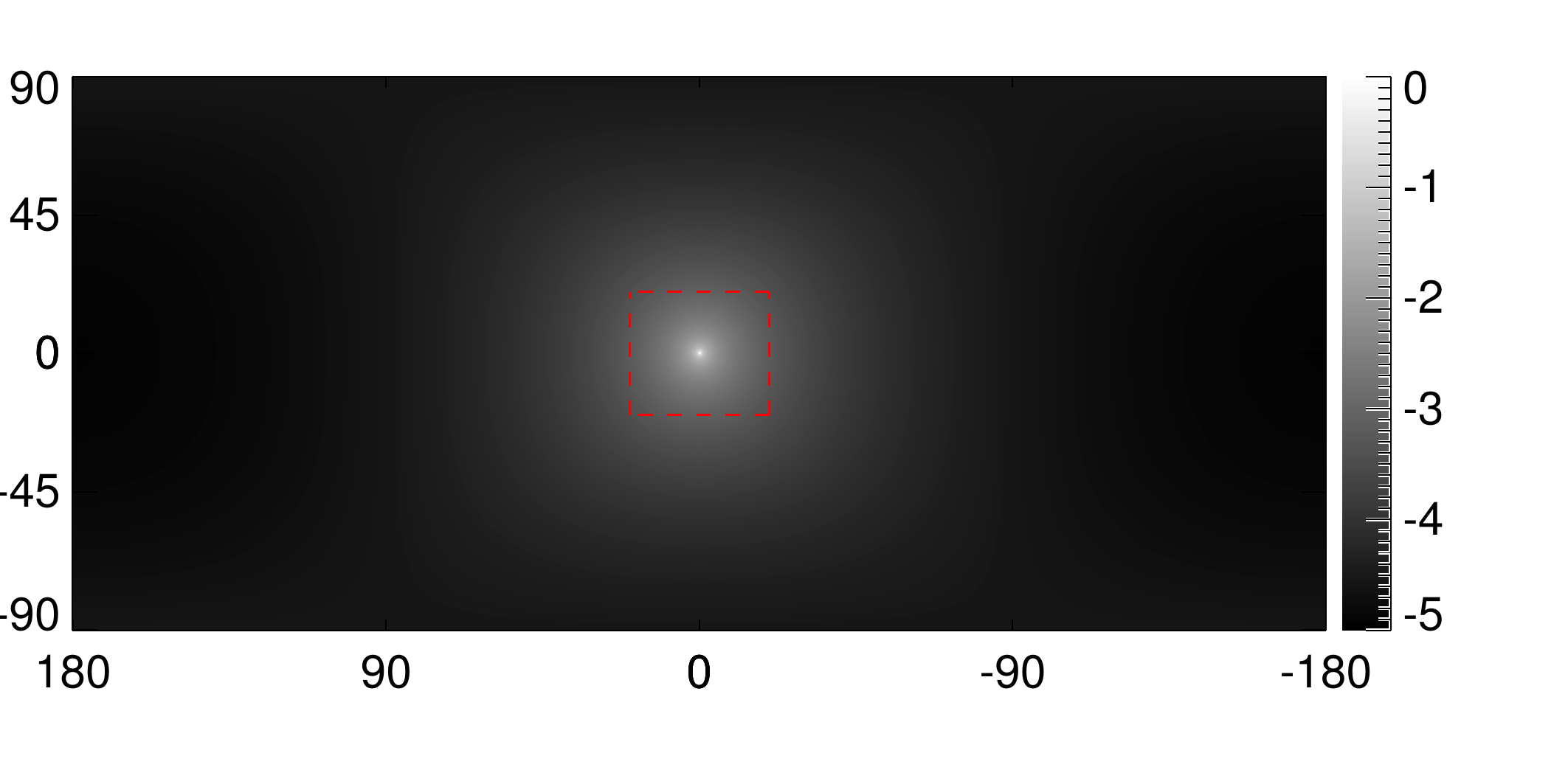}
\caption{The spatial templates (in galactic coordinates) for the Galactic diffuse model (upper left), the \textit{Fermi} bubbles (upper right), and dark matter annihilation products (lower), as used in our Inner Galaxy analysis. The scale is logarithmic (base 10), normalized to the brightest point in each map. The diffuse model template is shown as evaluated at 1 GeV, and the dark matter template corresponds to a generalized NFW profile with an inner slope of $\gamma=1.18$. Red dashed lines indicate the boundaries of our standard Region of Interest (we also mask bright point sources and the region of the Galactic plane with $|b| < 1^\circ$).}
\label{templates}
\end{figure*}

These cuts on CTBCORE have a substantial impact on \textit{Fermi}'s PSF, especially at low energies. In Fig.~\ref{psf}, we show the PSF for front-converting, Ultraclean events, at three representative energies, for different cuts on CTBCORE (all events, Q2, and Q1). 
%
%
%
Such a cut can be used to mitigate the leakage of astrophysical emission from the Galactic Plane and point sources into our regions of interest. This leakage is most problematic at low energies, where the PSF is quite broad and where the CTBCORE cut has the greatest impact. These new event classes and their characterization are further detailed in \cite{Portillo:2014aaa}, and accompanied by a data release of all-sky maps 
for each class, and the instrument response function files necessary for use with the \textit{Fermi} Science Tools. 

Throughout the remainder of this study, we will employ the Q2 event class by default, corresponding to the top 50\% (by CTBCORE) of \textit{Fermi}'s front-converting, Ultraclean photons, to maximize event quality. We select Q2 rather than Q1 to improve statistics, since as demonstrated in Fig.~\ref{psf}, the angular resolution improvement in moving from Q2 to Q1 is minimal. In Appendix \ref{app:consistency} we demonstrate that our results are stable upon removing the CTBCORE cut (thus doubling the dataset), or expanding the dataset to include lower-quality events.\footnote{An earlier version of this work found a number of apparent peculiarities in the results without the CTBCORE cut that were removed on applying the cut. However, we now attribute those peculiarities to an incorrect smoothing of the diffuse background model. When the background model is smoothed correctly, we find results that are much more stable to the choice of CTBCORE cut, and closely resemble the results previously obtained with Q2 events. Accordingly, the CTBCORE cut appears to be effective at separating signal from poorly-modeled background emission, but is less necessary when the background is well-modeled.}


\section{The Inner Galaxy}
\label{inner}

In this section, we follow the procedure previously pursued in Ref.~\cite{Hooper:2013rwa} (see also Refs.~\cite{Dobler:2009xz,Su:2010qj}) to study the gamma-ray emission from the Inner Galaxy. We use the term ``Inner Galaxy'' to denote the region of the sky that lies within several tens of degrees around the Galactic Center, excepting the Galactic Plane itself ($|b|<1^{\circ}$), which we mask in this portion of our analysis. 

Throughout our analysis, we make use of the Pass 7 (V15) reprocessed data taken between August 4, 2008 and December 5, 2013, using only front-converting, Ultraclean class events which pass the Q2 CTBCORE cut as described in Sec.~\ref{ctbcore}.  We also apply standard cuts to ensure data quality (zenith angle $<100^{\circ}$, instrumental rocking angle $<52^{\circ}$, \texttt{DATA\_QUAL} = 1, \texttt{LAT\_CONFIG}=1). Using this data set, we have generated a series of maps of the gamma-ray sky binned in energy. We apply the point source subtraction method described in Ref.~\cite{Su:2010qj}, updated to employ the 2FGL catalogue, and masking out the 300 brightest and most variable sources at a mask radius corresponding to $95\%$ containment. We then perform a pixel-based maximum likelihood analysis on the map, fitting the data in each energy bin to a sum of spatial templates. These templates consist of: 1) the \emph{Fermi} Collaboration \texttt{p6v11} Galactic diffuse model (which we refer to as the Pass 6 Diffuse Model),\footnote{Unlike more recently released Galactic diffuse models, the \texttt{p6v11} diffuse model does not implicitly include a component corresponding to the Fermi Bubbles. By using this model, we are free to fit the Fermi Bubbles component independently. See Appendix \ref{app:diffuse} for a discussion of the impact of varying the diffuse model.} 2) an isotropic map, intended to account for the extragalactic gamma-ray background and residual cosmic-ray contamination, and 3) a uniform-brightness spatial template coincident with the features known as the Fermi Bubbles, as described in Ref.~\cite{Su:2010qj}. In addition to these three background templates, we include an additional dark matter template, motivated by the hypothesis that the previously reported gamma-ray excess originates from annihilating dark matter. In particular, our dark matter template is taken to be proportional to the line-of-sight integral of the dark matter density squared, $J(\psi)$, for a generalized NFW density profile (see Eqs.~\ref{gennfw}--\ref{J}). The spatial morphology of the Galactic diffuse model (as evaluated at 1 GeV), Fermi Bubbles, and dark matter templates are each shown in Fig.~\ref{templates}.

We smooth the Galactic diffuse model template to match the data using the \texttt{gtsrcmaps} routine in the Fermi Science Tools, to ensure that the tails of the point spread function are properly taken into account.\footnote{We checked the impact of smoothing the diffuse model with a Gaussian and found no significant impact on our results.} Because the Galactic diffuse model template is much brighter than the other contributions in the region of interest, relatively small errors in its smoothing could potentially bias our results. However, the other templates are much fainter, and so we simply perform a Gaussian smoothing, with a FWHM matched to the FWHM of the \emph{Fermi} PSF at the minimum energy for the bin (since most of the counts are close to this minimum energy).

\begin{figure}[t!]
\includegraphics[width=3in]{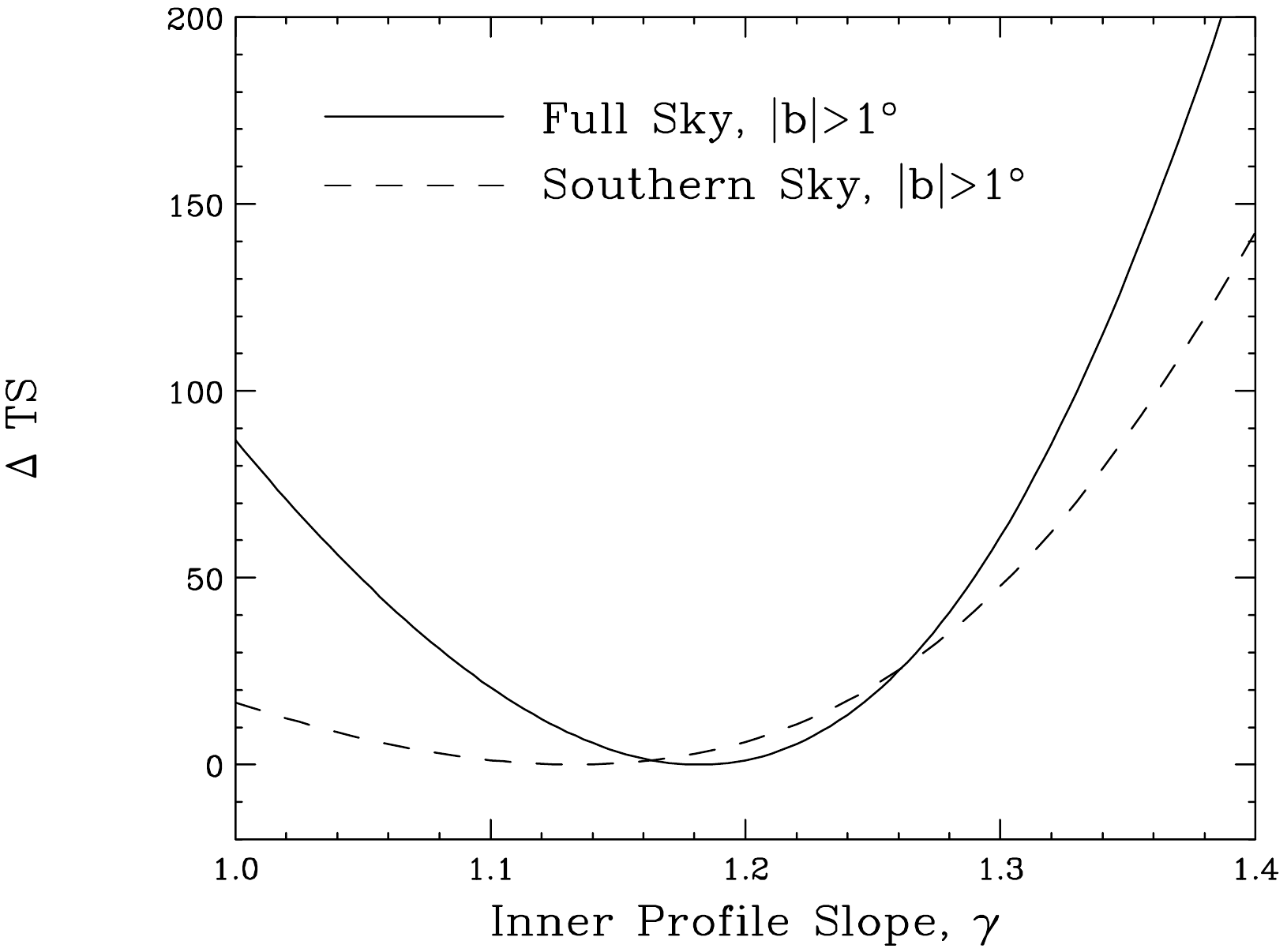}
\caption{The variation in the quantity $-2 \Delta \ln {\mathcal L}$ (referred to as TS) extracted from the likelihood fit, as a function of the inner slope of the dark matter halo profile, $\gamma$. All values are relative to the result for the best-fit (highest TS) template, and positive values thus indicate a reduction in TS. Results are shown using gamma-ray data from the full sky (solid line) and only the southern sky (dashed line). Unlike in the analysis of Ref.~\cite{Hooper:2013rwa}, we do not find any large north-south asymmetry in the preferred value of $\gamma$.}
\label{innerslope}
\end{figure}

\begin{figure*}[t!]
\includegraphics[width=3.5in]{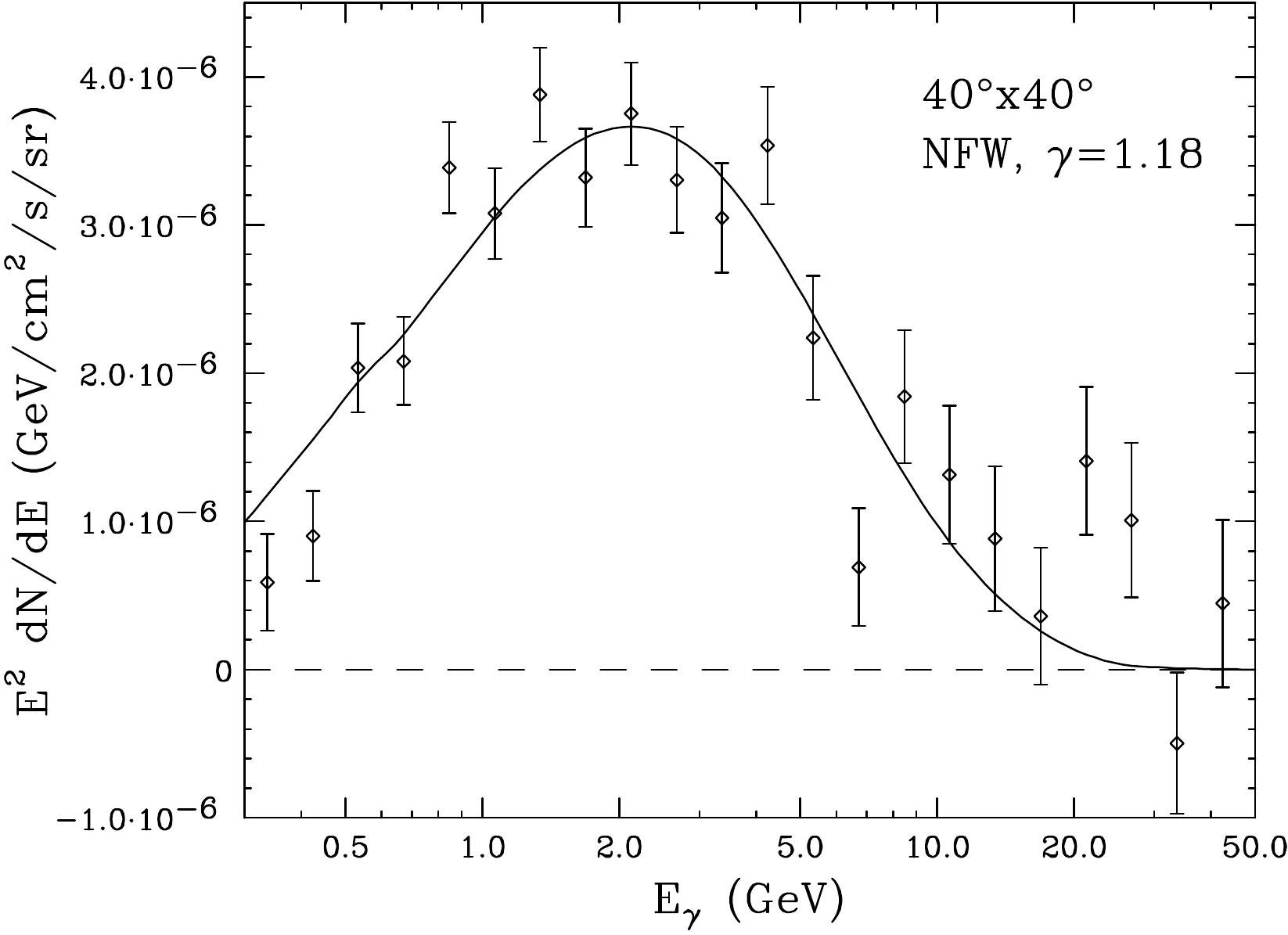}
\includegraphics[width=3.5in]{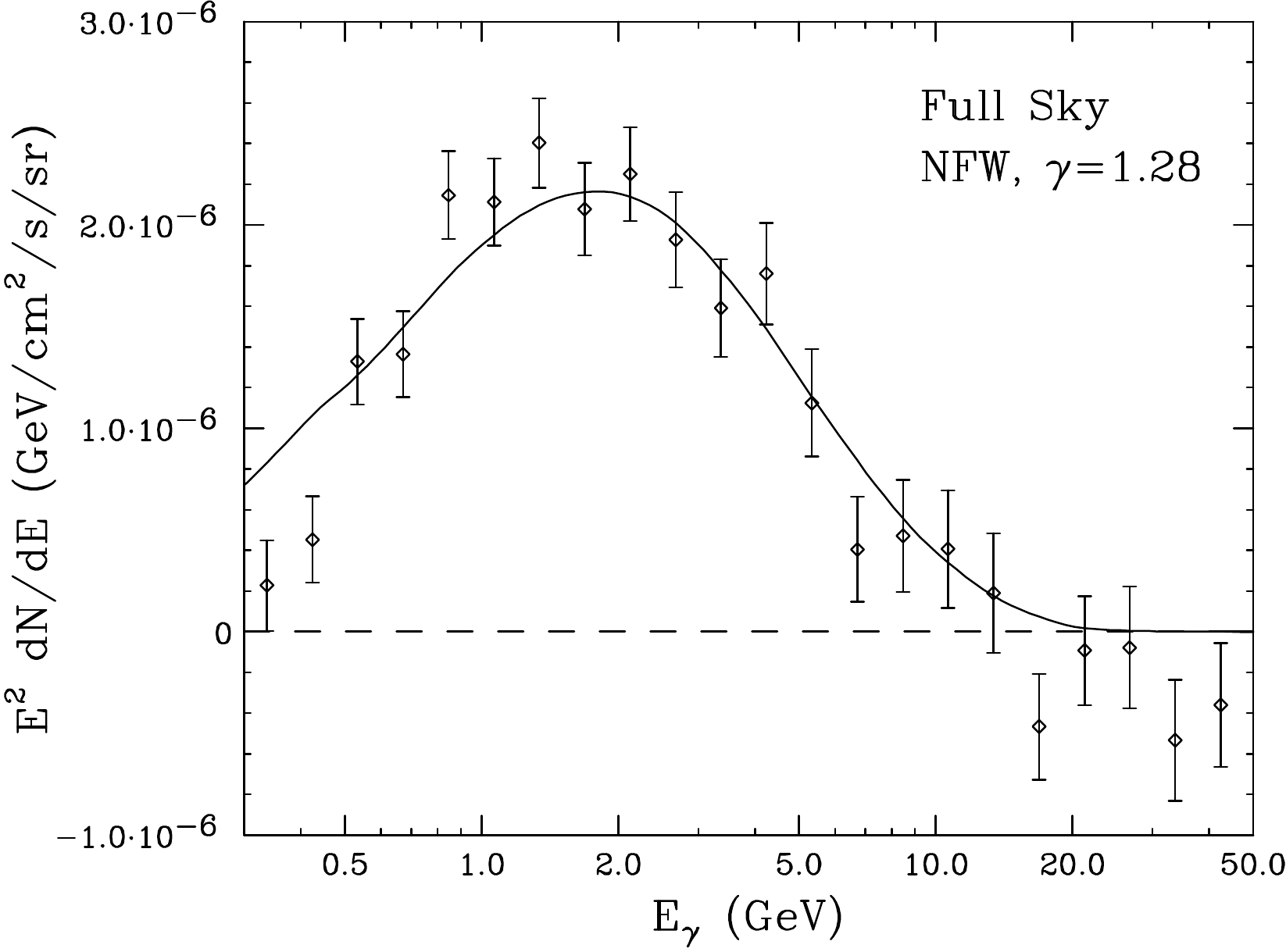}
\caption{Left frame:  The spectrum of the dark matter component, extracted from a fit in our standard ROI ($1^\circ < |b| < 20^\circ$, $|l| < 20^\circ$) for a template corresponding to a generalized NFW halo profile with an inner slope of $\gamma=1.18$ (normalized to the flux at an angle of 5$^{\circ}$ from the Galactic Center). Shown for comparison (solid line) is the spectrum predicted from a 43.0 GeV dark matter particle annihilating to $b\bar{b}$ with a cross section of $\sigma v = 2.25\times 10^{-26}$ cm$^3$/s $\, \times \, [(0.4 \,{\rm GeV}/{\rm cm}^3)/\rho_{\rm local}]^2$. Right frame: as left frame, but for a full-sky ROI ($|b| > 1^\circ$), with $\gamma=1.28$; shown for comparison (solid line) is the spectrum predicted from a 36.6 GeV dark matter particle annihilating to $b\bar{b}$ with a cross section of $\sigma v = 0.75\times 10^{-26}$ cm$^3$/s $\, \times \, [(0.4 \,{\rm GeV}/{\rm cm}^3)/\rho_{\rm local}]^2$.}
\label{innerspec}
\end{figure*}

\begin{figure*}[t!]
\hspace{0.5cm}
\includegraphics[width=6.5in]{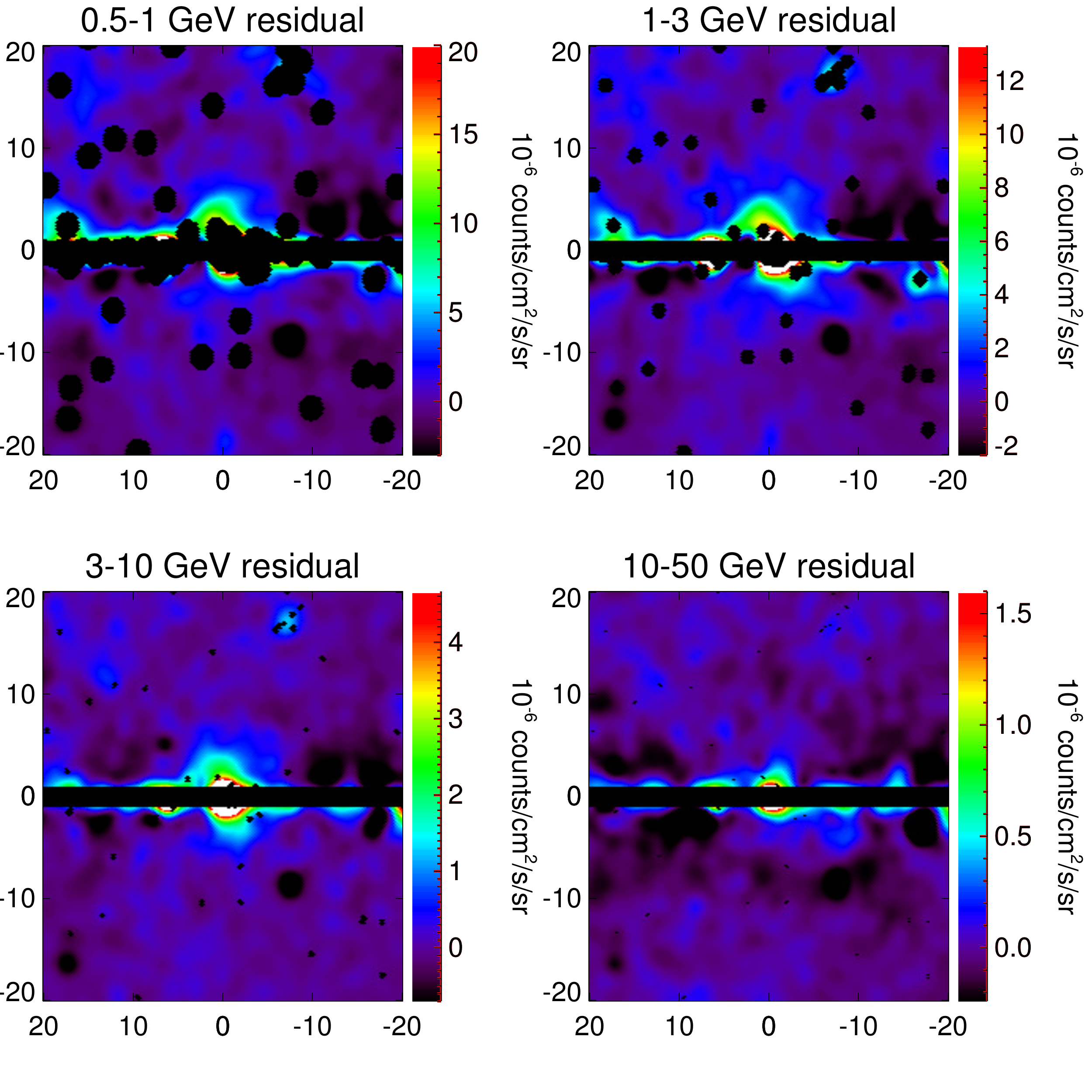}
\vspace{-0.5cm}
\caption{Intensity maps (in galactic coordinates) after subtracting the point source model and best-fit Galactic diffuse model, \textit{Fermi} bubbles, and isotropic templates. Template coefficients are obtained from the fit including these three templates and a $\gamma=1.3$ DM-like template. Masked pixels are indicated in black. All maps have been smoothed to a common PSF of 2 degrees for display, before masking (the corresponding masks have \emph{not} been smoothed; they reflect the actual masks used in the analysis). At energies between $\sim$0.5-10 GeV ({\it i.e.}~in the first three frames), the dark-matter-like emission is clearly visible around the Galactic Center.}
\label{innerresidual}
\end{figure*}

By default, we employ a Region of Interest (ROI) of $|\ell| < 20^\circ$, $1^\circ < |b| < 20^\circ$. An earlier version of this work used the full sky (with the plane masked at 1 degree) as the default ROI; we find that restricting to a smaller ROI alleviates oversubtraction in the inner Galaxy and improves the stability of our results.\footnote{This approach was in part inspired by the work presented in Ref.~\cite{Calore:2014xka}.} Thus we present ``baseline'' results for the smaller region, but show the impact of changing the ROI in Appendix \ref{app:consistency}, and in selected figures in the main text. Where we refer to the ``full sky'' analysis the Galactic plane is masked for $|b| < 1^\circ$ unless noted otherwise.

As found in previous studies~\cite{Hooper:2013rwa,Huang:2013pda}, the inclusion of the dark matter template dramatically improves the quality of the fit to the \textit{Fermi} data. For the best-fit spectrum and halo profile, we find that the inclusion of the dark matter template improves the formal fit by TS$\equiv -2 \Delta \ln \mathcal{L} \simeq 1100$ (here TS stands for ``test statistic''), corresponding to a statistical preference greater than 30$\sigma$. When considering this enormous statistical significance, one should keep in mind that in addition to statistical errors there is a degree of unavoidable and unaccounted-for systematic error, in that neither model (with or without a dark matter component) is a ``good fit'' in the sense of describing the sky to the level of Poisson noise. That being said, the data do very strongly prefer the presence of a gamma-ray component with a morphology similar to that predicted from annihilating dark matter (see Appendices \ref{app:consistency}-\ref{app:gc} for further details).

As in Ref.~\cite{Hooper:2013rwa}, we vary the value of the inner slope of the generalized NFW profile, $\gamma$, and compare the change in the log-likelihood, $\Delta \ln \mathcal{L}$, between the resulting fits in order to determine the preferred range for the value of $\gamma$.\footnote{Throughout, we describe the improvement in $-2 \Delta \ln \mathcal{L}$ induced by inclusion of a specific template as the ``test statistic'' or TS for that template.} The results of this exercise are shown in Fig.~\ref{innerslope}. We find that our default ROI has a best-fit value of $\gamma=1.18$, consistent with previous studies of the inner Galaxy (which did not employ any additional cuts on CTBCORE) that preferred an inner slope of $\gamma \simeq 1.2$~\cite{Hooper:2013rwa}. Fitting over the full sky, we find a preference for a slightly steeper value of $\gamma \simeq 1.28$. These results are quite stable to our mask of the Galactic plane; masking the region with $|b| < 2^\circ$ changes the preferred value to $\gamma=1.25$ in our default ROI, and $\gamma=1.29$ over the whole sky. In contrast to Ref.~\cite{Hooper:2013rwa}, we find no significant difference in the slope preferred by the fit over the standard ROI, and by a fit only over the southern half ($b<0$) of the ROI (we also find no significant difference between the fit over the full sky and the southern half of the full sky). This can be seen directly from Fig.~\ref{innerslope}, where the full-sky and southern-sky fits for the same level of masking are found to favor quite similar values of $\gamma$ (the southern sky distribution is broader than that for the full sky simply due to the difference in the number of photons). The best-fit values for gamma, from fits in the southern half of the standard ROI and the southern half of the full sky, are 1.13 and 1.26 respectively.

In Fig.~\ref{innerspec}, we show the spectrum of the emission correlated with the dark matter template in the default ROI and full-sky analysis, for their respective best-fit values of $\gamma=1.18$ and 1.28.\footnote{A  comparison between the two ROIs with $\gamma$ held constant is presented in Appendix \ref{app:consistency}.} We restrict to energies 50 GeV and lower to ensure numerical stability of the fit in the smaller ROI. While no significant emission is absorbed by this template at energies above $\sim$10 GeV, a bright and robust component is present at lower energies, peaking near $\sim$1-3 GeV. Relative to the analysis of Ref.~\cite{Hooper:2013rwa} (which used an incorrectly smoothed diffuse model), our spectrum is in both cases significantly harder at energies below 1 GeV, rendering it more consistent with that extracted at higher latitudes (see Appendix A).\footnote{An earlier version of this work found this improvement only in the presence of the CTBCORE cut; we now find this hardening independent of the CTBCORE cut.} Shown for comparison (as a solid line) is the spectrum predicted from (left panel) a 43.0 GeV dark matter particle annihilating to  $b\bar{b}$ with a cross section of $\sigma v = 2.25 \times 10^{-26}$ cm$^3$/s $\, \times \, [(0.4 \,{\rm GeV}/{\rm cm}^3)/\rho_{\rm local}]^2$, and (right panel) a 36.6 GeV dark matter particle annihilating to $b\bar{b}$ with a cross section of $\sigma v = 0.75 \times 10^{-26}$ cm$^3$/s $\, \times \, [(0.4 \,{\rm GeV}/{\rm cm}^3)/\rho_{\rm local}]^2$. The spectra extracted for this component are in moderately good agreement with the predictions of the dark matter models, yielding fits of $\chi^2=44$ and $64$ over the 22 error bars between 0.3 and 50 GeV. We emphasize that these uncertainties (and the resulting $\chi^2$ values) are purely statistical, and there are significant systematic uncertainties which are not accounted for here (see the discussion in the appendices). We also note that the spectral shape of the dark matter template is quite robust to variations in $\gamma$, within the range where good fits are obtained (see Appendix \ref{app:consistency}).

In Fig.~\ref{innerresidual}, we plot the maps of the gamma-ray sky in four energy ranges after subtracting the best-fit diffuse model, \textit{Fermi} Bubbles, and isotropic templates. In the 0.5-1 GeV, 1-3 GeV, and 3-10 GeV maps, the dark-matter-like emission is clearly visible in the region surrounding the Galactic Center. Much less central emission is visible at 10-50 GeV, where the dark matter component is absent, or at least significantly less bright.

\begin{figure*}[t!]
\includegraphics[width=3.40in]{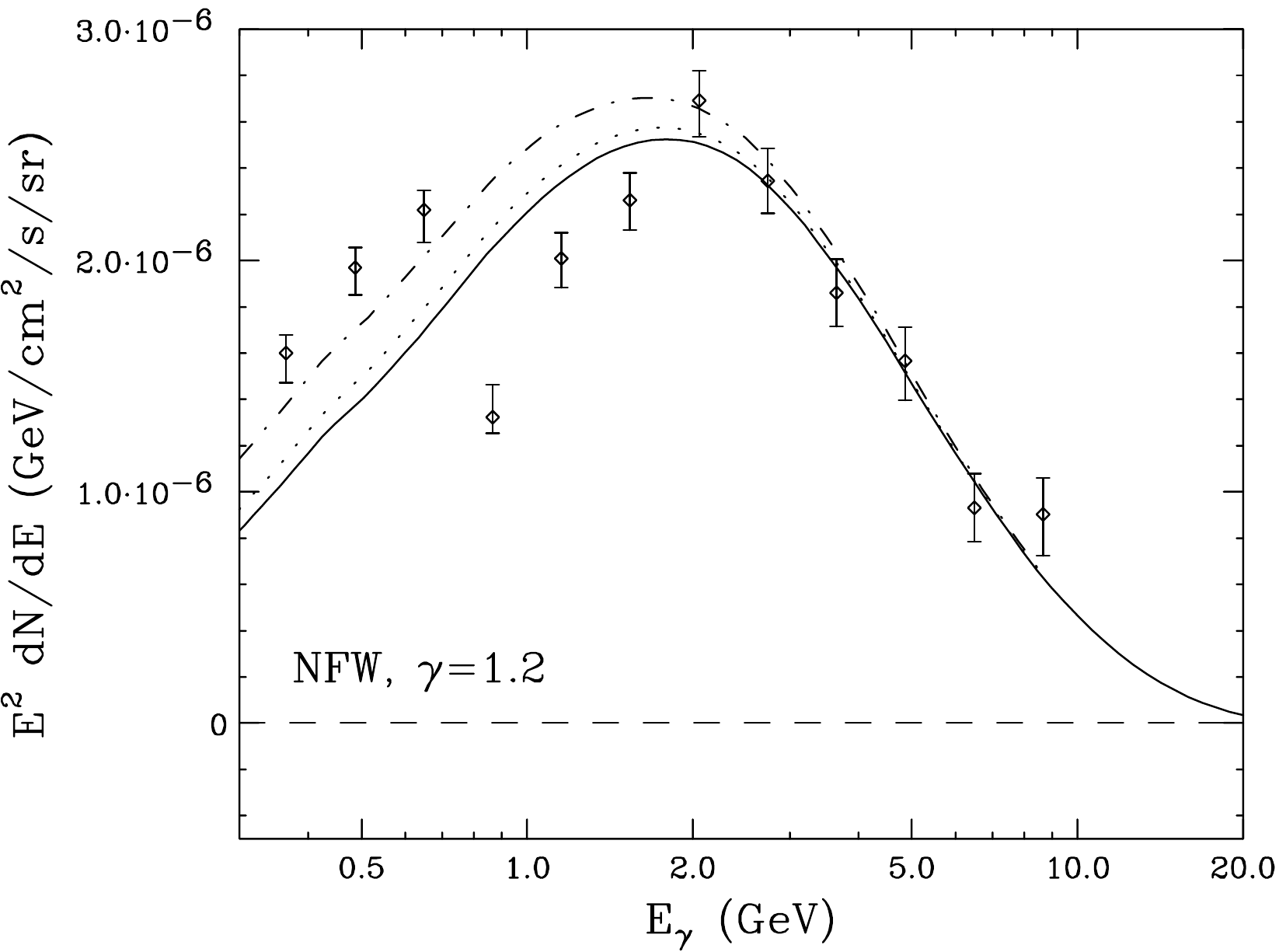}
\hspace{0.1in}
\includegraphics[width=3.40in]{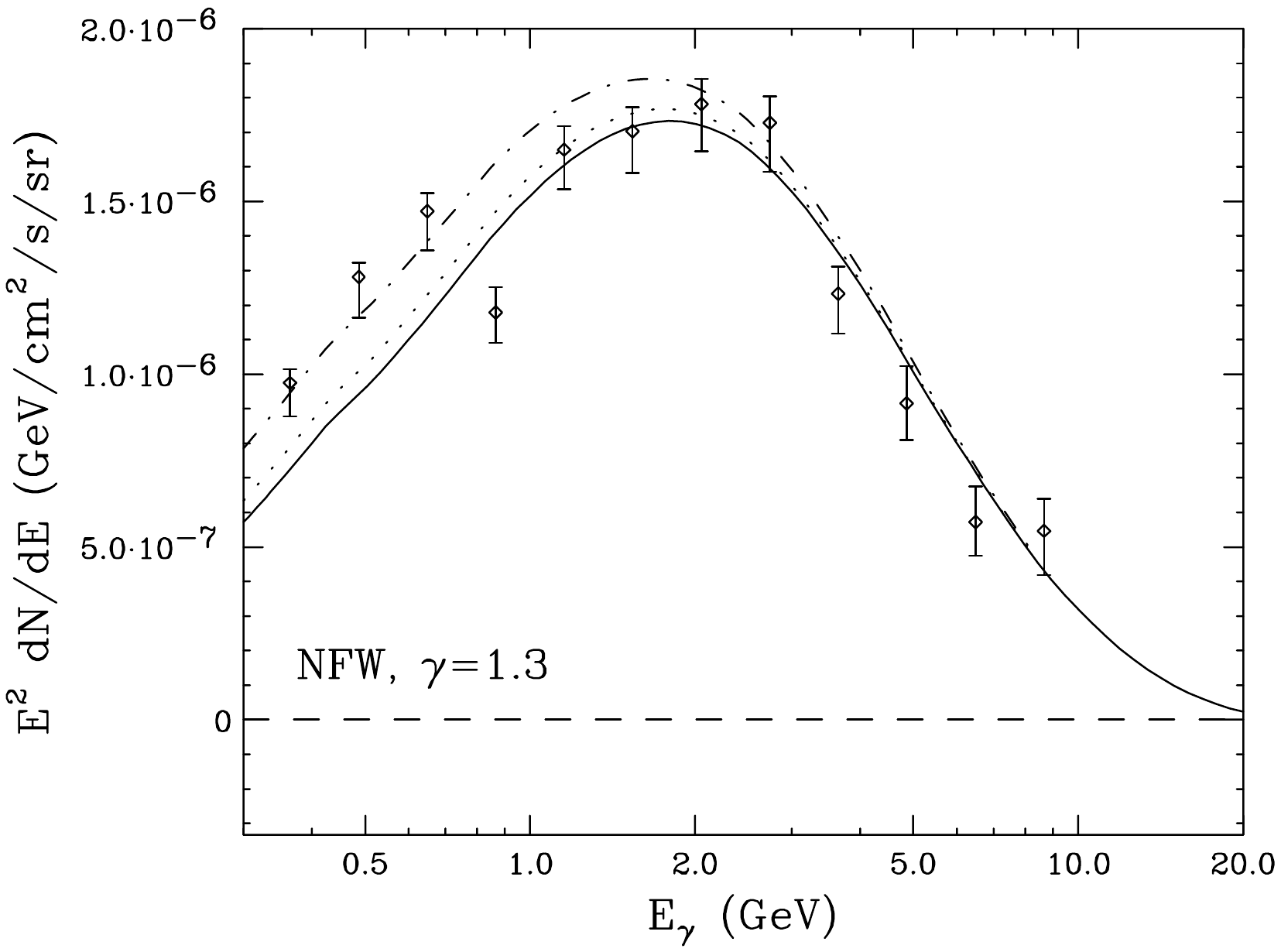}
\caption{The spectrum of the dark matter component derived in our Galactic Center analysis, for a template corresponding to an NFW halo profile with an inner slope of $\gamma=1.2$ (left) or 1.3 (right), normalized to the flux at an angle of 5$^{\circ}$ from the Galactic Center. We caution that significant and difficult to estimate systematic uncertainties exist in this determination, especially at energies below $\sim$1 GeV. Shown for comparison (solid line) is the spectrum predicted from a 35.25 GeV dark matter particle annihilating to $b\bar{b}$ with a cross section of $\sigma v = 1.21\times 10^{-26}$ cm$^3$/s $\, \times \, [(0.4 \,{\rm GeV}/{\rm cm}^3)/\rho_{\rm local}]^2$ (left) or $\sigma v = 0.56\times 10^{-26}$ cm$^3$/s $\, \times \, [(0.4 \,{\rm GeV}/{\rm cm}^3)/\rho_{\rm local}]^2$ (right). The dot-dash and dotted curves include an estimated contribution from bremsstrahlung, as shown in the right frame of Fig.~\ref{dnde}.}
\label{timspec}
\end{figure*}

\begin{figure}[t!]
\includegraphics[width=3.4in]{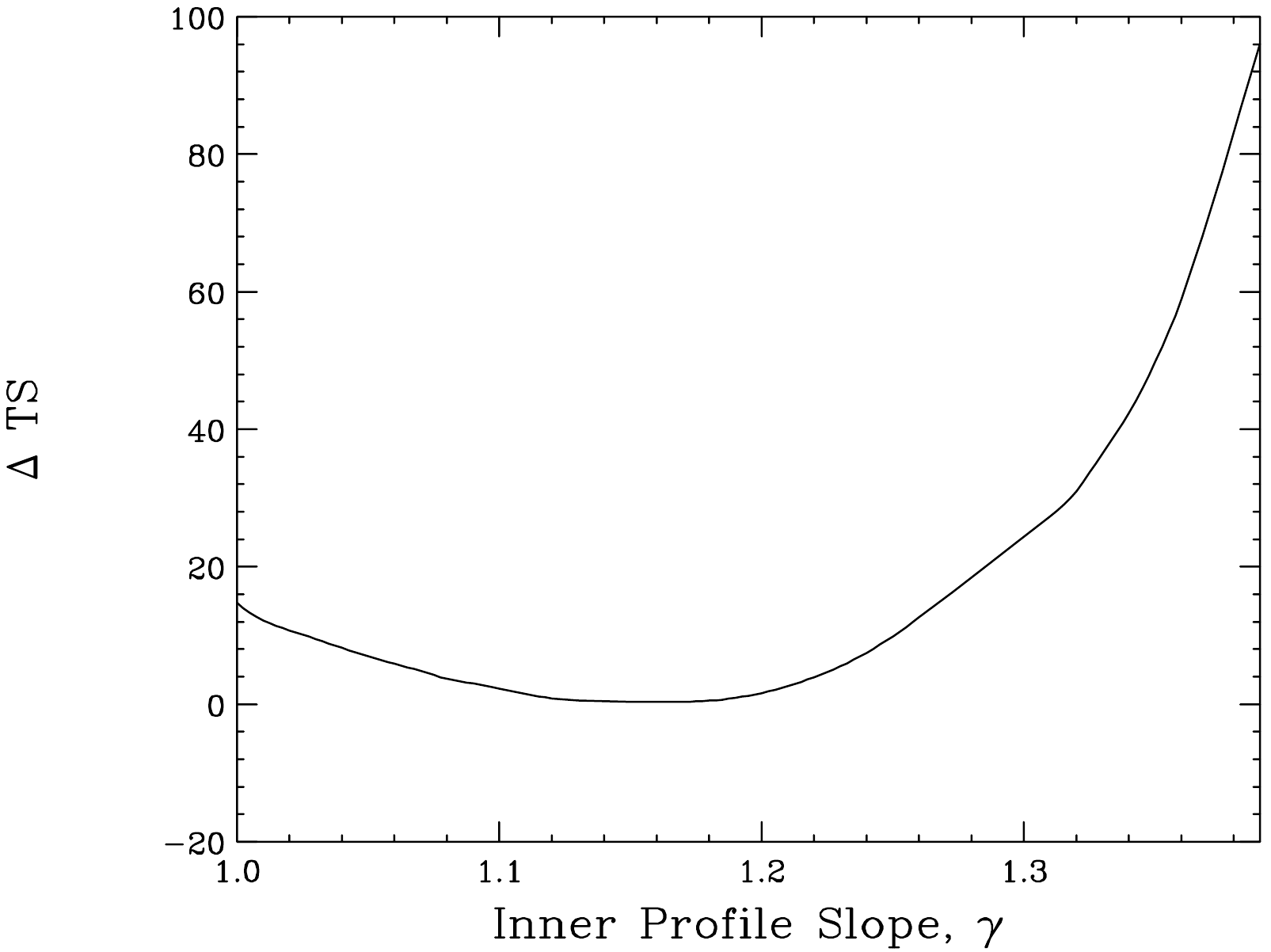}
\caption{The change in TS for the dark matter template as a function of the inner slope of the dark matter halo profile, $\gamma$, as found in our Galactic Center likelihood analysis. All values are relative to the result for the best-fit (highest TS) template, and positive values thus indicate a reduction in TS. The best-fit value is very similar to that found in our analysis of the larger Inner Galaxy region (in the default ROI), favoring $\gamma \sim 1.17$ (compared to $\gamma \simeq 1.18$ in the Inner Galaxy analysis).}
\label{timslope}
\end{figure}

\section{The Galactic Center}
\label{center}

In this section, we describe our analysis of the \textit{Fermi} data from the region of the Galactic Center, defined as $|b|<5^{\circ}$, $|l|<5^{\circ}$. We make use of the same Pass 7 data set, with Q2 cuts on CTBCORE, as described in the previous section. We performed a binned likelihood analysis to this data set using the \textit{Fermi} tool \texttt{gtlike}, dividing the region into 200$\times$200 spatial bins (each $0.05^{\circ}\times0.05^{\circ}$), and 12 logarithmically-spaced energy bins between 0.316-10.0 GeV. Included in the fit is a model for the Galactic diffuse emission, supplemented by a model spatially tracing the observed 20 cm emission~\cite{Law:2008uk}, a model for the isotropic gamma-ray background, and all gamma-ray sources listed in the 2FGL catalog~\cite{Fermi-LAT:2011iqa}, as well as the two additional point sources described in Ref.~\cite{YusefZadeh:2012nh}. We allow the flux and spectral shape of all high-significance ($\sqrt{{\rm TS}}>25$) 2FGL sources located within $7^{\circ}$ of the Galactic Center to vary. For somewhat more distant or lower significance sources ($\psi=7^{\circ}-8^{\circ}$ and $\sqrt{{\rm TS}}>25$, $\psi=2^{\circ}-7^{\circ}$ and $\sqrt{{\rm TS}}=10-25$, or $\psi<2^{\circ}$ and any TS), we adopt the best-fit spectral shape as presented in the 2FGL catalog, but allow the overall normalization to float. We additionally allow the spectrum and normalization of the two new sources from Ref.~\cite{YusefZadeh:2012nh}, the 20 cm template, and the extended sources W28 and W30~\cite{Fermi-LAT:2011iqa} to float. We fix the emission from all other sources to the best-fit 2FGL values. For the Galactic diffuse emission, we adopt the model \texttt{gal\_2yearp7v6\_v0}. Although an updated Galactic diffuse model has recently been released by the \textit{Fermi} Collaboration, that model includes additional empirically fitted features at scales greater than 2$^{\circ}$, and therefore is not recommended for studies of extended gamma-ray emission. For the isotropic component, we adopt the model of Ref.~\cite{Abdo:2010nz}.  We allow the overall normalization of the Galactic diffuse and isotropic emission to freely vary. In our fits, we found that the isotropic component prefers a normalization that is considerably brighter than the extragalactic gamma-ray background. In order to account for this additional isotropic emission in our region of interest, we attempted simulations in which we allowed the spectrum of the isotropic component to vary, but found this to have a negligible impact on the fit.

\begin{figure*}[t!]
\includegraphics[width=6.5in]{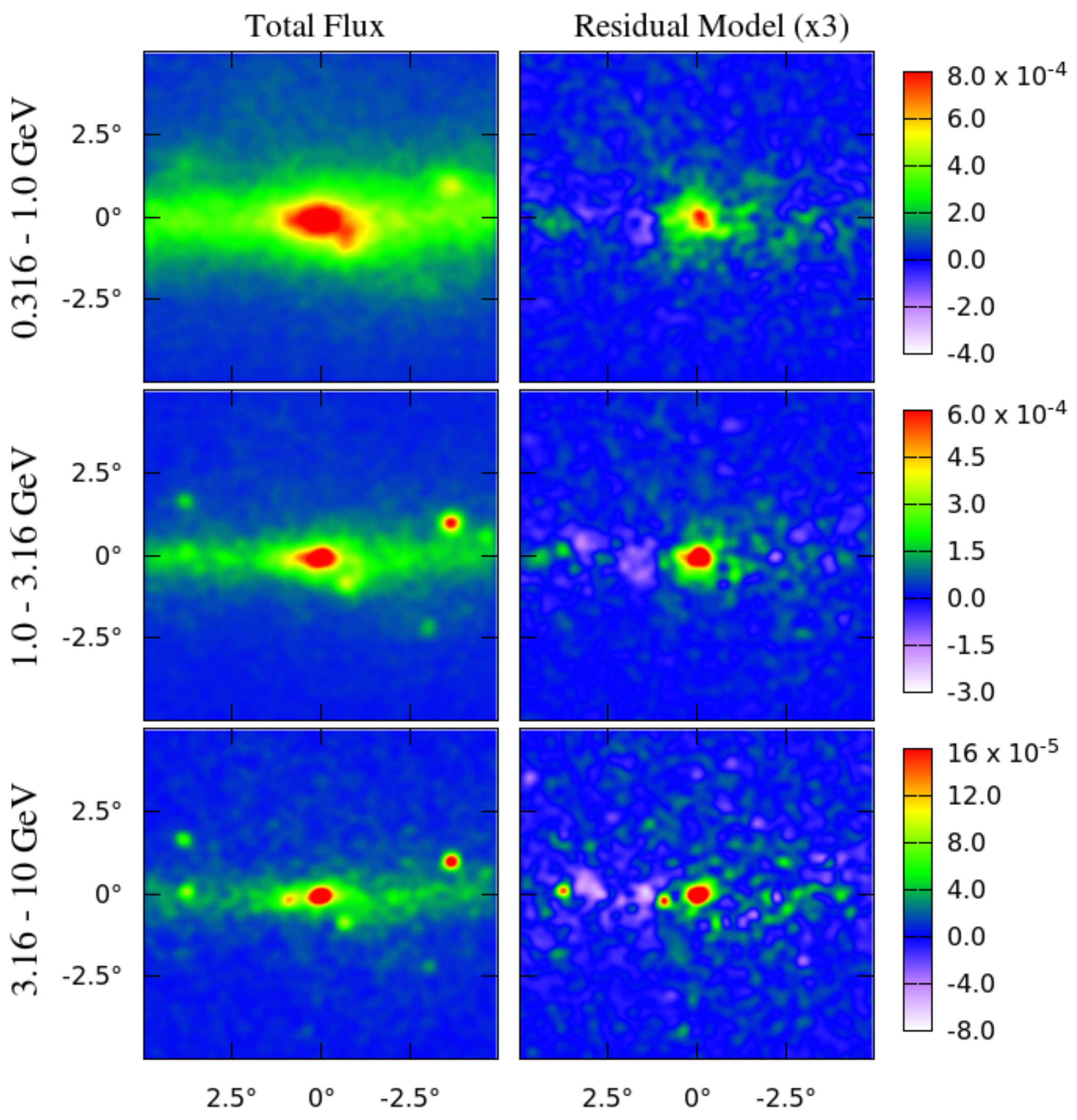}
\caption{The raw gamma-ray maps (left) and the residual maps after subtracting the best-fit Galactic diffuse model, 20 cm template, point sources, and isotropic template (right), in units of photons/cm$^2$/s/sr. The right frames clearly contain a significant central and spatially extended excess, peaking at $\sim$1-3 GeV. Results are shown in galactic coordinates, and all maps have been smoothed by a 0.25$^\circ$~Gaussian.}
\label{dmmap}
\end{figure*}

In addition to these astrophysical components, we include a spatially extended model in our fits motivated by the possibility of annihilating dark matter. The morphology of this component is again taken to follow the line-of-sight integral of the square of the dark matter density, as described in Sec.~\ref{intro2}. We adopt a generalized NFW profile centered around the location of Sgr A$^*$ ($b=-0.04608^{\circ}$, $l=-0.05578^{\circ}$~\cite{1999ApJ...518L..33Y}), and allow the inner slope ($\gamma$) and overall normalization (set by the annihilation cross section) to freely float.

In Figs.~\ref{timspec} and~\ref{timslope}, we show the main results of our Galactic Center likelihood analysis. In Fig.~\ref{timslope}, we plot the change of the log-likelihood of our fit as a function of the inner slope of the halo profile, $\gamma$. For our best-fit value of $\gamma=1.17$, the inclusion of the dark matter component can improve the overall fit with TS $\simeq 300$, corresponding to a statical preference for such a component at the level of $\sim$17$\sigma$. In Fig.~\ref{timspec}, we show the spectrum of the dark-matter-like component, for values of $\gamma=1.2$ (left frame) and $\gamma=1.3$ (right frame).  Shown for comparison is the spectrum predicted from a 35.25 GeV WIMP annihilating to $b\bar{b}$. The solid line represents the contribution from prompt emission, whereas the dot-dashed and dotted lines also include an estimate for the contribution from bremsstrahlung (for the $z=0.15$ and 0.3 kpc cases, as shown in the right frame of Fig.~\ref{dnde}, respectively). The normalizations of the Galactic Center and Inner Galaxy signals are compatible (see Figs.~\ref{innerspec} and ~\ref{timspec}), although the details of this comparison depend on the precise morphology that is adopted.

We note that the \textit{Fermi} tool \texttt{gtlike} determines the quality of the fit assuming a given spectral shape for the dark matter template, but does not generally provide a model-independent spectrum for this or other components. In order to make a model-independent determination of the dark matter component's spectrum, we adopt the following procedure. First, assuming a seed spectrum for the dark matter component, the normalization and spectral shape of the various astrophysical components are each varied and set to their best-fit values. Then, the fit is performed again, allowing the spectrum of the dark matter component to vary in each energy bin. The resultant dark matter spectrum is then taken to be the new seed, and this procedure is repeated iteratively until convergence is reached.

In Fig.~\ref{dmmap}, we plot the gamma-ray count maps of the Galactic Center region. In the left frames, we show the raw maps, while in the right frames we have subtracted the best-fit contributions from each component in the fit except for that corresponding to the dark matter template (the Galactic diffuse model, 20 cm template, point sources, and isotropic template). In each frame, the map has been smoothed by a 0.25$^{\circ}$ Gaussian (0.59$^{\circ}$ full-width-half-maximum). The excess emission is clearly present in the right frames, and most evidently in the 1.0-3.16 GeV range, where the signal is most significant. 

The slope favored by our Galactic Center analysis ($\gamma \simeq 1.04$--1.24) is very similar to that found in the Inner Galaxy analysis ($\gamma$~$\simeq$~1.15-1.22). Our results are also broadly consistent with those of the recent analysis of Ref.~\cite{Abazajian:2014fta}, which studied a smaller region of the sky ($|b|<3.5^{\circ}$, $|l|<3.5^{\circ}$), and found a preference for $\gamma \simeq 1.12 \pm 0.05$. We discuss this question further in Sec.~\ref{morphology}.

As mentioned above, in addition to the Galactic diffuse model, we include a spatial template in our Galactic Center fit with a morphology tracing the 20 cm (1.5 GHz) map of Ref.~\cite{Law:2008uk}. This map is dominated by synchrotron emission, and thus traces a convolution of the distribution of cosmic-ray electrons and magnetic fields in the region. As cosmic-ray electrons also generate gamma rays via bremsstrahlung and inverse Compton processes, the inclusion of the 20 cm template in our fit is intended to better account for these sources of gamma rays. And although the Galactic diffuse model already includes contributions from bremsstrahlung and inverse Compton emission, the inclusion of this additional template allows for more flexibility in the fit. In actuality, however, we find that this template has only a marginal impact on the results of our fit, absorbing some of the low energy emission that (without the 20 cm template) would have been associated with our dark matter template.

\begin{figure*}[t!]
\hspace{-0.15in}
\includegraphics[width=3.4in]{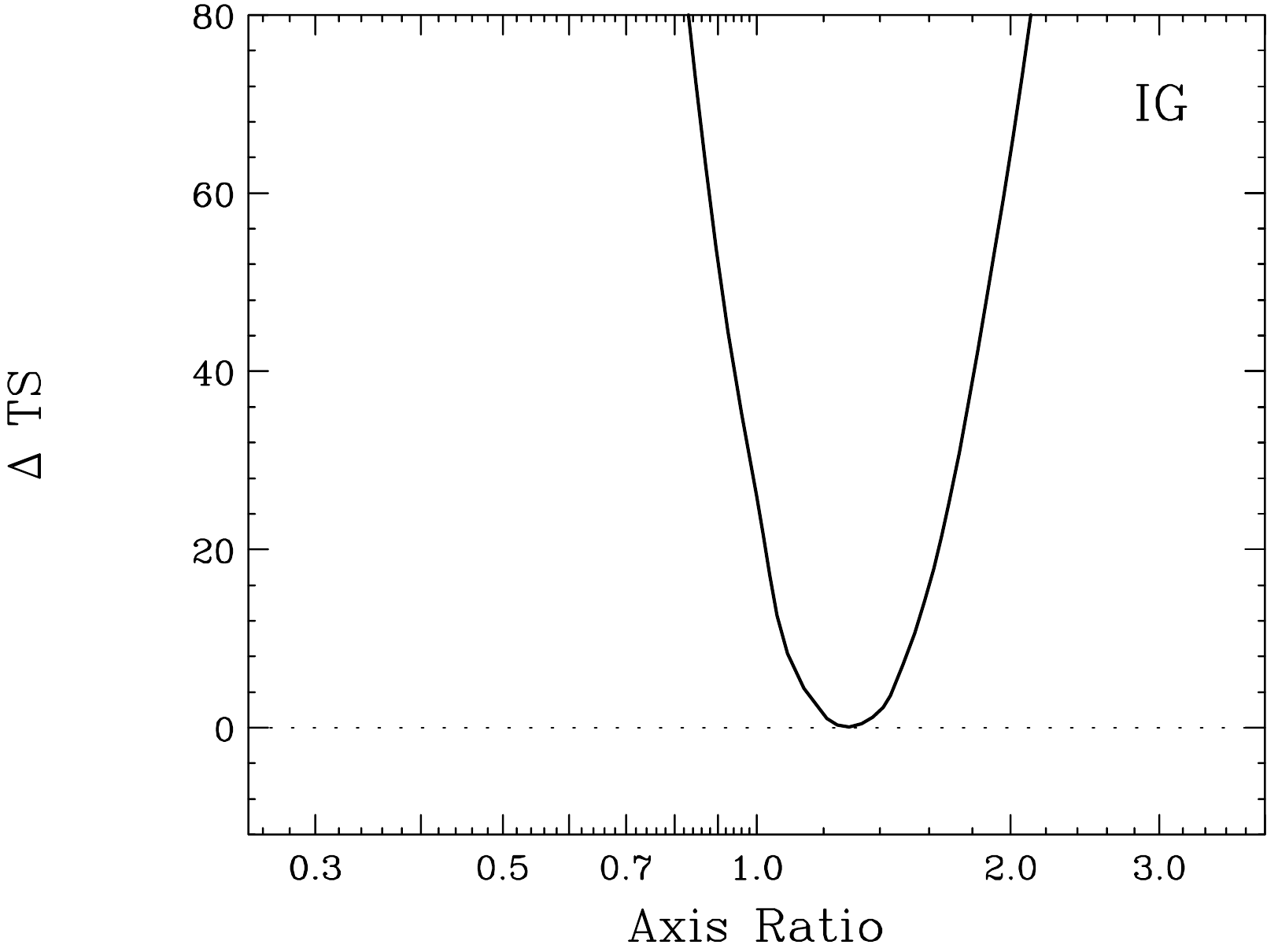}
\hspace{0.2in}
\includegraphics[width=3.4in]{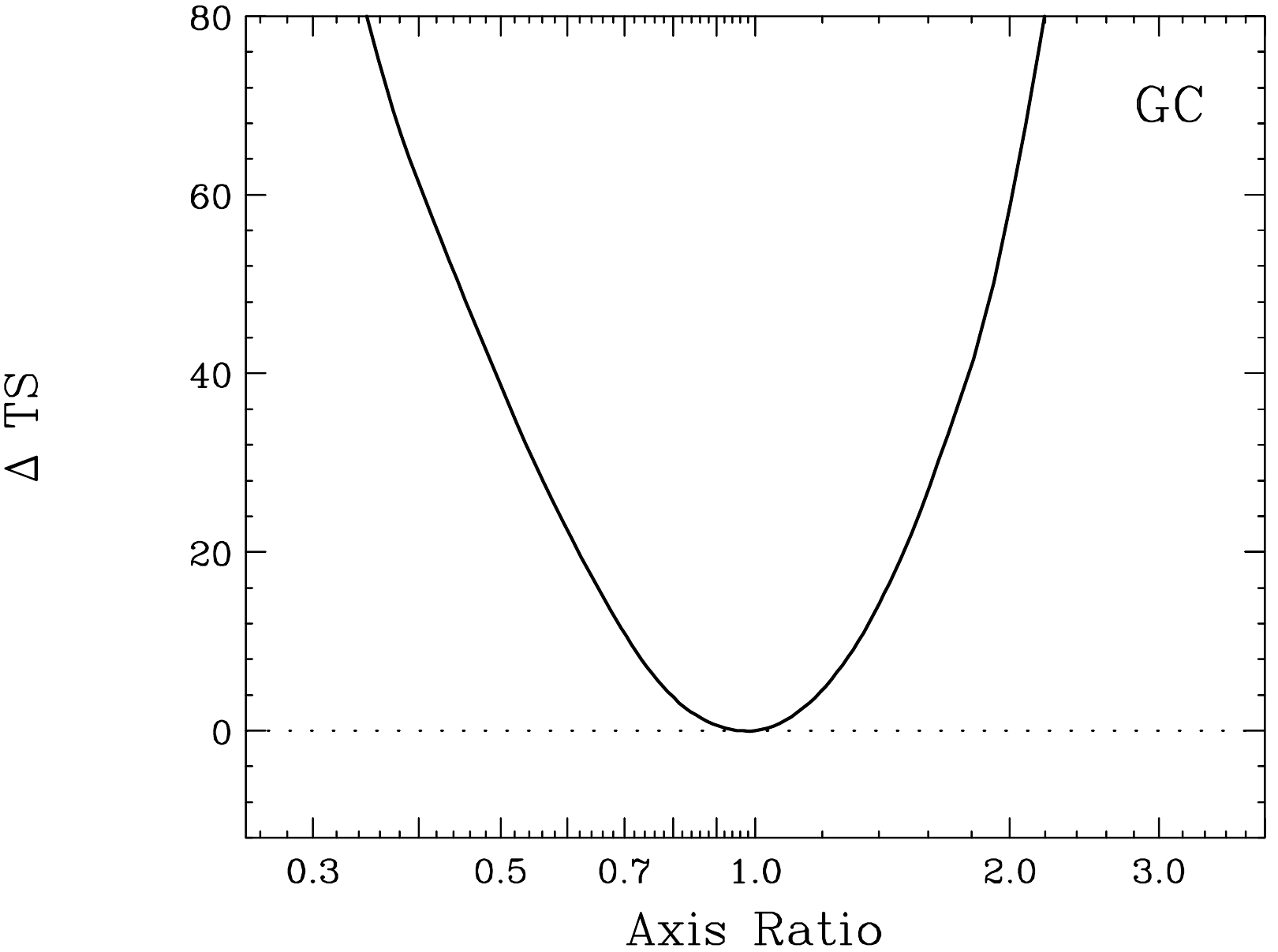}
\caption{The variation in TS for the dark matter template, as performed in Sec.~\ref{inner}'s Inner Galaxy analysis (left frame) and Sec.~\ref{center}'s Galactic Center analysis (right frame), when breaking our assumption of spherical symmetry for the dark matter template. All values shown are relative to the choice of axis ratio with the highest TS; positive values thus indicate a reduction in TS. The axis ratio is defined such that values less than one are elongated along the Galactic Plane, whereas values greater than one are elongated with Galactic latitude. The fit strongly prefers a morphology for the anomalous component that is approximately spherically symmetric, with an axis ratio near unity.}
\label{asymmetry}
\end{figure*}

\begin{figure}
\includegraphics[width=3.695in]{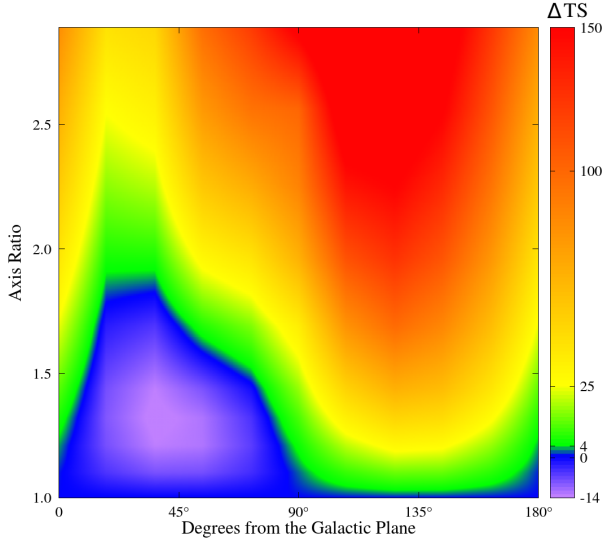}
\caption{The change in the quality of the fit in our Galactic Center analysis, for a dark matter template that is elongated along an arbitrary orientation (x-axis) and with an arbitrary axis ratio (y-axis). As shown in Fig.~\ref{asymmetry}, the fit worsens if the this template is significantly stretched either along or perpendicular to the direction of the Galactic Plane (corresponding to $0^{\circ}$ or $90^{\circ}$ on the x-axis, respectively). A mild statistical preference, however, is found for a morphology with an axis ratio of $\sim$1.3-1.4 elongated along an axis rotated $\sim$35$^{\circ}$ clockwise from the Galactic Plane.}
\label{alldirections}
\end{figure}

\begin{figure}
\includegraphics[width=3.695in]{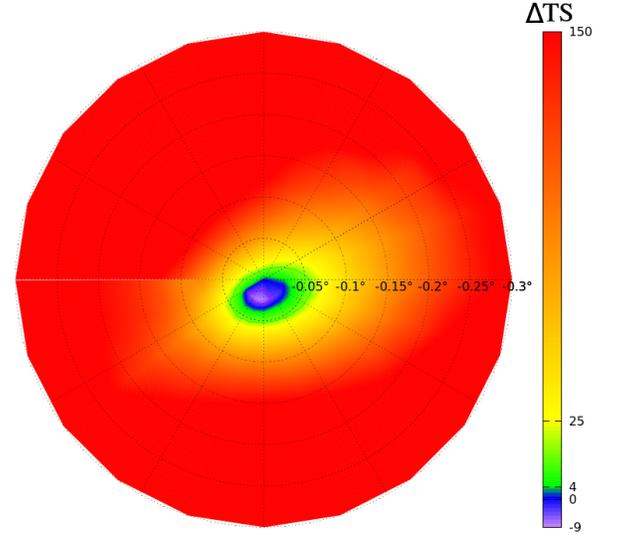}
\caption{To test whether the excess emission is centered around the dynamical center of the Milky Way (Sgr A$^*$), we plot the change in the TS associated with the dark matter template found in our Galactic Center analysis, as a function of the center of the template. Positive values correspond to a worse fit (lower TS). The fit clearly prefers this template to be centered within $\sim$$0.05^{\circ}$ of the location of Sgr A$^*$.} 
\label{offset}
\end{figure}










\section{Further Constraining the Morphology of the Anomalous Gamma-Ray Emission}
\label{morphology}

In the previous two sections, we showed that the gamma-ray emission observed from the regions of the Inner Galaxy and Galactic Center is significantly better fit when we include an additional component with an angular distribution that follows that predicted from annihilating dark matter. In particular, our fits favor a morphology for this component that follows the square of a generalized NFW halo profile with an inner slope of $\gamma \simeq 1.1-1.3$. Implicit in those fits, however, was the assumption that the angular distribution of the anomalous emission is spherically symmetric with respect to the dynamical center of the Milky Way. In this section, we challenge this assumption and test whether other morphologies might provide a better fit to the observed emission.


We begin by considering templates which are elongated either along or perpendicular to the direction of the Galactic Plane. In Fig.~\ref{asymmetry}, we plot the change in the TS of the Inner Galaxy (left) and Galactic Center (right) fits with such an asymmetric template, relative to the case of spherical symmetry. The axis ratio is defined such that values less than unity are elongated in the direction of the Galactic Plane, while values greater than one are preferentially extended perpendicular to the plane. The profile slope averaged over all orientations is taken to be $\gamma=1.2$ in both cases. From this figure, it is clear that the gamma-ray excess in the GC prefers to be fit by an approximately spherically symmetric distribution, and disfavors any axis ratio which departs from unity by more than approximately 20\%. In the Inner Galaxy there is a preference for a stretch \emph{perpendicular} to the plane, with an axis ratio of $\sim 1.3$. As we will discuss in Appendix \ref{app:consistency}, however, there are reasons to believe this may be due to the oversubtraction of the diffuse model along the plane, and this result is especially sensitive to the choice of ROI.

In Fig.~\ref{alldirections}, we generalize this approach within our Galactic Center analysis to test morphologies that are not only elongated along or perpendicular to the Galactic Plane, but along any arbitrary orientation. Again, we find that that the quality of the fit worsens if the the template is significantly elongated either along or perpendicular to the direction of the Galactic Plane. A mild statistical preference is found, however, for a morphology with an axis ratio of $\sim$1.3-1.4 elongated along an axis rotated $\sim$35$^{\circ}$ clockwise from the Galactic Plane in galactic coordinates.\footnote{We define a ``clockwise" rotation such that a 90$^\circ$ rotation turns +l into +b.} While this may be a statistical fluctuation, or the product of imperfect background templates, it could also potentially reflect a degree of triaxiality in the underlying dark matter distribution.

\begin{figure}[t!]
\includegraphics[width=3.4in,trim=1cm 11cm 1cm 0]{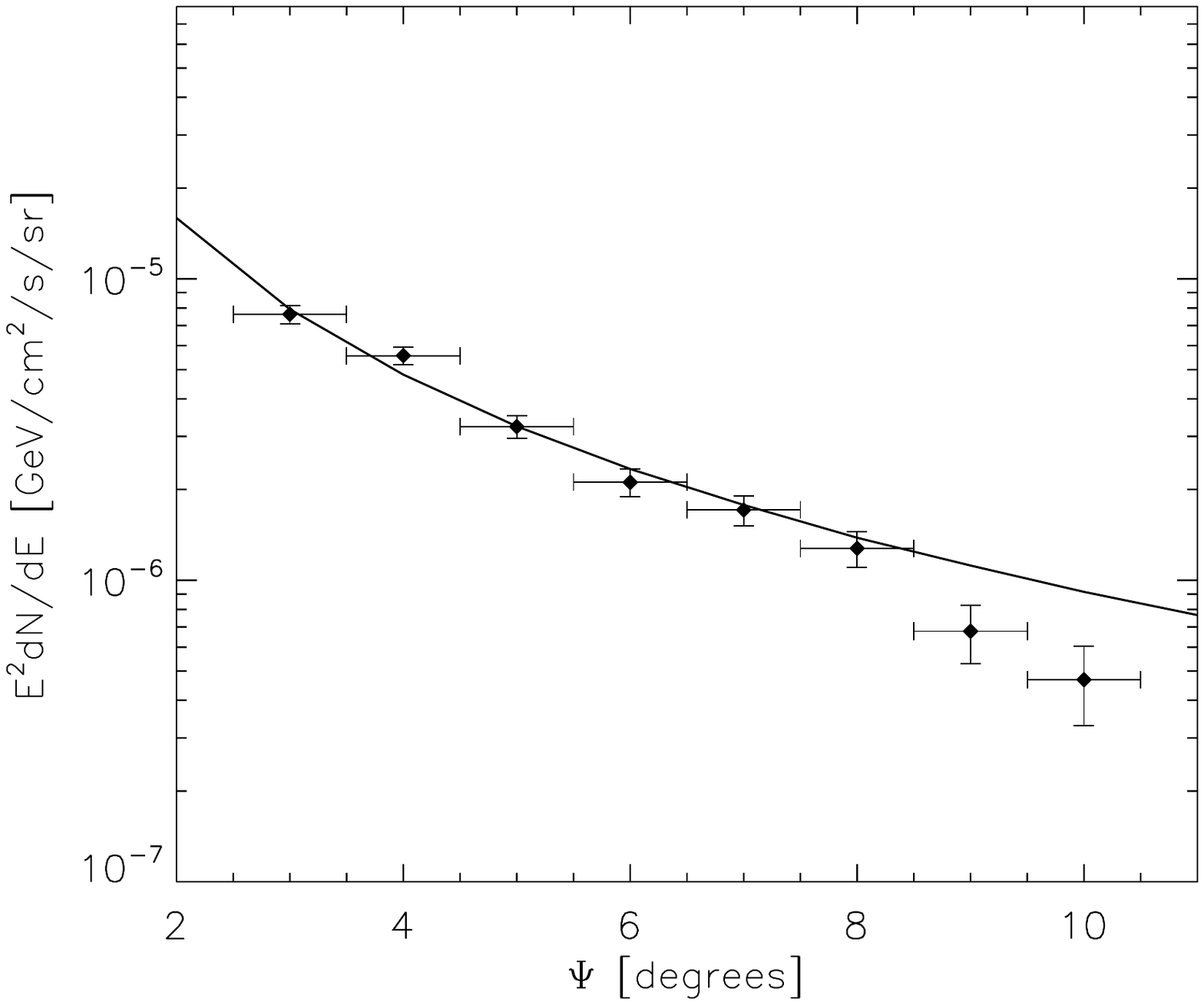}
\caption{To constrain the degree to which the gamma-ray excess is spatially extended, we have repeated our Inner Galaxy analysis, replacing the dark matter template with a series of concentric ring templates centered around the Galactic Center. The dark-matter-like emission is clearly and consistently present in each ring template out to $\sim$$10^{\circ}$, beyond which systematic and statistical limitations make such determinations difficult. For comparison, we also show the predictions for a generalized NFW profile with $\gamma=1.3$. The spectrum of the rings is held fixed at that of Fig. \ref{innerspec}, and the fluxes displayed in the plot correspond to an energy of 2.67 GeV.}
\label{ringfit}
\end{figure}

\begin{figure*}[t!]
\includegraphics[width=3.4in]{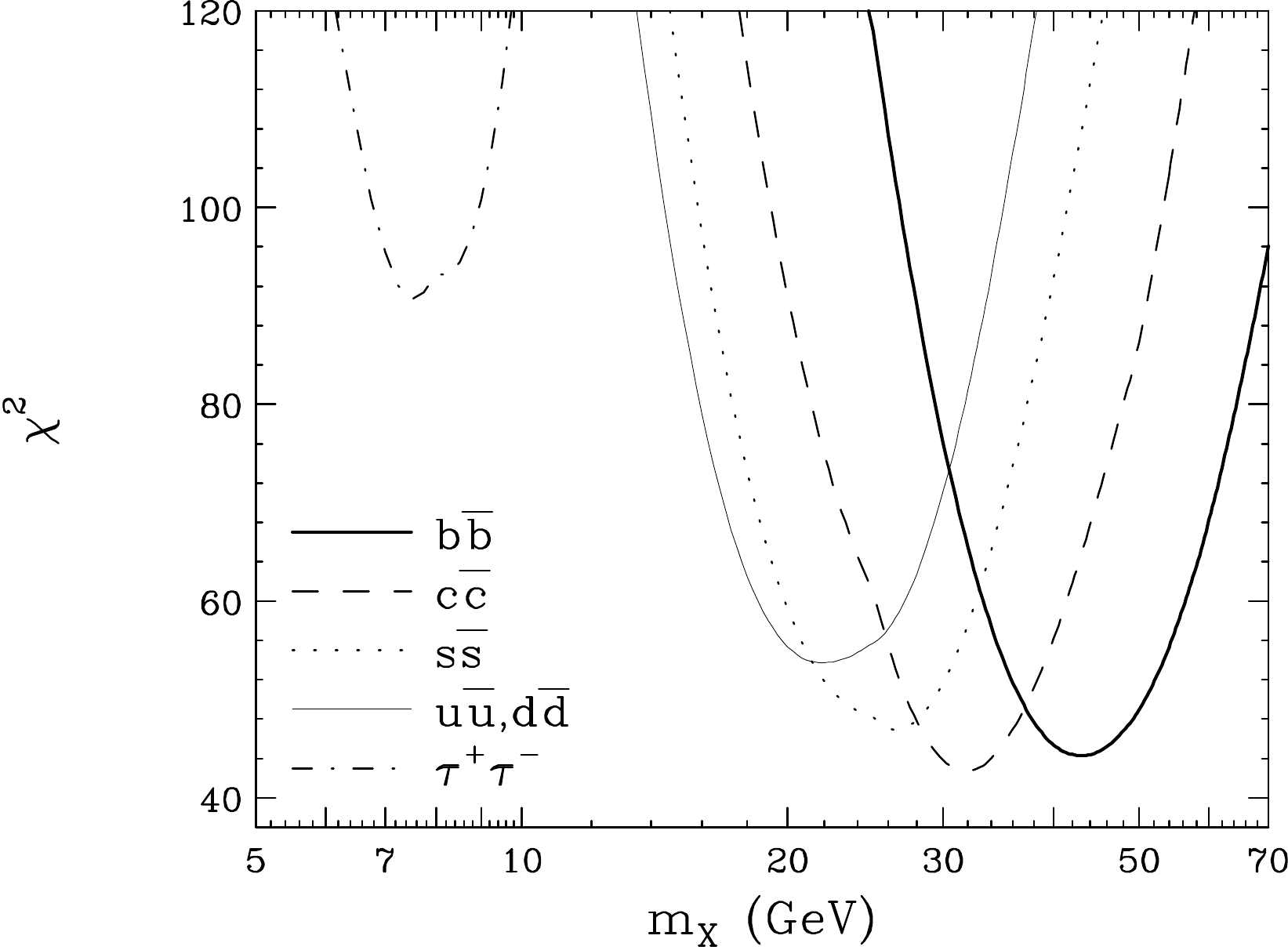}
\hspace{0.15in}
\includegraphics[width=3.4in]{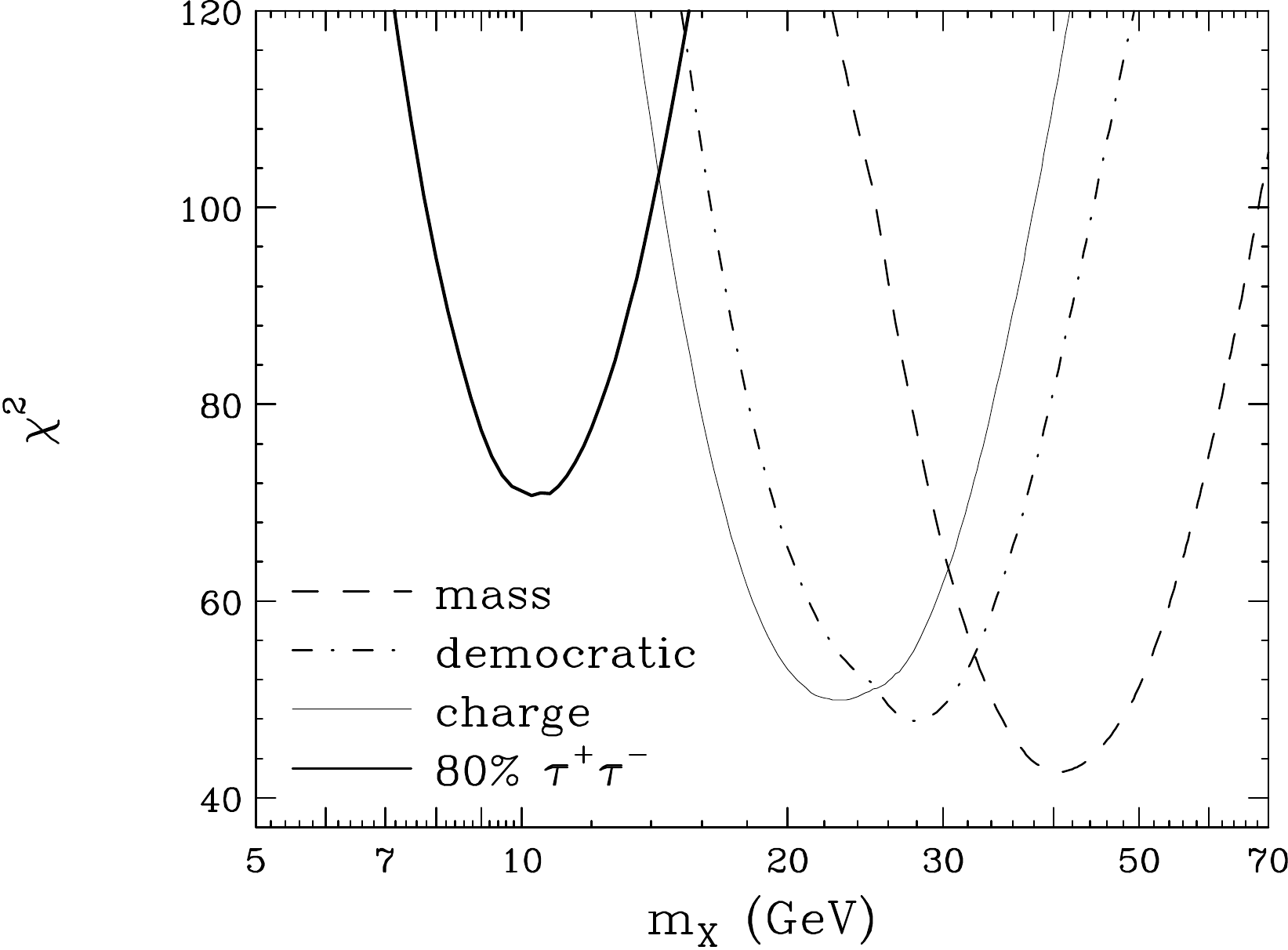}
\caption{The quality of the fit ($\chi^2$, over 25-1 degrees-of-freedom) for various annihilating dark matter models to the spectrum of the anomalous gamma-ray emission from the Inner Galaxy (as shown in the left frame of Fig.~\ref{innerspec}) as a function of mass, and marginalized over the value of the annihilation cross section. In the left frame, we show results for dark matter particles which annihilate uniquely to $b\bar{b}$, $c\bar{c}$, $s\bar{s}$, light quarks ($u\bar{u}$ and/or $d\bar{d}$), or $\tau^+ \tau^-$. In the right frame, we consider models in which the dark matter annihilates to a combination of channels, with cross sections proportional to the square of the mass of the final state particles, the square of the charge of the final state particles, democratically to all kinematically accessible Standard Model fermions, or 80\% to $\tau^+ \tau^-$ and 20\% to $b\bar{b}$. The best fits are found for dark matter particles with masses in the range of $\sim$20-60 GeV and which annihilate mostly to quarks.}
\label{chisq}
\end{figure*}

We have also tested whether the excess emission is, in fact, centered around the dynamical center of the Milky Way (Sgr A$^*$), as we have thus far assumed. In Fig.~\ref{offset}, we plot the change in TS of the dark-matter-motivated template, as found in our Galactic Center analysis, when we vary the center of the template. The fit clearly prefers this template to be centered within $\sim$$0.05^{\circ}$ of the location of Sgr A$^*$.

An important question to address is to what degree the gamma-ray excess is spatially extended, and over what range of angles from the Galactic Center can it be detected? To address this issue, we have repeated our Inner Galaxy analysis, replacing the dark matter template with 8 concentric, rotationally symmetric ring templates, each 1$^{\circ}$ wide, and centered around the Galactic Center. However instead of allowing the spectrum of the ring templates to each vary freely (which would have introduced an untenable number of free parameters), we fix their spectral shape between 0.3 GeV - 30 GeV to that found for the dark matter component in the single template fit. By floating the ring coefficients with a fixed spectral dependence, we obtain another handle on the spatial extent and morphology of the excess. In order to be self-consistent we inherit the background modeling and ROI from the Inner Galaxy analysis (except that we mask the plane for $|b| < 2^\circ$ rather than $|b| < 1^\circ$) and fix the spectra of all the other templates to the best fit values from the Inner Galaxy fit. We also break the template associated with the \textit{Fermi} Bubbles into two sub-templates, in 10$^{\circ}$ latitude slices (each with the same spectrum, but with independent normalizations). We smooth the templates to the Fermi PSF.

The results of this fit are shown in Fig.~\ref{ringfit}. The dark-matter-like emission is clearly and consistently present in each ring template out to $\sim 10^{\circ}$, beyond which systematic and statistical limitations make such determinations difficult. In order to compare the radial dependence with that expected from a generalized NFW profile, we weight the properly smoothed NFW squared/projected template with each ring to obtain ring coefficients expected from an ideal NFW distribution. We then perform a minimum $\chi^2$ fit on the data-driven ring coefficients taking as the template the coefficients obtained from an NFW profile with $\gamma=1.3$. We exclude the two outermost outlier ring coefficients from this fit in order to avoid systematic bias on the preferred $\gamma$ value. Since the ring templates spatially overlap upon smoothing, we take into account the correlated errors of the maximum likelihood fit, which add to the spectral errors in quadrature. We show an interpolation of the best fit NFW ring coefficients with the solid line on the same figure.

We caution that systematic uncertainties associated with the diffuse model template may be biasing this fit toward somewhat steeper values of $\gamma$ (we discuss this question further in Appendix \ref{app:consistency}, in the context of the increased values of $\gamma$ found for larger ROIs). It is also plausible that the dark matter slope could vary with distance from the Galactic Center, for example as exhibited by an Einasto profile~\cite{Springel:2008cc}.


To address the same question within the context of our Galactic Center analysis, we have re-performed our fit using dark matter templates which are based on density profiles which are set to zero beyond a given radius.  We find that templates corresponding to density profiles set to zero outside of 800 pc (600 pc, 400 pc) provide a fit that is worse relative to that found using an untruncated template at the level of $\Delta$ TS=10.7 (57.6,108, respectively).

We have also tested our Galactic Center fit to see if a cored dark matter profile could also provide a good fit to the data. We find, however, that the inclusion of even a fairly small core is disfavored. Marginalizing over the inner slope of the dark matter profile, we find that flattening the density profile within a radius of 10 pc (30 pc, 50 pc, 70 pc, 90 pc) worsens the overall fit by $\Delta$ TS=3.6 (12.2, 22.4, 30.6, 39.2, respectively). The fit thus strongly disfavors any dark matter profile with a core larger than a few tens of parsecs.


Lastly, we confirm that the morphology of the anomalous emission does not significantly vary with energy. If we fit the inner slope of the dark matter template in our Inner Galaxy analysis one energy bin at a time, we find a similar value of $\gamma\sim$1.1-1.3 for all bins between 0.7 and 13 GeV. At energies $\sim 0.5$ GeV and lower, the fit prefers somewhat steeper slopes ($\gamma \sim 1.6$ or higher) and a corresponding spectrum with a very soft spectral index, probably reflecting contamination from the Galactic Plane. At energies above $\sim 13$ GeV, the fit again tends to prefers a steeper profile.

The results of this section indicate that the gamma-ray excess exhibits a morphology which is both approximately spherically symmetric and steeply falling (yet detectable) over two orders of magnitude in galactocentric distance (between $\sim$20 pc and $\sim$2 kpc from Sgr A*). This result is to be expected if the emission is produced by annihilating dark matter particles, but is not anticipated for any proposed astrophysical mechanisms or sources of this emission.

\begin{figure*}[t!]
\hspace{-0.3in}
\includegraphics[width=3.45in]{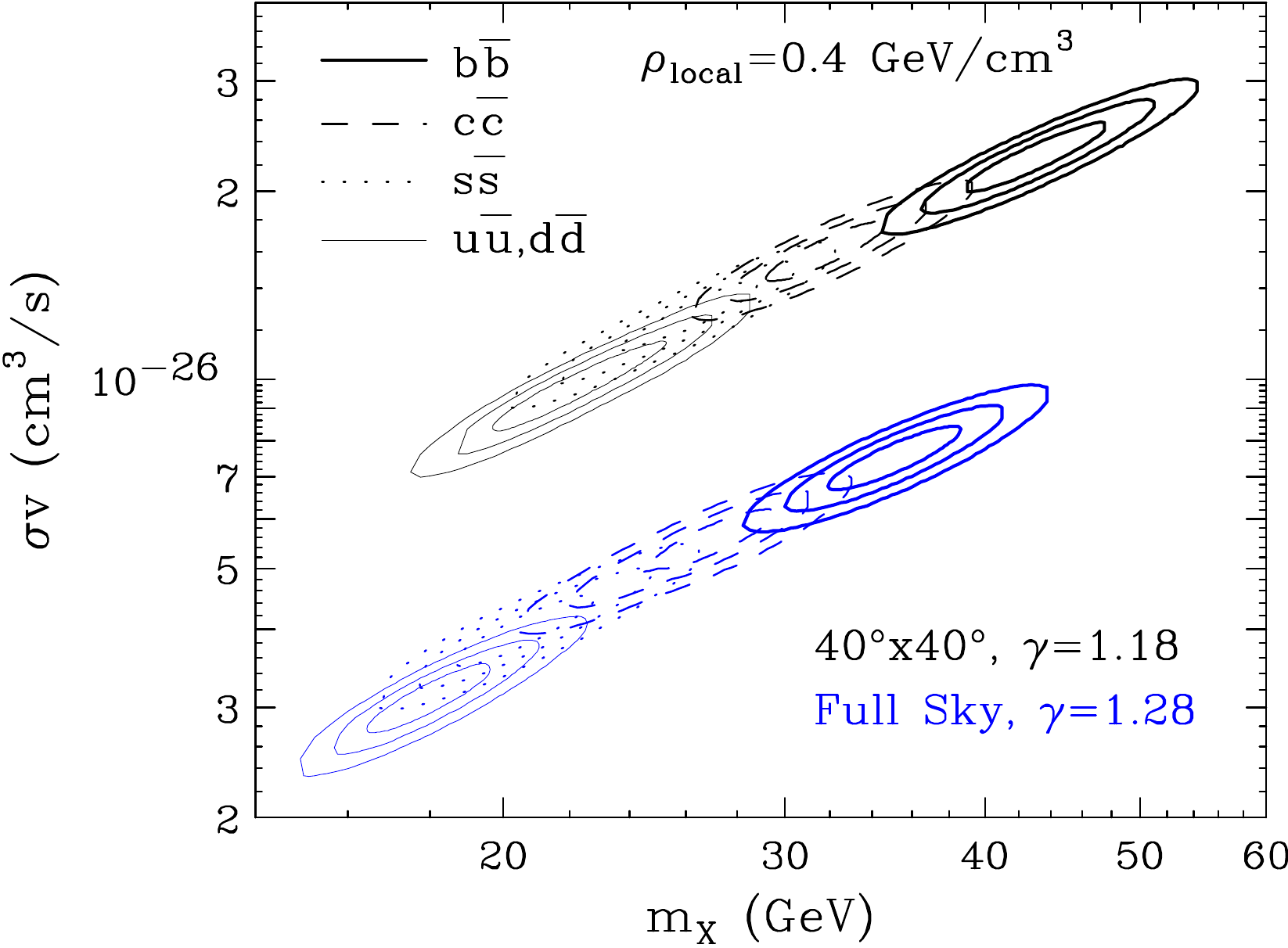}
\hspace{0.1in}
\includegraphics[width=3.45in]{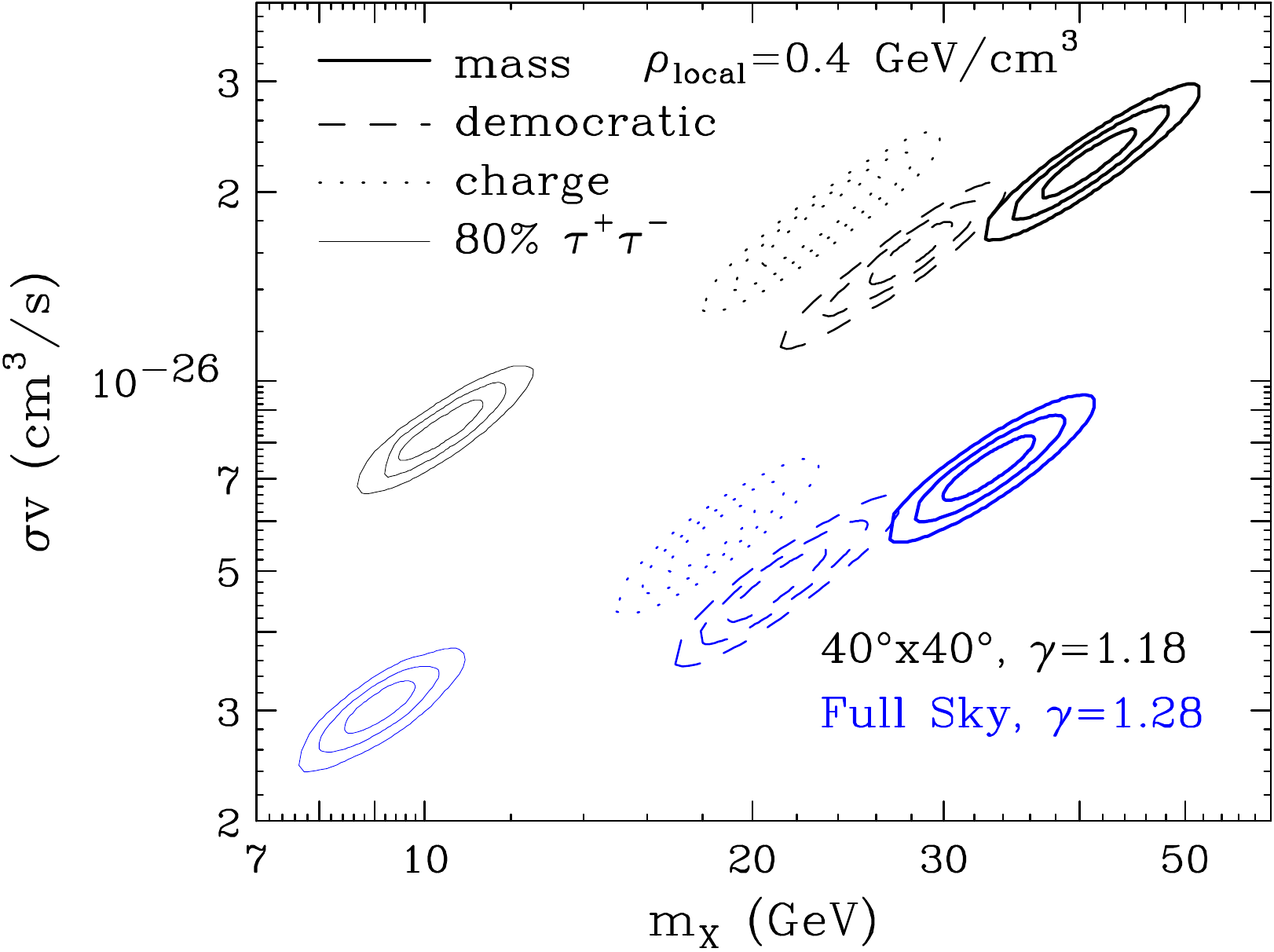}
\caption{The range of the dark matter mass and annihilation cross section required to fit the gamma-ray spectrum observed from the Inner Galaxy, for a variety of annihilation channels or combination of channels (see Fig.~\ref{chisq}). We show results for our standard ROI (black) and as fit over the full sky (blue). The observed gamma-ray spectrum is generally best fit by dark matter particles with a mass of $\sim$20-50 GeV and that annihilate to quarks with a cross section of $\sigma v\sim 10^{-26}$ cm$^3$/s. Note that the cross-section for each model is computed for the best-fit slope $\gamma$ in that ROI and the assumed dark matter densities at 5$^\circ$ from the Galactic Center (where the signal is normalized) are different for different values of $\gamma$. This is responsible for roughly half of the variation between the best-fit cross-sections. Figures~\ref{fig:igresults} and~\ref{regioncompare2} show the impact of changing the ROI when holding the assumed DM density profile constant.}
\label{regions}
\end{figure*}

\section{Implications for Dark Matter}
\label{darkmatter}

In this section, we use the results of the previous sections to constrain the characteristics of the dark matter particle species potentially responsible for the observed gamma-ray excess. 
%
%
We begin by fitting various dark matter models to the spectrum of the gamma-ray excess as found in our Inner Galaxy analysis (as shown in the left frame of Fig.~\ref{innerspec}). In Fig.~\ref{chisq}, we plot the quality of this fit ($\chi^2$) as a function of the WIMP mass, for a number of dark matter annihilation channels (or combination of channels), marginalized over the value of the annihilation cross section. Given that this fit is performed over 22-1 degrees-of-freedom, a goodness-of-fit with a $p$-value of 0.05 (95\% CL) corresponds to a $\chi^2$ of approximately 36.8. Given the systematic uncertainties associated with the choice of background templates, we take any value of $\chi^2 \lsim 50$ to constitute a reasonably ``good fit'' to the Inner Galaxy spectrum. Good fits are found for dark matter that annihilates to bottom, strange, or charm quarks. The fits are slightly worse for annihilations to light quarks, or to combinations of fermions proportional to the square of the mass of the final state, the square of the charge of the final state, or equally to all fermonic degrees of freedom (democratic). In the light mass region ($m_X$$\sim$7-10 GeV) motivated by various direct detection anomalies~\cite{Aalseth:2010vx,Aalseth:2011wp,Agnese:2013rvf,Angloher:2011uu,Bernabei:2008yi,Bernabei:2010mq}, the best fit we find is for annihilations which proceed mostly to $\tau^+ \tau^-$, with an additional small fraction to quarks, such as $b\bar{b}$. Even this scenario, however, provides a somewhat poor fit, significantly worse that that found for heavier ($m_X\sim 20-60$ GeV) dark matter particles annihilating mostly to quarks.

In Fig.~\ref{regions}, we show the regions of the dark matter mass-annihilation cross section plane that are best fit by the gamma-ray spectrum shown in Fig.~\ref{innerspec}. For each annihilation channel (or combination of channels), the 1, 2 and 3$\sigma$ contours are shown around the best-fit point (corresponding to $\Delta \chi^2=2.30$, 6.18, and 11.83, respectively). Again, in the left frame we show results for dark matter particles which annihilate entirely to a single final state, while the right frame considers instead combinations of final states. Generally speaking, the best-fit models are those in which the dark matter annihilates to second or third generation quarks with a cross section of $\sigma v\sim 10^{-26}$ cm$^3$/s.\footnote{The cross sections shown in Fig.~\ref{regions} were normalized assuming a local dark matter density of 0.4 GeV/cm$^3$. Although this value is near the center of the range preferred by the combination of dynamical and microlensing data (for $\gamma=1.18$), there are non-negligible uncertainties in this quantity. The analysis of Ref.~\cite{Iocco:2011jz}, for example, finds a range of $\rho_{\rm local}=0.26-0.49$ GeV/cm$^3$ at the 2$\sigma$ level. This range of densities corresponds to a potential rescaling of the y-axis of Fig.~\ref{regions} by up to a factor of 0.7-2.4.}

This range of values favored for the dark matter's annihilation cross section is quite interesting from the perspective of early universe cosmology. For the mass range being considered here, a WIMP with an annihilation cross section of $\sigma v \simeq 2.2 \times 10^{-26}$ cm$^3$/s (as evaluated at the temperature of freeze-out) will freeze-out in the early universe with a relic abundance equal to the measured cosmological dark matter density (assuming the standard thermal history)~\cite{Steigman:2012nb}. The dark matter annihilation cross section evaluated in the low-velocity limit (as is relevant for indirect searches), however, is slightly lower than the value at freeze-out in many models. For a generic $s$-wave annihilation process, for example, one generally expects dark matter in the form of a thermal relic to annihilate at low-velocities with a cross section near $\sigma v_{v=0} \simeq (1-2) \times 10^{-26}$ cm$^3$/s, in good agreement with the range of values favored by the observed gamma-ray excess.

\begin{figure}[t!]
\includegraphics[width=3.45in]{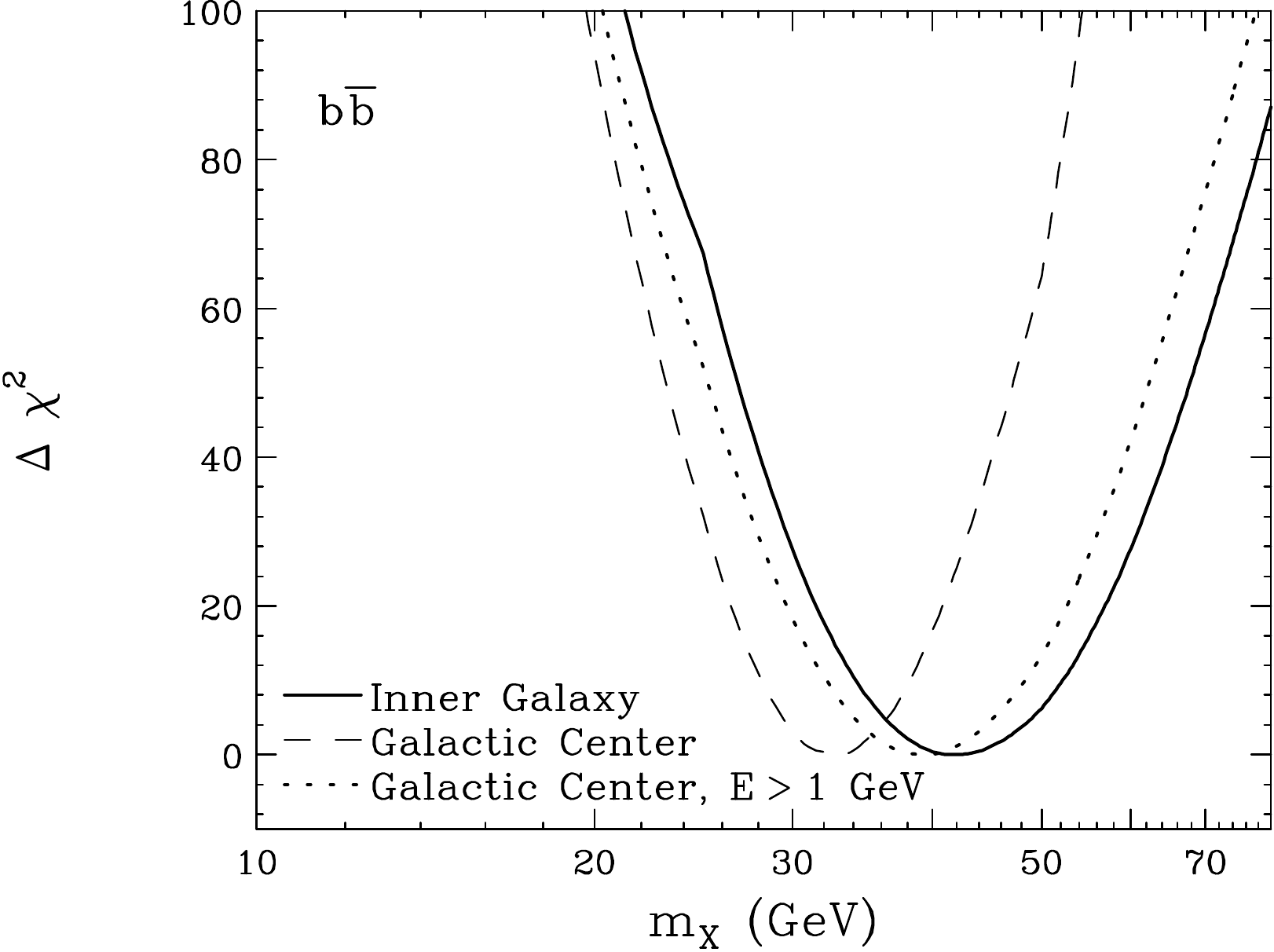}
\caption{A comparison of the dark matter mass determination using the spectrum derived from our Inner Galaxy analysis (solid line) and using the spectrum derived from our Galactic Center analysis (dashed and dotted lines). For each case shown, we have considered a profile with an inner slope of $\gamma=$1.2 and annihilations to $b\bar{b}$.}
\label{compare}
\end{figure}

Thus far in this section, we have fit the predictions of various dark matter models to the gamma-ray spectrum derived from our Inner Galaxy analysis. In Fig.~\ref{compare}, we compare the mass range best fit to the Inner Galaxy spectrum to that favored by our Galactic Center analysis. Overall, these two analyses favor a similar range of dark matter masses and annihilation channels, although the Galactic Center spectrum does appear to be slightly softer, and thus prefers WIMP masses that are a few GeV lower than favored by the Inner Galaxy analysis. This could, however, be the result of bremsstrahlung, which can soften the gamma-ray spectrum from dark matter in regions near the Galactic Plane (see Fig.~\ref{timspec} and the right frame of Fig.~\ref{dnde}). Such emission could plausibly cause a $\sim$40-45 GeV WIMP, for example, to produce a gamma-ray spectrum along the Galactic Plane that resembles the prompt emission predicted from a $\sim$35-40 GeV WIMP.

\section{Discussion}
\label{discussion}

In this paper (and in previous studies~\cite{Goodenough:2009gk,Hooper:2010mq,Hooper:2011ti,Abazajian:2012pn,Gordon:2013vta,Hooper:2013rwa,Huang:2013pda,Abazajian:2014fta}), it has been shown that the gamma-ray excess observed from the Inner Galaxy and Galactic Center is compatible with that anticipated from annihilating dark matter particles. This is not, however, the first time that an observational anomaly has been attributed to dark matter. Signals observed by numerous experiments, including \textit{INTEGRAL}~\cite{Boehm:2003bt}, \textit{PAMELA}~\cite{Adriani:2008zr}, \textit{ATIC}~\cite{aticlatest}, \textit{Fermi}~\cite{Weniger:2012tx,Su:2012ft}, \textit{WMAP}~\cite{Finkbeiner:2004us,Hooper:2007kb}, \textit{DAMA/LIBRA}~\cite{Bernabei:2008yi,Bernabei:2010mq}, \textit{CoGeNT}~\cite{Aalseth:2010vx,Aalseth:2011wp}, \textit{CDMS}~\cite{Agnese:2013rvf}, and \textit{CRESST}~\cite{Angloher:2011uu}, among others, have received a great deal of attention as possible detections of dark matter particles. Most, if not all, of these signals, have nothing to do with dark matter, but instead result from some combination of astrophysical, environmental, and instrumental backgrounds (see e.g.~\cite{Hooper:2008kg, Profumo:2008ms, Dobler:2011rd, Dobler:2012ef, Ackermann:2013uma,Aalseth:2012if,Aprile:2012nq,Akerib:2013tjd}). Given the frequency of such false alarms, we would be wise to apply a very high standard before concluding that any new signal is, in fact, the result of annihilating dark matter. 

There are significant reasons to conclude, however, that the gamma-ray signal described in this paper is far more likely to be a detection of dark matter than any of the previously reported anomalies. Firstly, this signal consists of a very large number of events, and has been detected with overwhelming statistical significance. The the excess consists of $\sim$$10^4$ gamma rays per square meter, per year above 1 GeV (from within 10$^{\circ}$~of the Galactic Center). Not only does this large number of events enable us to conclude with confidence that the signal is present, but it also allows us to determine its spectrum and morphology in some detail. And as shown, the measured spectrum, angular distribution, and normalization of this emission does indeed match well with that expected from annihilating dark matter particles. 

Secondly, the gamma-ray signal from annihilating dark matter can be calculated straightforwardly, and generally depends on only a few unknown parameters. The morphology of this signal, in particular, depends only on the distribution of dark matter in the Inner Galaxy (as parameterized in our study by the inner slope, $\gamma$). The spectral shape of the signal depends only on the mass of the dark matter particle and on what Standard Model particles are produced in its annihilations. The Galactic gamma-ray signal from dark matter can thus be predicted relatively simply, in contrast to, {\it e.g}.,  dark matter searches using cosmic rays, where putative signals are affected by poorly constrained diffusion and energy-loss processes. In other words, for the gamma-ray signal at hand, there are relatively few ``knobs to turn'', making it less likely that one would be able to mistakenly fit a well-measured astrophysical signal with that of an annihilating dark matter model.


Thirdly, we once again note that the signal described in this study can be explained by a very simple dark matter candidate, without any baroque or otherwise unexpected features. After accounting for uncertainties in the overall mass of the Milky Way's dark matter halo profile~\cite{Iocco:2011jz}, our results favor dark matter particles with an annihilation cross section of $\sigma v = (0.4-6.6) \times 10^{-26}$ cm$^3$/s (for annihilations to $b\bar{b}$, see Fig.~\ref{regions}). This range covers the long predicted value that is required of a thermal relic that freezes-out in the early universe with an abundance equal to the measured cosmological dark matter density ($2.2 \times 10^{-26}$ cm$^3$/s). No substructure boost factors, Sommerfeld enhancements, or non-thermal histories are required. Furthermore, it is not difficult to construct simple models in which a $\sim$30-50 GeV particle annihilates to quarks with the required cross section without violating constraints from direct detection experiments, colliders, or other indirect searches (for work related to particle physics models capable of accommodating this signal, see Refs.~\cite{Boehm:2014hva,Hardy:2014dea,Modak:2013jya,Huang:2013apa,Okada:2013bna,Hagiwara:2013qya,Buckley:2013sca,Anchordoqui:2013pta,Buckley:2011mm,Boucenna:2011hy,Marshall:2011mm,Zhu:2011dz,Buckley:2010ve,Logan:2010nw}).

And lastly, the dark matter interpretation of this signal is strengthened by the absence of plausible or well motivated alternatives. There is no reason to expect that any diffuse astrophysical emission processes would exhibit either the spectrum or the morphology of the observed signal. In particular, the spherical symmetry of the observed emission with respect to the Galactic Center does not trace any combination of astrophysical components ({\it i.e.} radiation, gas, dust, star formation, etc.), but does follow the square of the anticipated dark matter density. 

The astrophysical interpretation most often discussed within the context of this signal is that it might originate from a large population of unresolved millisecond pulsars. The millisecond pulsars observed within the Milky Way are largely located either within globular clusters or in or around the Galactic Disk (with an exponential scale height of $z_s \sim$~1 kpc~\cite{Gregoire:2013yta,Hooper:2013nhl}). This pulsar population would lead to a diffuse gamma-ray signal that is highly elongated along the disk, and would be highly incompatible with the constraints described in Sec.~\ref{morphology}. For example, the best-fit model of Ref.~\cite{Gregoire:2013yta}, which is based on the population of presently resolved gamma-ray millisecond pulsars, predicts a morphology for the diffuse gamma-ray emission exhibiting an axis ratio of $\sim$1-to-6. Within $10^{\circ}$ of the Galactic Center, this model predicts that millisecond pulsars should account for $\sim$1\% of the observed diffuse emission, and less than $\sim$5-10\% of the signal described in this paper.

\begin{figure}[t]
\includegraphics[width=3.4in]{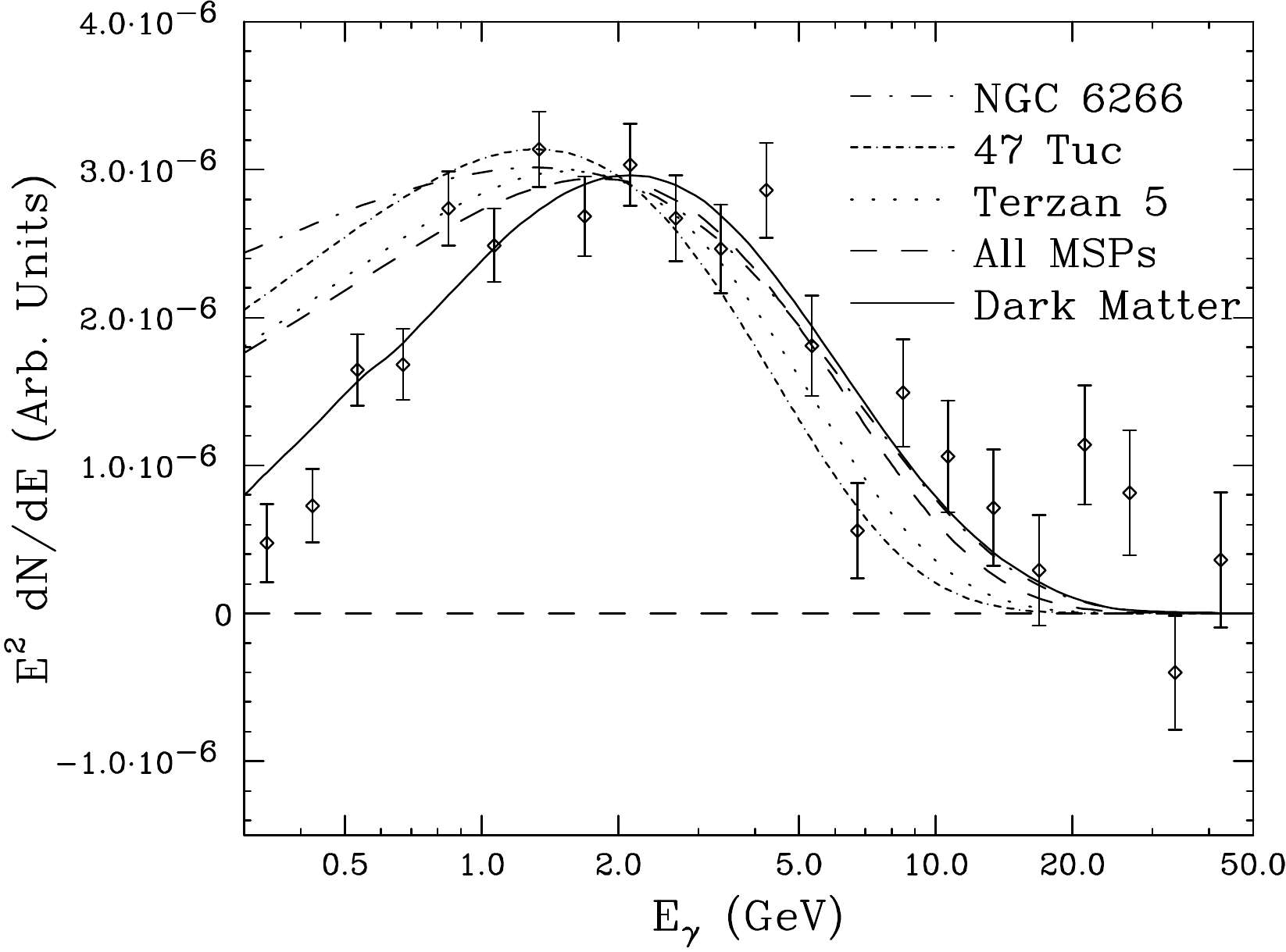}
\caption{A comparison of the spectral shape of the gamma-ray excess described in this paper (error bars) to that measured from a number of high-significance globular clusters (NGC 6266, 47 Tuc, and Terzan 5), and from the sum of all millisecond pulsars detected as individual point sources by \textit{Fermi}. The gamma-ray spectrum measured from millisecond pulsars and from globular clusters (whose emission is believed to be dominated by millisecond pulsars) is consistently softer than that of the observed excess at energies below $\sim$1 GeV. See text for details.}
\label{pulsarspec}
\end{figure}

To evade this conclusion, however, one could contemplate an additional (and less constrained) millisecond pulsar population associated with the Milky Way's central stellar cluster. This scenario can be motivated by the fact that globular clusters are known to contain large numbers of millisecond pulsars, presumably as a consequence of their very high stellar densities. If our galaxy's central stellar cluster contains a large number of millisecond pulsars with an extremely concentrated distribution (with a number density that scales approximately as $n_{\rm MSP} \propto r^{-2.4}$), those sources could plausibly account for much of the gamma-ray excess observed within the inner $\sim$1$^{\circ}$ around the Galactic Center~\cite{Hooper:2010mq,Abazajian:2010zy,Hooper:2011ti,Abazajian:2012pn,Gordon:2013vta,Abazajian:2014fta}. It is much more challenging, however, to imagine that millisecond pulsars could account for the more extended component of this excess, which we have shown to be present out to at least $\sim$$10^{\circ}$ from the Galactic Center. Expectations for the Inner Galaxy's pulsar population are not consistent with such an extended distribution. Furthermore, if the required number of millisecond pulsars were present $\sim$$10^{\circ}$ ($\sim$1.5 kpc) north or south of the Galactic Center, a significant number of these sources would have been resolved by \emph{Fermi} and appeared within the 2FGL catalog (assuming that the pulsars in question have a similar luminosity function to other observed millisecond pulsars)~\cite{Fermi-LAT:2011iqa,Gregoire:2013yta,Hooper:2013nhl}. The lack of such resolved sources strongly limits the abundance of millisecond pulsars in the region of the Inner Galaxy. Furthermore, the shape of the gamma-ray spectrum observed from resolved millisecond pulsars and from globular clusters (whose emission is believed to be dominated by millisecond pulsars) appears to be not-insignificantly softer than that of the gamma-ray excess observed from the Inner Galaxy. In Fig.~\ref{pulsarspec}, we compare the spectral shape of the gamma-ray excess to that measured from a number of globular clusters, and from the sum of all resolved millisecond pulsars. Here, we have selected the three highest significance globular clusters (NGC 6266, 47 Tuc, and Terzan 5), and plotted their best fit spectra as reported by the \textit{Fermi} Collaboration~\cite{collaboration:2010bb}. For the emission from resolved millisecond pulsars, we include the 37 sources as described in Ref.~\cite{Hooper:2013nhl}. Although each of these spectral shapes provides a reasonably good fit to the high-energy spectrum, they also each significantly exceed the amount of emission that is observed at energies below $\sim$1 GeV.  This comparison further disfavors millisecond pulsars as the source of the observed gamma-ray excess.

The near future offers encouraging prospects for detecting further evidence in support of a dark matter interpretation of this signal. The dark matter mass and annihilation cross section implied by the gamma-ray excess is similar to \emph{Fermi}'s sensitivity from observations of dwarf spheroidal galaxies. In fact, the \emph{Fermi} Collaboration has reported a modestly statistically significant excess ($\sim$2-3$\sigma$) in their search for annihilating dark matter particles in dwarf galaxies. If interpreted as a detection of dark matter, this observation would imply a similar mass and cross section to that favored by our analysis~\cite{Ackermann:2013yva}.  A similar ($\sim$$3\sigma$) excess has also been reported from the direction of the Virgo Cluster~\cite{Han:2012uw,MaciasRamirez:2012mk}. With the full dataset anticipated from \emph{Fermi}'s 10 year mission, it may be possible to make statistically significant detections of dark matter annihilation products from a few of the brightest dwarf galaxies, galaxy clusters, and perhaps nearby dark matter subhalos~\cite{Berlin:2013dva}.  Anticipated measurements of the cosmic-ray antiproton-to-proton ratio by \emph{AMS} may also be sensitive to annihilating dark matter with the characteristics implied by our analysis~\cite{Cirelli:2013hv,Fornengo:2013xda}.

\section{Summary and Conclusions}
\label{summary}

In this study, we have revisited and scrutinized the gamma-ray emission from the central regions of the Milky Way, as measured by the \textit{Fermi} Gamma-Ray Space Telescope. In doing so, we have confirmed a robust and highly statistically significant excess, with a spectrum and angular distribution that is in excellent agreement with that expected from annihilating dark matter. The signal is distributed with approximate spherical symmetry around the Galactic Center, with a flux that falls off as $F_{\gamma} \propto r^{-(2.2-2.6)}$, implying a dark matter distribution of $\rho \propto r^{-\gamma}$, with $\gamma \simeq 1.1-1.3$. The spectrum of the excess peaks at $\sim$1-3 GeV, and is well fit by 36-51 GeV dark matter particles annihilating to $b\bar{b}$. The annihilation cross section required to normalize this signal is $\sigma v = (1.9-2.8) \times 10^{-26}$ cm$^3$/s (for a local dark matter density of 0.4 GeV/cm$^3$), in good agreement with the value predicted for a simple thermal relic. In particular, a dark matter particle with this cross section will freeze-out of thermal equilibrium in the early universe to yield an abundance approximately equal to the measured cosmological dark matter density (for the range of masses and cross sections favored for other annihilation channels, see Sec.~\ref{darkmatter}).

In addition to carrying out two different analyses (as described in Secs.~\ref{inner} and~\ref{center}), subject to different systematic uncertainties, we have applied a number of tests to our results in order to more stringently determine whether the characteristics of the observed excess are in fact robust and consistent with the signal predicted from annihilating dark matter. These tests uniformly confirm that the signal is present throughout the Galactic Center and Inner Galaxy (extending out to angles of at least $10^{\circ}$ from the Galactic Center), without discernible spectral variation or significant departures from spherical symmetry. No known, anticipated, or proposed astrophysical diffuse emission mechanisms can account for this excess. And while a population of several thousand millisecond pulsars could have plausibly been responsible for much of the anomalous emission observed from within the innermost $\sim1^{\circ}-2^{\circ}$ around the Galactic Center, the extension of this signal into regions well beyond the confines of the central stellar cluster strongly disfavors such objects as the primary source of this signal. In light of these considerations, we consider annihilating dark matter particles to be the leading explanation for the origin of this signal, with potentially profound implications for cosmology and particle physics.

\bigskip

\bigskip

{\it Acknowledgements}: We would like to thank Keith Bechtol, Eric Charles and Alex Drlica-Wagner for their 
help with the Fermi-LAT likelihood analysis, Oscar Macias-Ramirez and
Farhad Yusef-Zadeh for providing the 20 cm templates, and Simona Murgia, Jesse Thaler and Neal Weiner for helpful discussions. We particularly thank Francesca Calore, Ilias Cholis and Christoph Weniger for providing an independent cross-check of some of our sphericity results. We acknowledge the University of Chicago Research Computing Center for providing support for this work. DPF is supported in part by the NASA Fermi Guest Investigator Program, DH is supported by the Department of Energy, and TL is supported by NASA through Einstein Postdoctoral Award Number PF3-140110. TRS is supported by the U.S. Department of Energy under cooperative research agreement Contract Number DE-FG02-05ER41360.

\bibliography{gevexcess_resubmit}

\newpage

\begin{figure*}
\includegraphics[width=0.49\textwidth ]{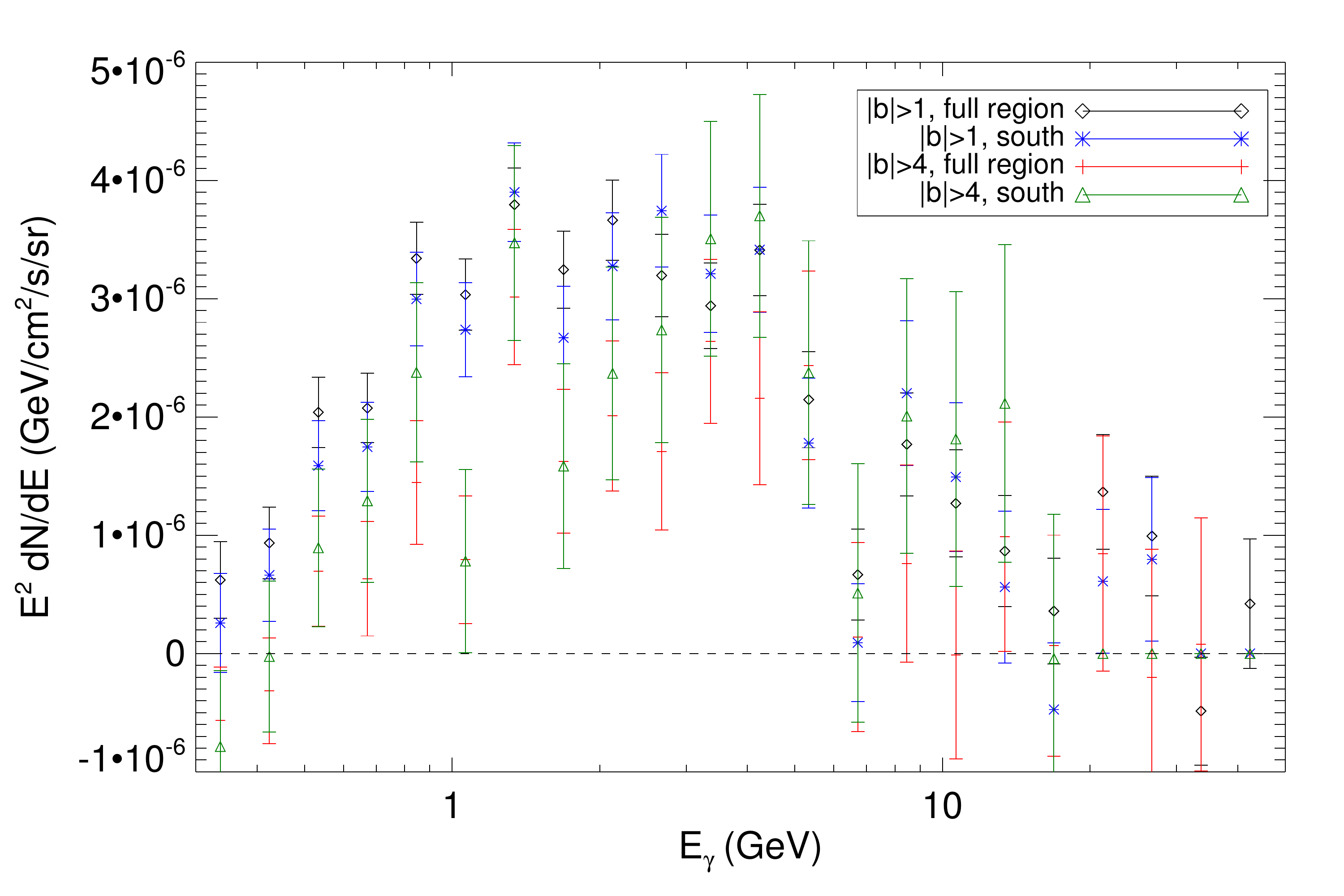}
\includegraphics[width=0.49\textwidth ]{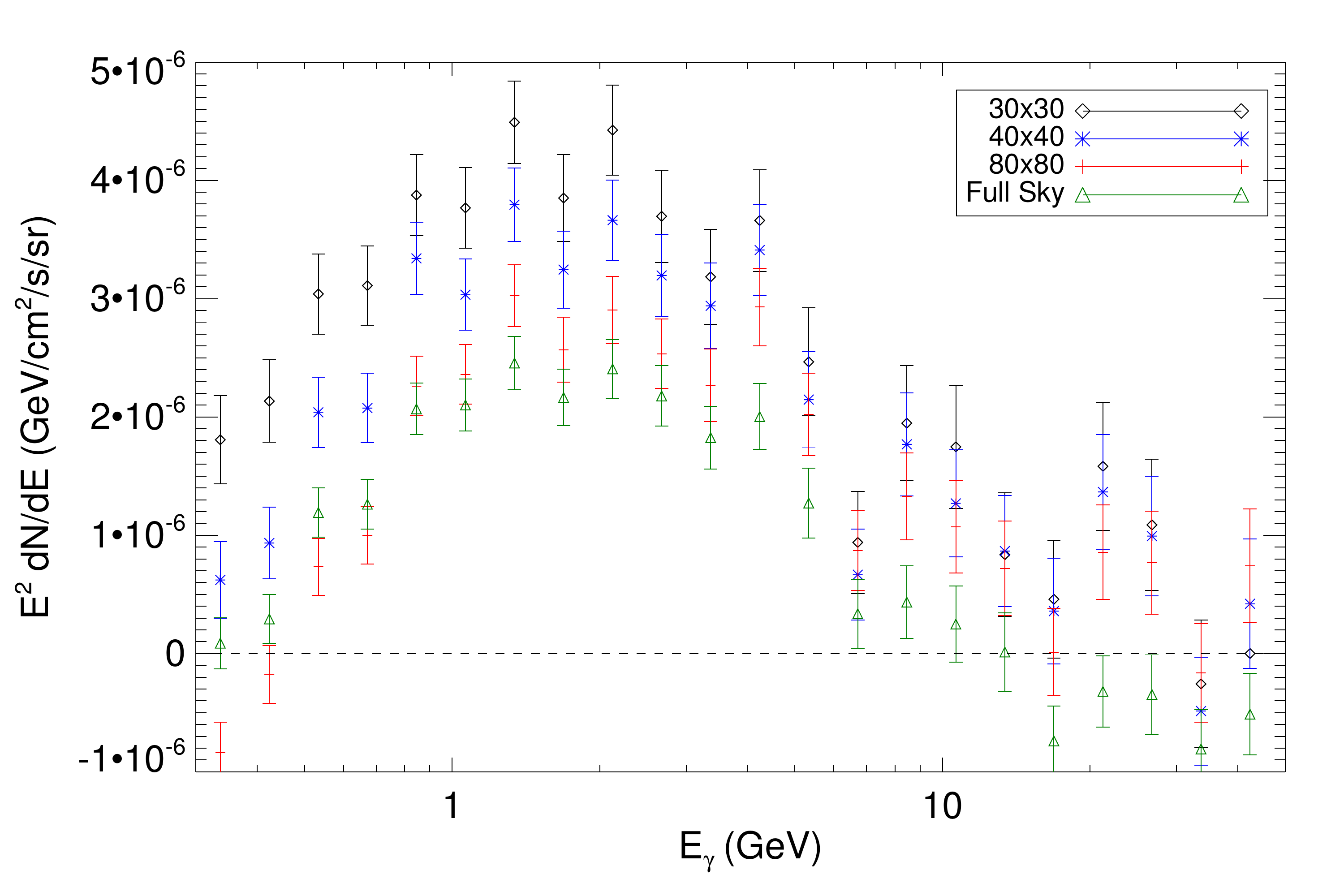}
\caption{The spectrum of the dark matter template found in our Inner Galaxy analysis when performing the fit over different regions of the sky. Using our standard ROI as a baseline, in the left panel we show variations of the Galactic plane mask and fits restricted to the southern sky, where backgrounds are typically somewhat lower, i.e. $|b| > 1^\circ$, $b < -1^\circ$, $|b| > 4^\circ$, and $b < -4^\circ$. All fits employ a single template for the Bubbles, the \texttt{p6v11} \emph{Fermi} diffuse model, and a dark matter motivated signal template with an inner profile slope of $\gamma=1.2$. In the right frame, we show the impact of varying the region over which the fit is performed. All ROIs have $|b| > 1^\circ$; aside from this Galactic plane mask, the ROIs are $|b| < 15^\circ, |l| < 15^\circ$ (``$30\times30$''), $|b| < 20^\circ, |l| < 20^\circ$ (``$40\times40$'', standard ROI),  $|b| < 40^\circ, |l| < 40^\circ$ (``$80\times80$''), and the full sky.}
\label{fig:igresults}
\end{figure*}

\begin{figure}
\includegraphics[width=0.45\textwidth ]{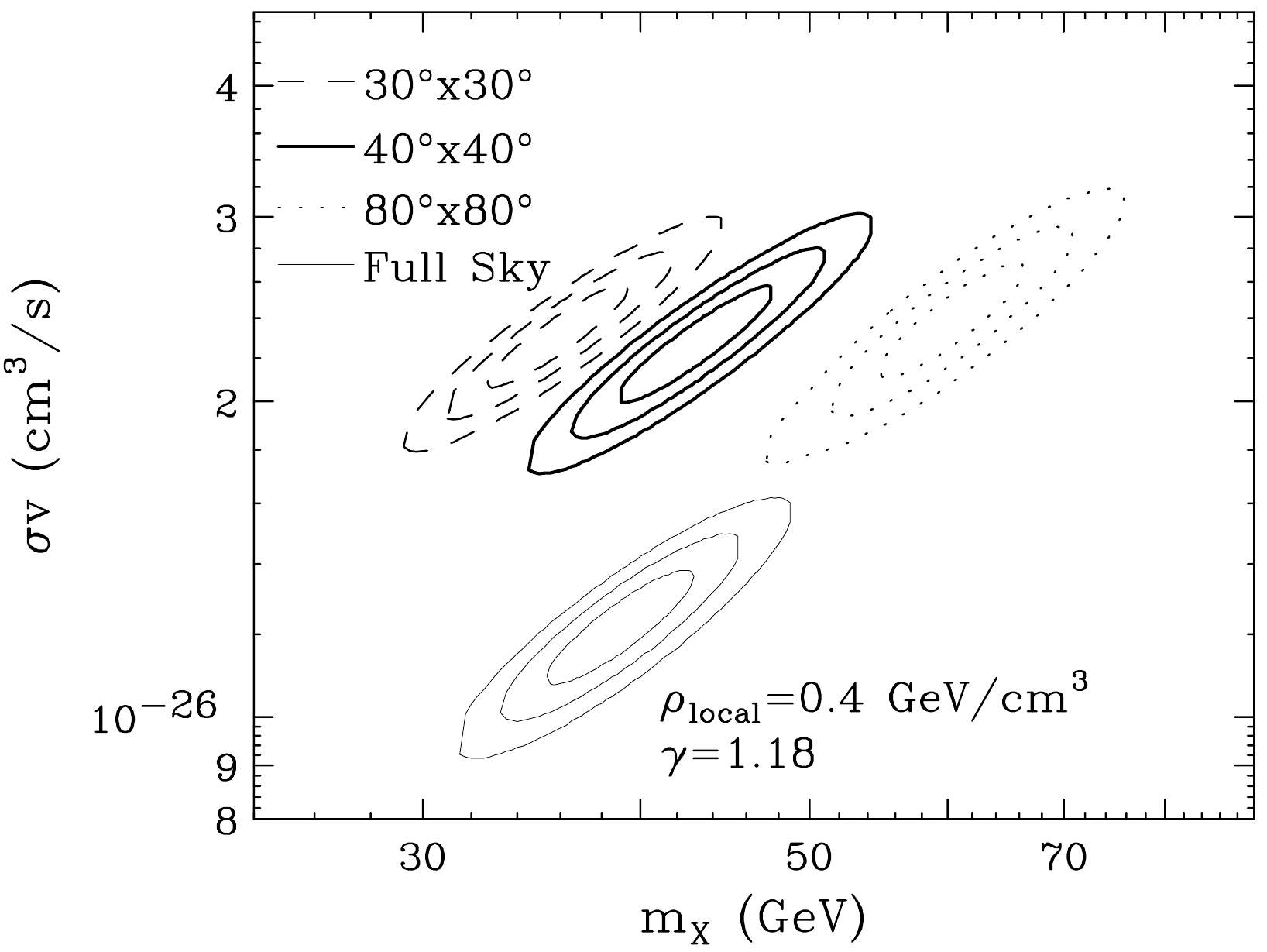}
\caption{A comparison of the regions of the dark matter mass-annihilation cross section plane (for annihilations to $b\bar{b}$ and an inner slope of 1.18) best fit by the spectrum found in our default Inner Galaxy analysis (fit over the central $40^{\circ} \times 40^{\circ}$ region), to that found for fits to other ROIs. See text for details.}
\label{regioncompare2}
\end{figure}

\begin{appendix}

\section{Stability Under Modifications to the Analysis}
\label{app:consistency}

\subsection{Changing the Region of Interest}
\label{app:roi}

In Fig.~\ref{fig:igresults}, we compare the spectrum correlated with the dark matter template (with $\gamma=1.2$) for variations of the ROI. In the left panel, we study different degrees of masking the Galactic Plane ($|b|>1^{\circ}$ and $|b|>4^{\circ}$), and the impact of performing the fit only in the southern sky (where the diffuse backgrounds are somewhat fainter) rather than in the full ROI. In the right panel, we show the impact of expanding or shrinking the ROI.

There is no evidence of asymmetry between the southern sky and the overall signal. Masking at $4^\circ$ gives rise to a similar spectral shape but a lower overall normalization than obtained with the $1^\circ$ mask, albeit with large error bars. As discussed in Sec.~\ref{morphology}, this may reflect a steepening of the spatial profile at larger distances from the GC, although the fainter emission at these larger radii is likely also more sensitive to mismodeling of the diffuse gamma-ray background.

Shrinking or expanding the size of the ROI also changes the height of the peak, while preserving a ``bump''-like spectrum that rises steeply at low energies and peaks around $\sim 2$ GeV. In general, larger ROIs give rise to lower normalizations for the signal. This effect appears to be driven by a higher normalization of the diffuse background model for larger ROIs; when the fit is confined to the inner Galaxy, the diffuse model prefers a lower coefficient than when fitted over the full sky, suggesting that the Pass 6 model has a tendency to overpredict the data in this region. This may also explain why larger ROIs prefer a somewhat steeper slope for the profile (higher $\gamma$); subtracting a larger background will lead to a greater relative decrease in the signal at large radii, where it is fainter. We also find evidence for substantial oversubtraction of the Galactic plane in larger ROIs, consistent with this hypothesis, as we will discuss in Appendix \ref{app:spherical}.

In Fig.~\ref{regioncompare2}, we show the regions of the dark matter mass-annihilation cross section plane favored by our fit, for several choices of the ROI (for annihilations to $b\bar{b}$ and an inner slope of 1.18). The degree of variation shown in this figure provides a measure of the systematic uncertainties involved in this determination; we see that the cross section is always very close to the thermal relic value, but the best-fit mass can shift substantially (from $\sim 35-60$ GeV). As previously, the contours are based on statistical errors only.

\begin{figure}
\includegraphics[width=0.49\textwidth ]{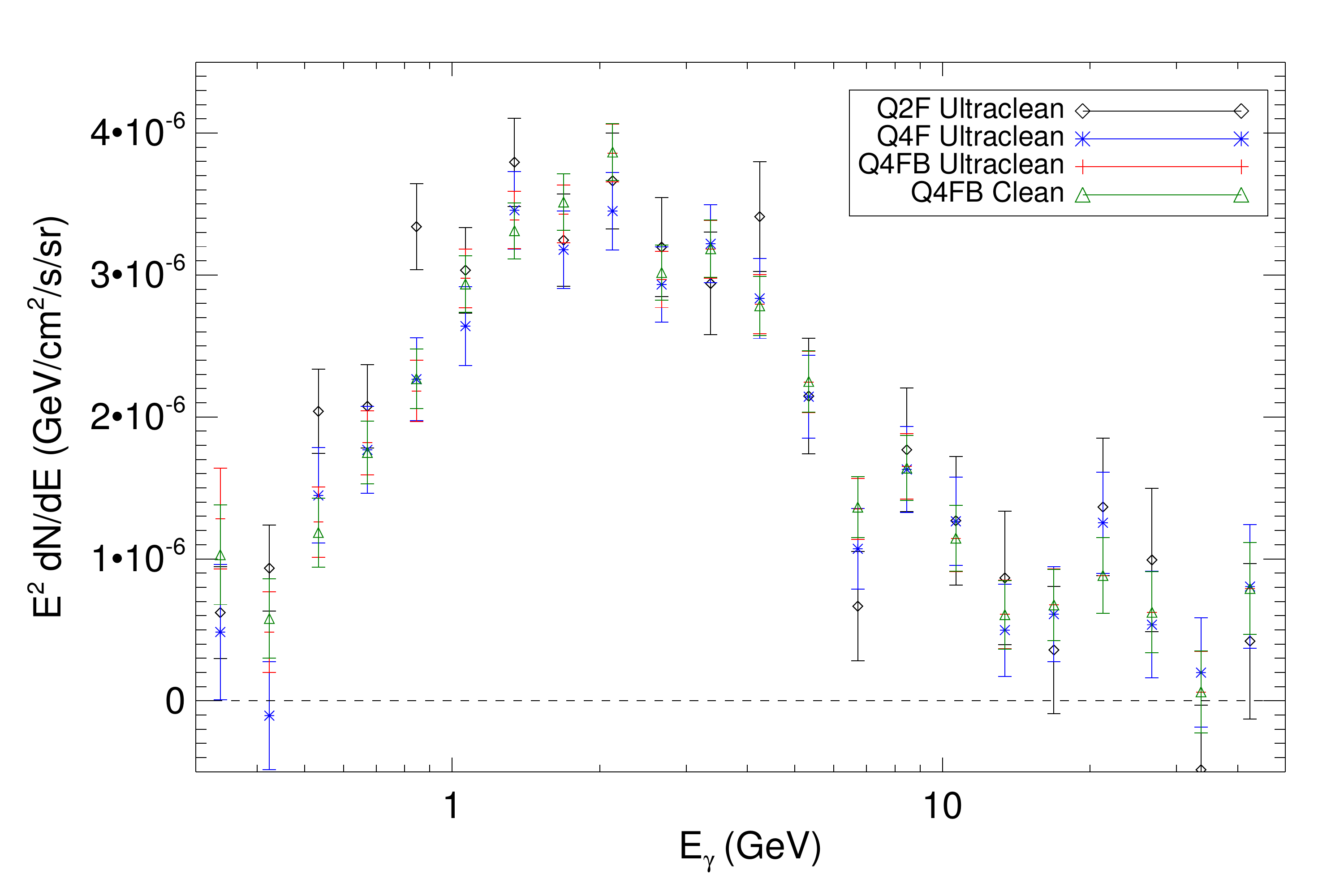}
\caption{The spectrum of emission associated with the dark matter template, corresponding to a generalized NFW profile with an inner slope of $\gamma=1.2$, as performed for four different event selections. Black diamonds indicate the spectrum extracted from the usual fit. The blue stars, red crosses and green triangles represent the spectra extracted from repeating our analysis on datasets without a CTBCORE cut, for (respectively) ULTRACLEAN front-converting events, all ULTRACLEAN events, and all CLEAN events.}
\label{fig:eventselec}
\end{figure}

\subsection{Varying the Event Selection}
\label{app:selections}

By default, we employ cuts on the CTBCORE parameter to improve angular resolution and minimize cross-leakage between the background and the signal. In an earlier version of this work, this resulted in a pronounced improvement in the consistency of the spectrum between different regions (in particular, in the hardness of the low-energy spectrum); however, this appears to have been due to a mismodeling of the background emission.\footnote{We suggested in that earlier work that the soft low-energy spectrum observed in the absence of a CTBCORE cut was likely due to contamination by mismodeled diffuse emission from the Galactic plane; our current results support that interpretation.} We now find that when the backgrounds are treated correctly, the spectrum has a consistent shape independent of the CTBCORE cut, and the significant changes in the tails of the point spread function (PSF) associated with a CTBCORE cut do not materially affect our results. Similarly, we find that our results are robust to the choice of ULTRACLEAN or CLEAN event selection, and to the inclusion or exclusion of back-converting events. We show the spectra extracted for several different event selections in Fig.~\ref{fig:eventselec}. Systematics associated with these choices are therefore unlikely to affect the observed excess.

\begin{figure*}
\includegraphics[width=0.4\textwidth ]{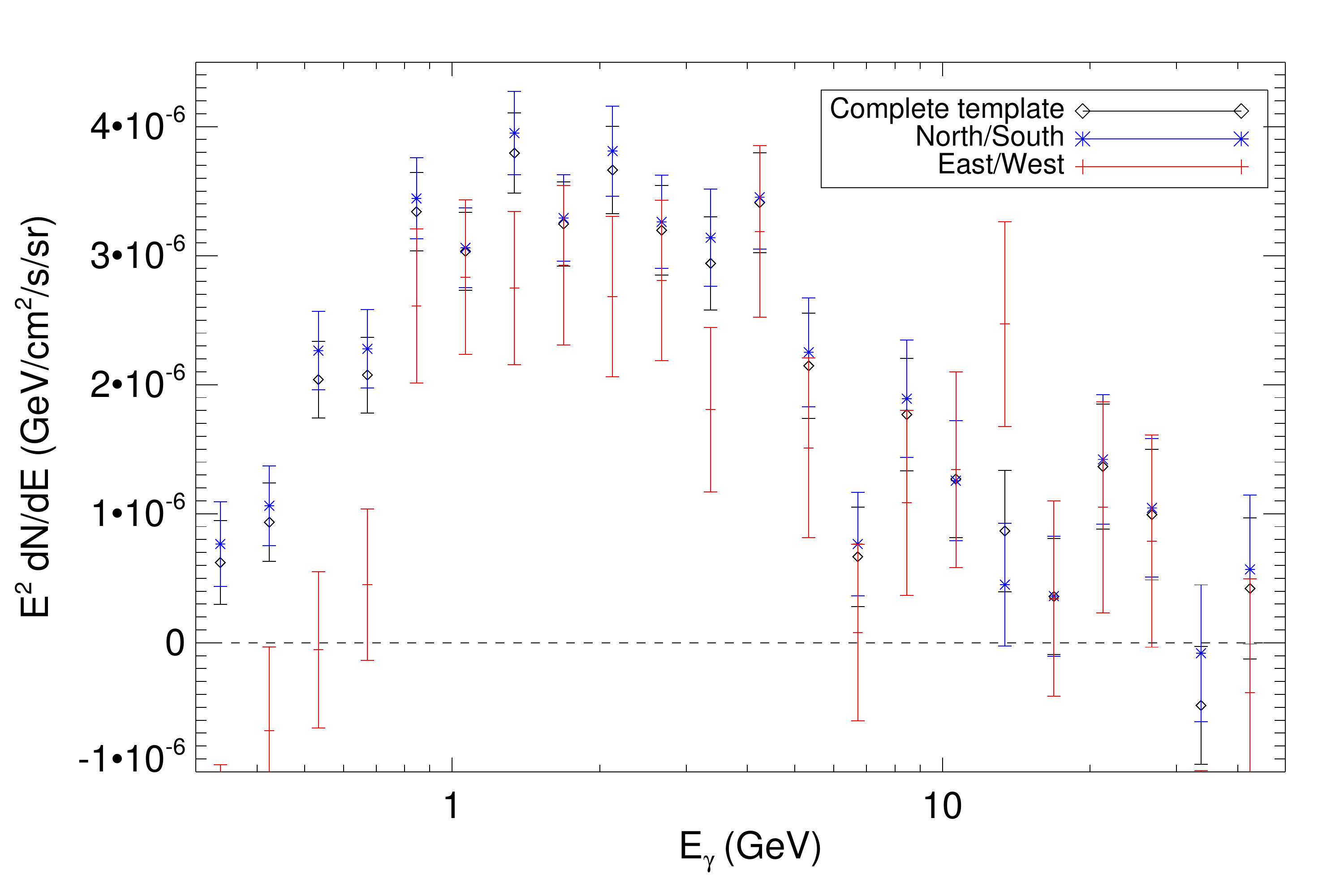} \\
\includegraphics[width=0.45\textwidth ]{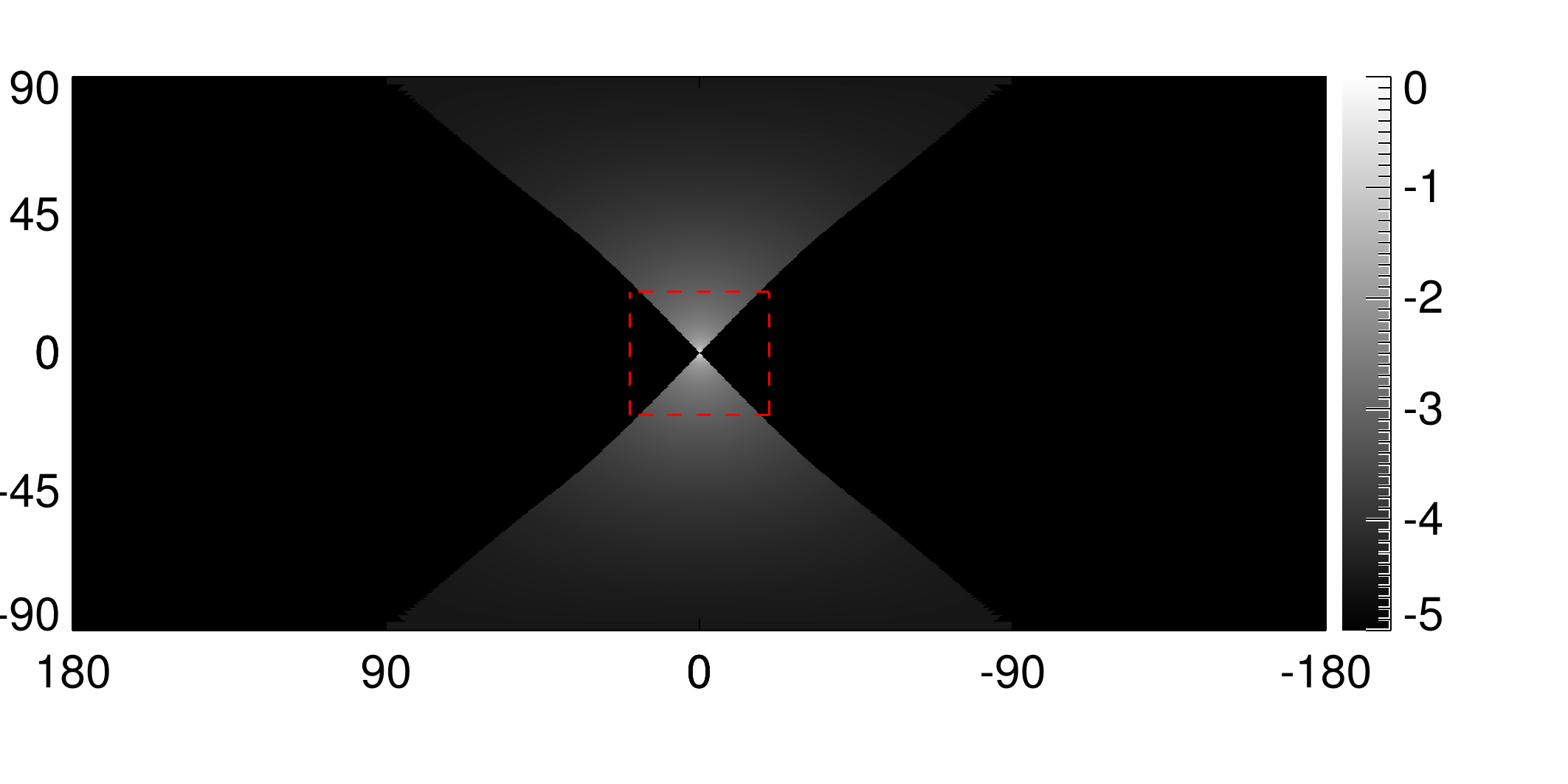}
\includegraphics[width=0.45\textwidth ]{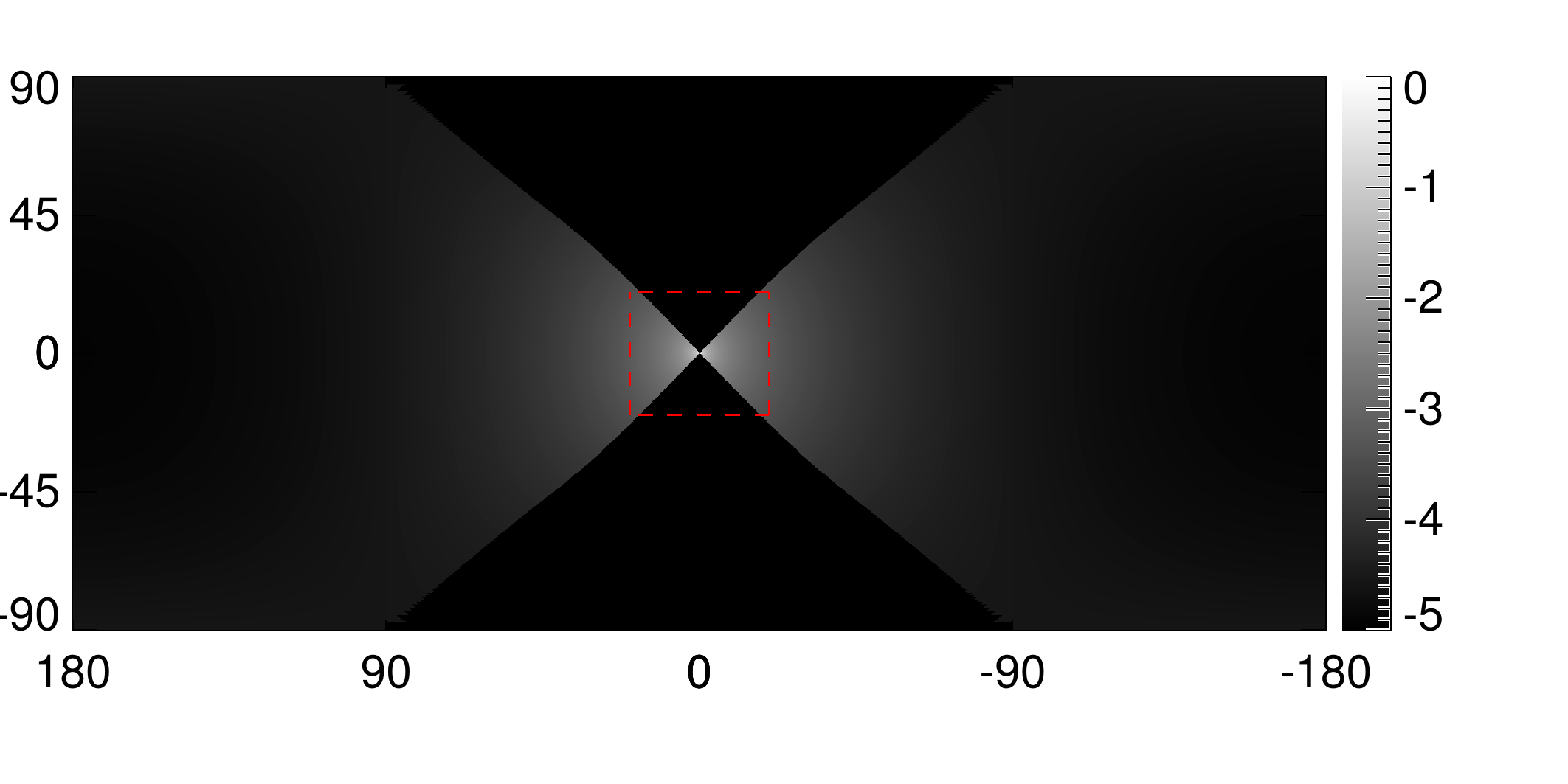}
\caption{In the upper frame, we show the spectra of the emission associated with the dark matter template, corresponding to a generalized NFW profile with an inner slope of $\gamma=1.2$, as performed over three regions of the sky. Black diamonds indicate the spectrum extracted from the usual fit, whereas the blue stars and red crosses represent the spectra correlated with the parts of the template in which $|b| > |l|$ and $|b| < |l|$, respectively (when the two are allowed to vary independently). The corresponding spatial templates are shown in the lower row, in logarithmic (base 10) units, normalized to the brightest point in each map. Red dashed lines indicate the boundaries of our standard ROI.}
\label{fig:igspherical}
\vskip 10mm
%
\includegraphics[width=0.4\textwidth ]{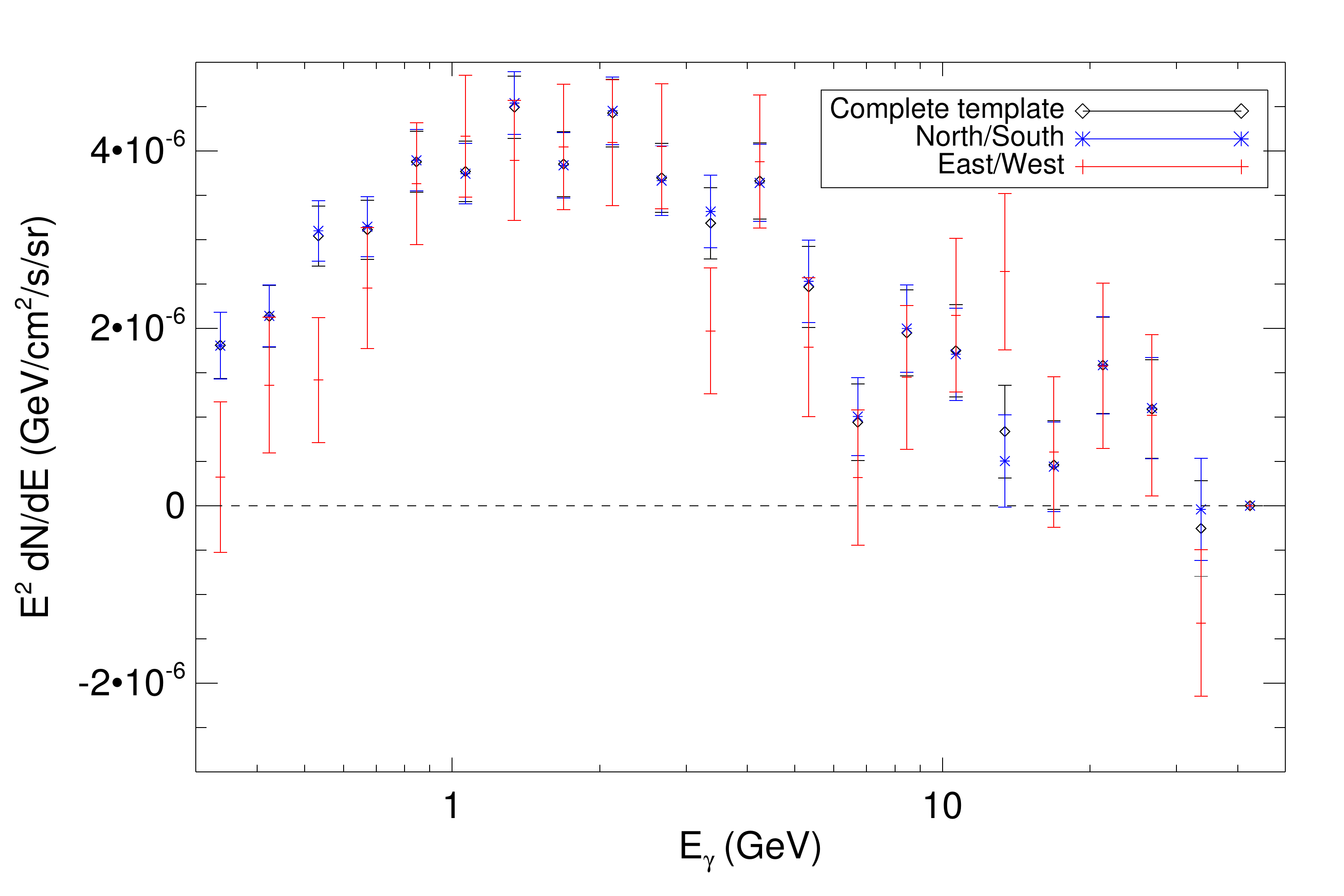}
\includegraphics[width=0.4\textwidth ]{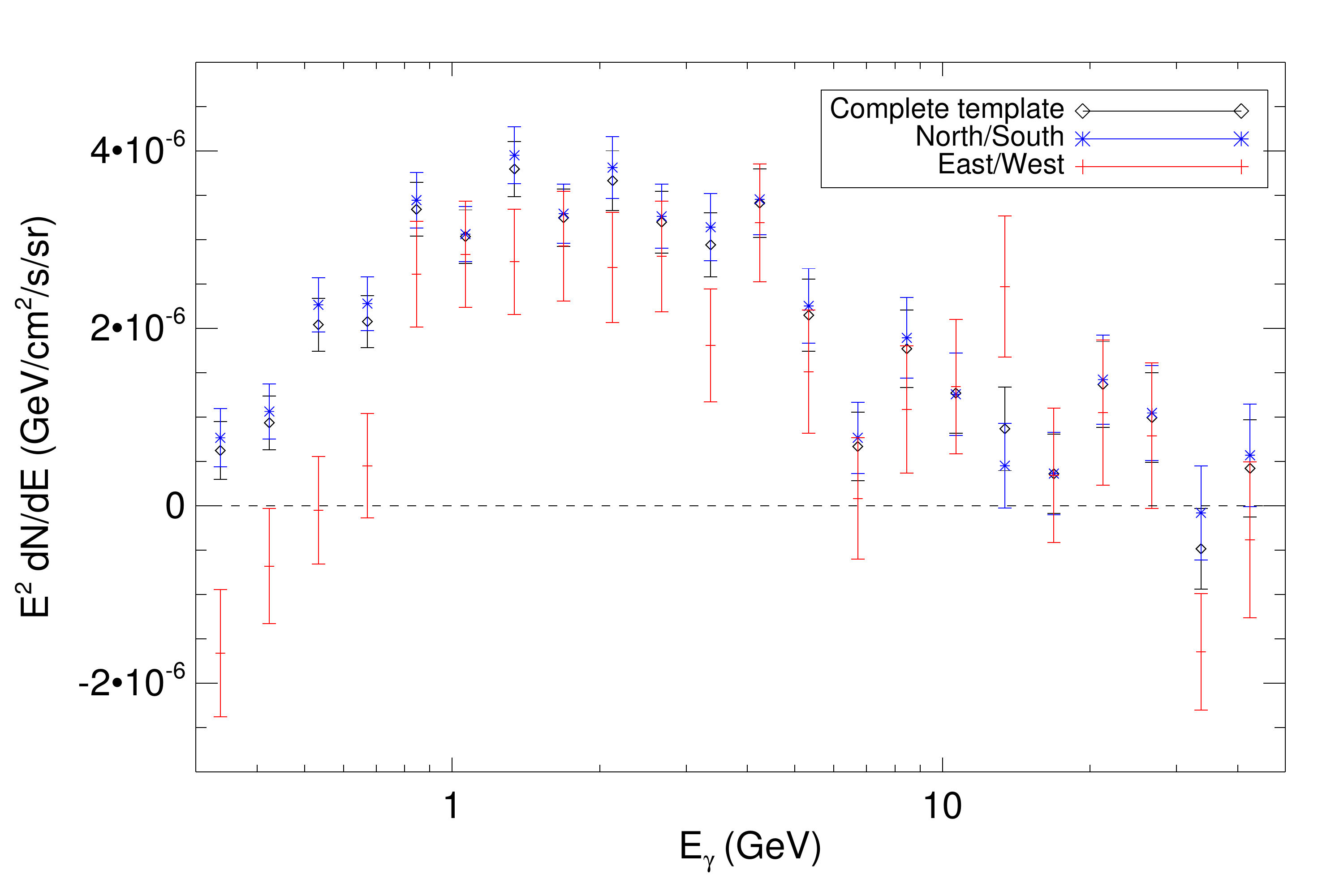} 
\\
\includegraphics[width=0.4\textwidth ]{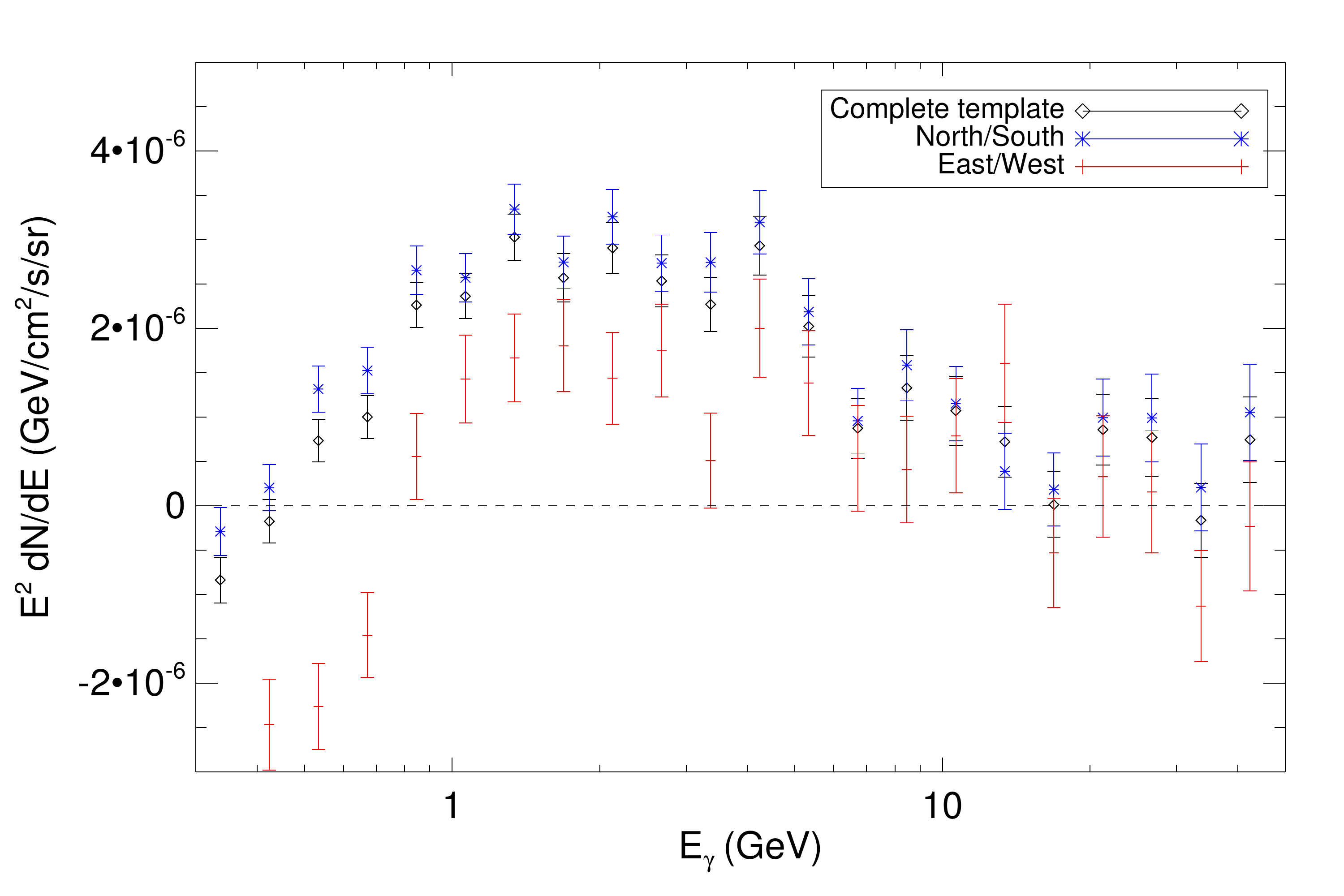}
\includegraphics[width=0.4\textwidth ]{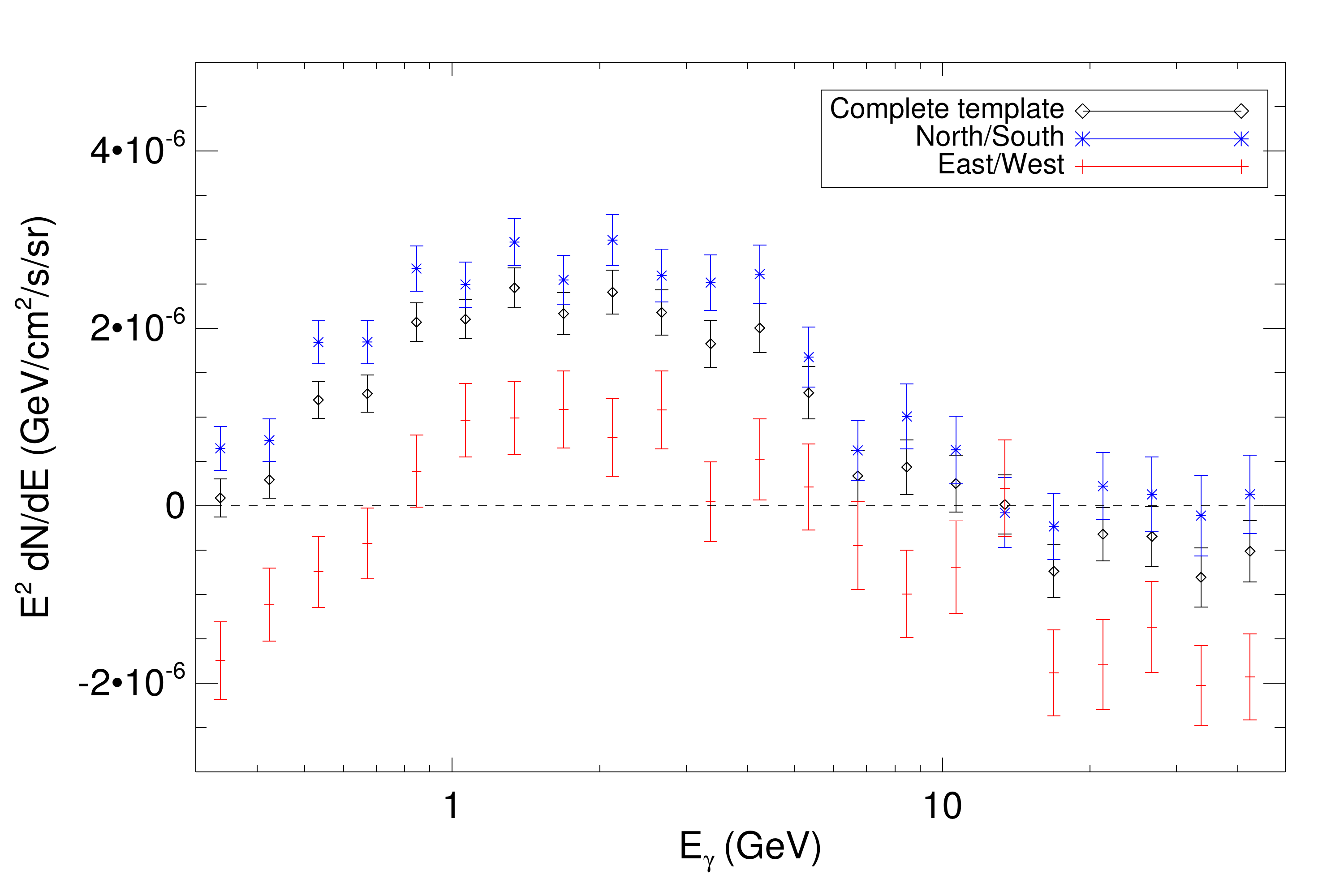}
\caption{As the upper panel of Fig.~\ref{fig:igspherical}, but for ROIs given by (upper left frame) $|b|,|l| < 15^\circ$, (upper right frame) $|b|, |l| < 20^\circ$, (lower left frame) $|b|, |l| < 40^\circ$, (lower right frame) full sky. In all cases the Galactic plane is masked for $|b| < 1^\circ$. We attribute the lower emission in the East/West quadrants in the larger ROIs to oversubtraction by the Galactic diffuse model along the Galactic plane. The slope parameter for the dark matter template is set to $\gamma=1.2$ in all cases.}
\label{fig:nsewroi}
\end{figure*}


 
\subsection{A Simplified Test of Elongation}
\label{app:spherical}

Probing the morphology of the Inner Galaxy excess is complicated by the bright emission correlated with the Galactic Plane. In Ref.~\cite{Hooper:2013rwa}, it proved difficult to robustly determine whether any signal was present outside of the regions occupied by the \emph{Fermi} Bubbles, as the regions both close to the Galactic Center and outside of the Bubbles were dominated by the bright emission from the Galactic Plane. The improved analysis presented in this paper mitigates this issue.

In addition to the detailed study of morphology described in Sec.~\ref{morphology}, we perform here a fit dividing the signal template into two independent templates, one with $|l| > |b|$ and the other with $|b| > |l|$. The former template favors the Galactic Plane, while the latter contains the \emph{Fermi} Bubbles. As previously, the fit also includes a single template for the Bubbles in addition to the \emph{Fermi} diffuse model and an isotropic offset. The extracted spectra of the signal templates are shown in Fig. \ref{fig:igspherical}. Both regions exhibit a clear spectral feature with broadly consistent shape and normalization, although the best-fit spectrum for the region with $|l| > |b|$ is generally slightly lower and has larger uncertainties. A lower normalization in these quadrants is expected, from the preference for a slight stretch perpendicular to the Galactic plane noted for the inner Galaxy in Sec.~\ref{morphology}.

As shown in Appendix \ref{app:roi}, the impact of the choice of ROI on the overall shape of the spectrum is modest. However, upon repeating this analysis in each of the ROIs, we find that the spectrum extracted from the quadrants lying along the Galactic plane ($|l| > |b|$) is much more sensitive to this choice. While a spectral ``bump'' peaked at $\sim 2$ GeV is always present, it appears to be superimposed on a negative offset which grows larger as the size of the ROI is increased. As discussed above, we believe this is due to oversubtraction along the plane by the Galactic diffuse model, which is most acute when the diffuse model normalization is determined by regions outside the inner Galaxy. We display this progression explicitly in Fig. \ref{fig:nsewroi}.

The relative heights of the spectra in the $|l| > |b|$ and $|b| > |l|$ regions are a reasonable proxy for sphericity of the signal; the former will be higher if the signal is elongated along the plane, and lower if the signal has perpendicular extension. Increased oversubtraction along the plane thus induces an apparent elongation of the signal perpendicular to the plane; we suspect this may be the origin of the apparent stretch perpendicular to the plane shown in Fig.~\ref{asymmetry}.

One might wonder whether this oversubtraction might give rise to apparent sphericity even if the true signal were elongated perpendicular to the plane. We argue that this is unlikely, as our results appear to converge to sphericity as the size of the ROI is reduced and the constraint on the normalization of the diffuse background is relaxed; the Galactic Center analysis, which includes the peak of the excess and the region where the signal-to-background ratio is largest, also prefers a spherical excess. 

We also performed the additional test of \emph{not} including any model for the point sources in the fit, allowing their flux to be absorbed by the NFW template. Since many point sources are clustered along the plane, over-subtracting them could bias the extracted morphology of the signal and hide an elongation along the plane. However, we found that even when no sources were subtracted, there was no ROI in which the spectrum extracted from the $|l| > |b|$ quadrants exceeded that for the $|b| > |l|$ quadrants.


\begin{figure}
\includegraphics[width=0.49\textwidth ]{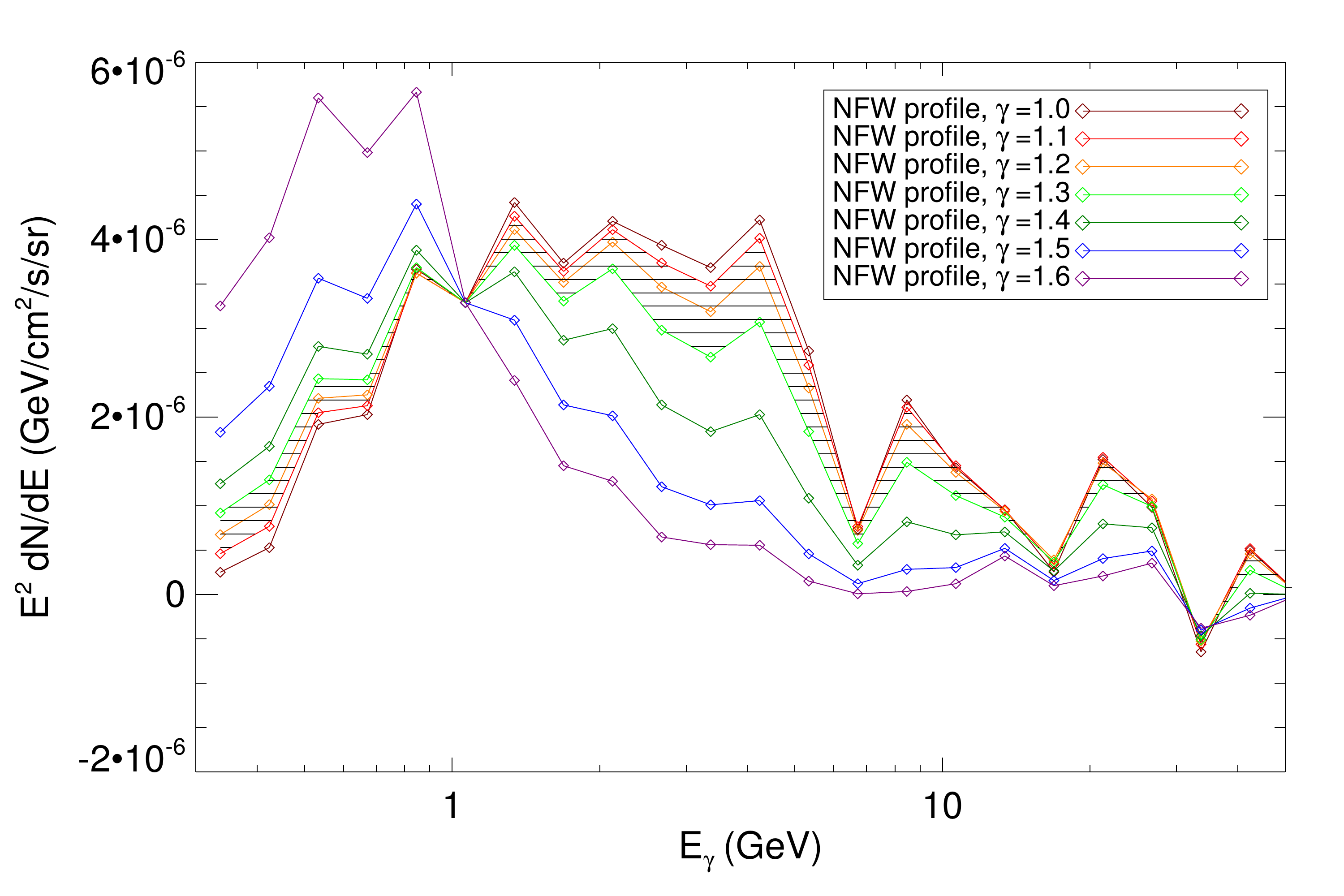}
\caption{The central values of the spectra of the dark matter templates for different values of the dark matter profile's inner slope, $\gamma$. To better facilitate comparison, each curve has been rescaled to match the $\gamma=1.0$ curve at 1 GeV. All fits have been performed with the \texttt{p6v11} \emph{Fermi} diffuse model, a single flat template for the Bubbles, and the dark matter signal template. The region between the $\gamma=1.1$ and $\gamma=1.3$ lines, preferred by the fit, is cross-hatched. Error bars are not shown to avoid cluttering the plot. In this preferred range, the spectra are remarkably consistent. Allowing very high values of $\gamma$ seems to pick up a much softer spectrum, likely due to contamination by the Galactic plane, but these high values of $\gamma$ provide commensurately worse fits to the data.}
\label{fig:slopebias}
\end{figure}

\subsection{Sensitivity of the Spectral Shape to the Assumed Morphology}

In our main analyses, we have derived spectra for the component associated with the dark matter template assuming a dark matter density profile with a given inner slope, $\gamma$. One might ask, however, to what degree uncertainties in the morphology of the template might bias the spectral shape extracted from our analysis. In Fig.~\ref{fig:slopebias}, we plot the (central values of the) spectrum found for the dark matter template in our Inner Galaxy analysis, for a number of values of $\gamma$. The shapes of the spectra are quite consistent, within the range of slopes favored by our fits ($\gamma=1.1-1.3$); the extracted spectrum is not highly sensitive to the specified signal morphology. However, for $\gamma \gtrsim 1.5$ this statement is no longer true: higher values of $\gamma$ pick up a much softer spectrum, which we ascribe to contamination from the Galactic plane at the edge of the mask. Of course, such high values of $\gamma$ also have much worse TS.

\section{Modeling of Background Diffuse Emission in the Inner Galaxy}
\label{app:diffuse}

\subsection{The \emph{Fermi} Bubbles}

The fit described in Sec.~\ref{inner} is a simplified version of the analysis performed in Ref.~\cite{Hooper:2013rwa}, where the spectrum of the Bubbles was allowed to vary with latitude. From the results in Ref.~\cite{Hooper:2013rwa}, it appears that this freedom is not necessary -- the spectrum and normalization of the Bubbles varies only slightly with Galactic latitude. 

\begin{figure*}
\includegraphics[width=0.7\textwidth ]{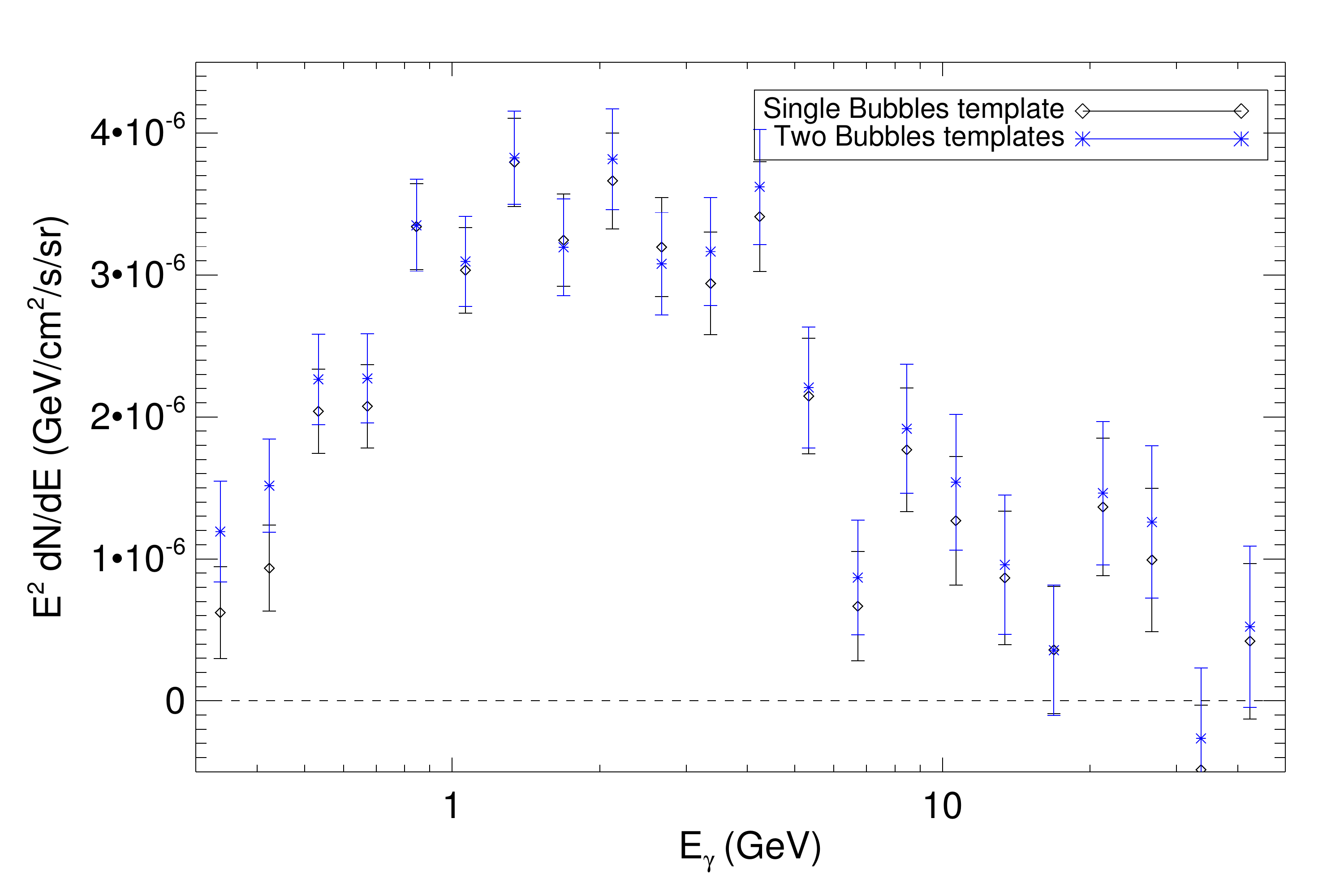}
\caption{The spectrum of the emission correlated with a dark matter template, corresponding to a generalized NFW profile with an inner slope of $\gamma=1.2$, obtained by a fit containing either a single template for the \emph{Fermi} Bubbles (black diamonds) or two templates for 10-degree-wide slices in Galactic latitude through the Bubbles (blue stars). The latter allows the spectrum of the \emph{Fermi} Bubbles to vary somewhat with Galactic latitude (there are only two templates, in contrast to the five employed in \cite{Hooper:2013rwa}, because the ROI only extends to $\pm 20$ degrees). }
\label{fig:simplebubble}
\end{figure*}

It is straightforward to reintroduce this freedom, and we show in Fig.~\ref{fig:simplebubble} the spectrum correlated with the dark matter template if this is done. Above 0.5 GeV, the spectrum of the excess is not significantly altered by fixing the Bubbles to have a single spectrum; at low energies, reintroducing this freedom slightly raises the extracted spectrum for the dark matter template.

%
%

\begin{figure*}
\includegraphics[width=0.7\textwidth ]{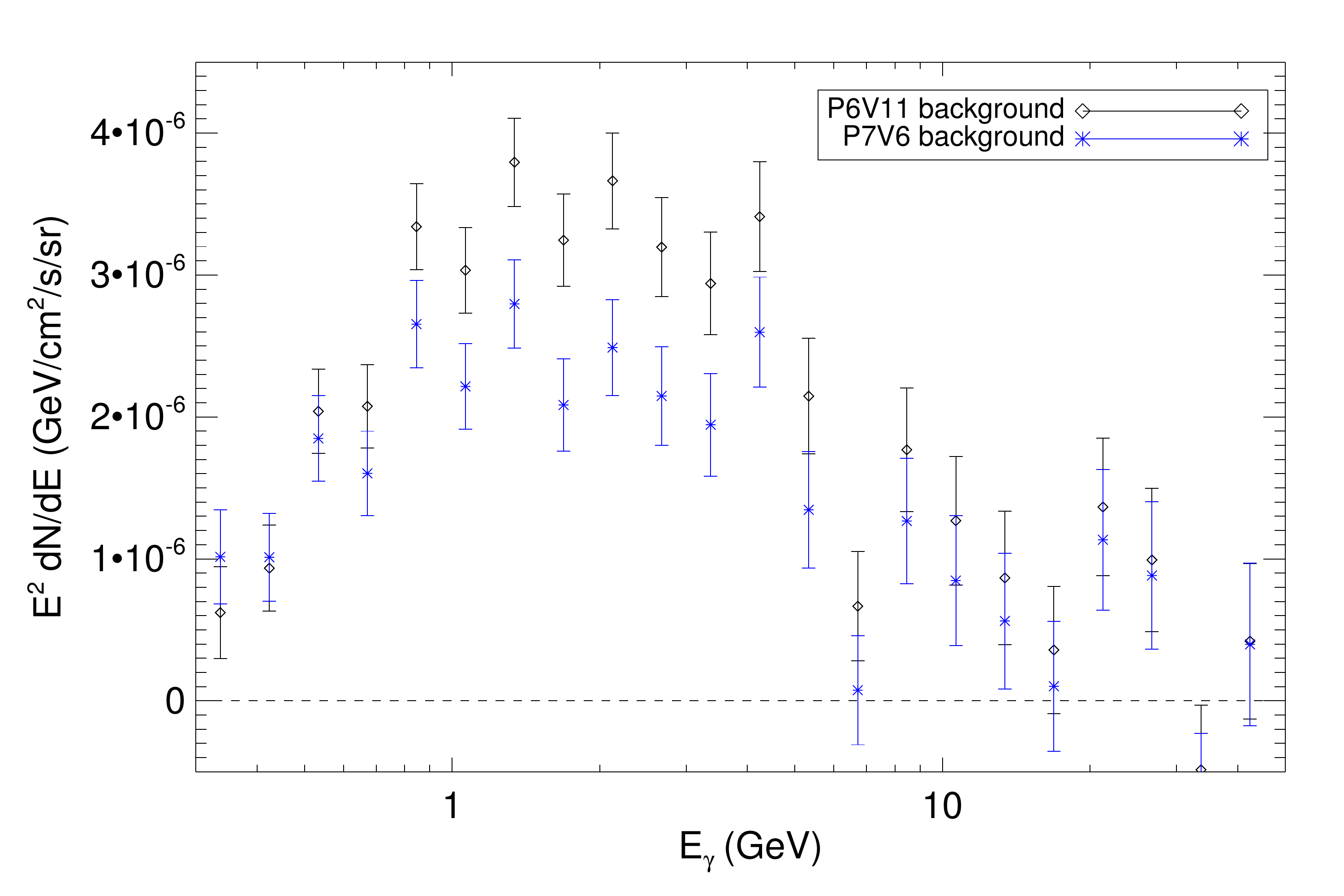}
\caption{The spectra of the emission correlated with a dark matter template, corresponding to a generalized NFW profile with an inner slope of $\gamma=1.2$, with the background modeled by the \texttt{p6v11} diffuse model (black diamonds) or the \texttt{p7v6} diffuse model (blue stars). In both cases, the fit also contains an isotropic offset and a template for the \emph{Fermi} Bubbles.}
\label{fig:p7diffmodel}
\end{figure*}

\begin{figure}
\includegraphics[width=0.45\textwidth ]{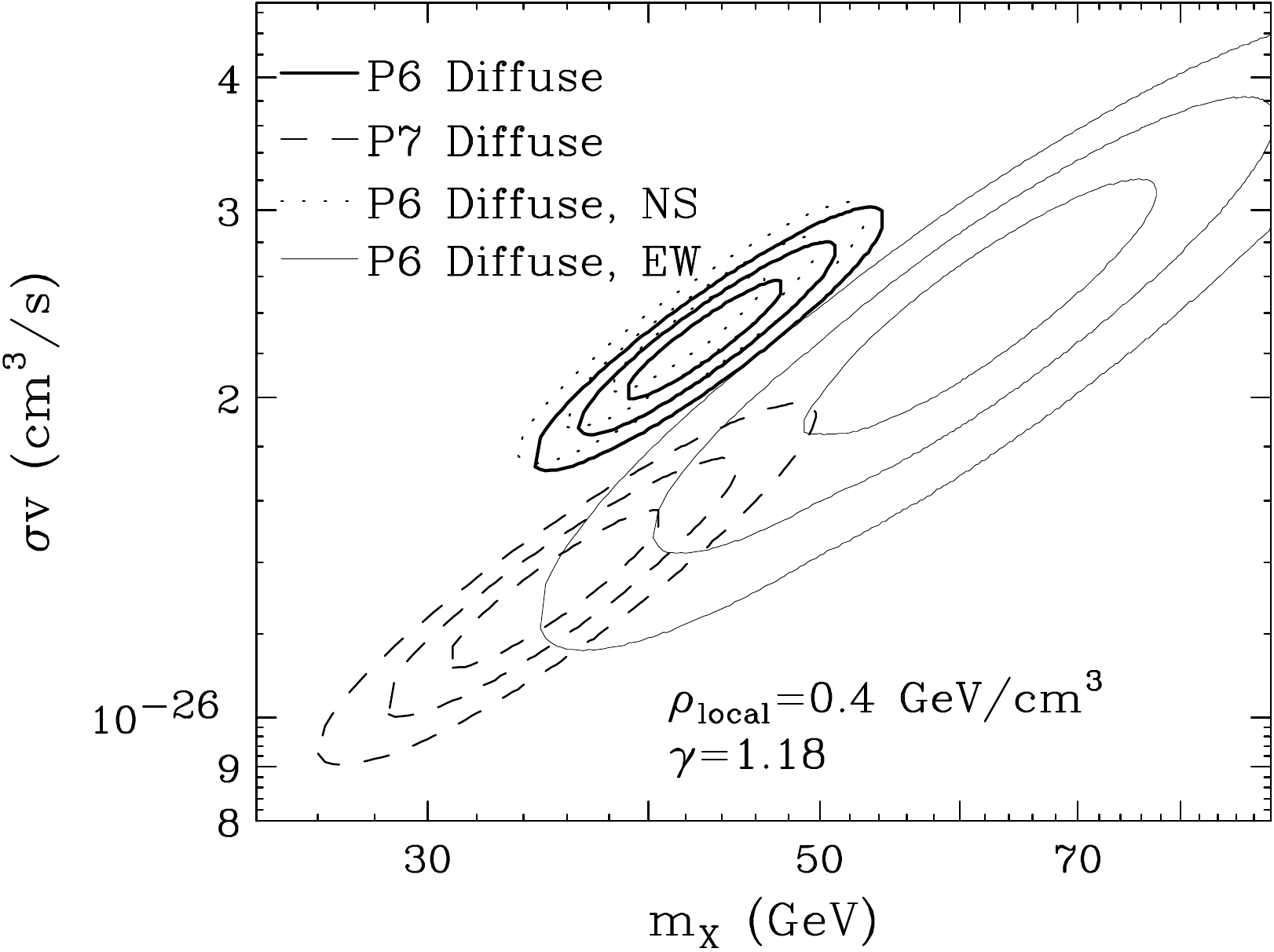}
\caption{A comparison of the regions of the dark matter mass-annihilation cross section plane (for annihilations to $b\bar{b}$) best fit by the spectrum found in our default Inner Galaxy analysis (using the Pass 6 Galactic diffuse model, and fit over the standard ROI), to that found for the spectra shown in Figs.~\ref{fig:igspherical} and~\ref{fig:p7diffmodel}. See text for details.}
\label{regioncompare}
\end{figure}

\subsection{The Choice of Diffuse Model}

Throughout our Inner Galaxy analysis, we employed the \texttt{p6v11} diffuse model released by the \textit{Fermi} Collaboration, rather than the more up-to-date \texttt{p7v6}  model. As noted earlier, this choice was made because the \texttt{p7v6} model contains artificial templates for the \emph{Fermi} Bubbles and other large-scale features (with fixed spectra), making it more difficult to interpret any residuals.

Having shown that a single flat-luminosity template for the Bubbles is sufficient to capture their contribution without biasing the spectrum of the signal template, one might also employ the \texttt{p7v6} model in \emph{addition} to an independent template for the Bubbles, in order to absorb any deviations between the true spectrum of the Bubbles and their description in the model. Unfortunately, the template for the \emph{Fermi} Bubbles employed in constructing the \texttt{p7v6} diffuse model (which is not separately characterized from the overall Galactic diffuse emission) is different to the one employed in our analysis, especially in the regions close to the Galactic plane. Consequently, this approach gives rise to residuals correlated with the spatial differences between these templates. For this reason, we employ the \texttt{p6v11} diffuse model for our principal analysis. However, using the \texttt{p7v6} model does not quantitatively change our results, although the peak of the spectrum is somewhat lower (yielding results more comparable to that obtained from the full-sky ROI with the \texttt{p6v11} model). A direct comparison of these two results is shown in Fig.~\ref{fig:p7diffmodel}.

In Fig.~\ref{regioncompare}, we compare the regions of the dark matter mass-annihilation cross section plane (for annihilations to $b\bar{b}$) that are best fit by the spectrum found in our default Inner Galaxy analysis (using the Pass 6 Galactic diffuse model, and fit over the $|l| < 20^\circ$, $20^\circ > |b|>1^{\circ}$ ROI), to that found for the spectra shown in Figs.~\ref{fig:igspherical} and~\ref{fig:p7diffmodel}. The excess is still clearly present and consistent with a dark matter interpretation, and the qualitative results do not change with choice of diffuse model.


\begin{figure*}
\includegraphics[width=0.40\textwidth ]{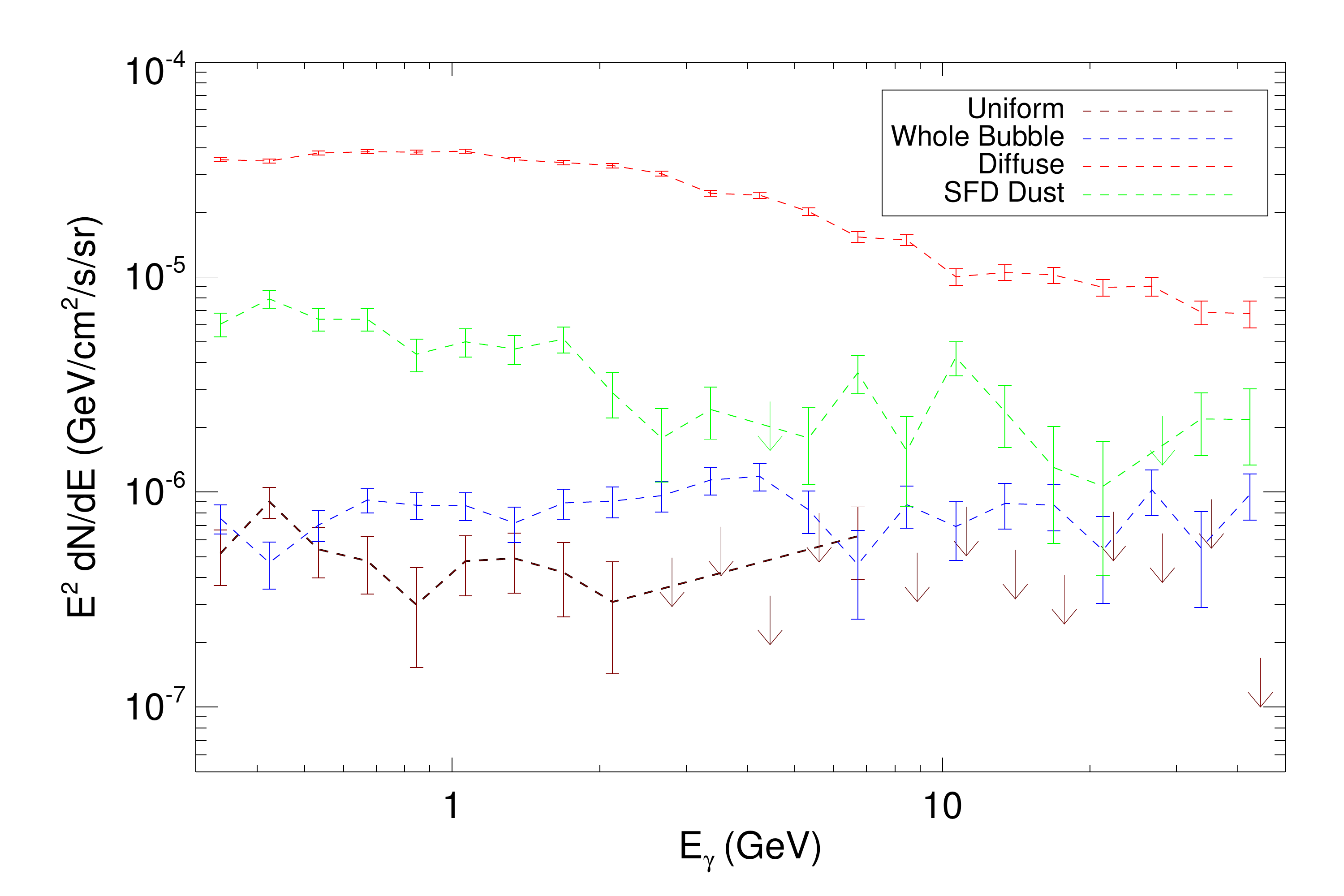}
\includegraphics[width=0.40\textwidth ]{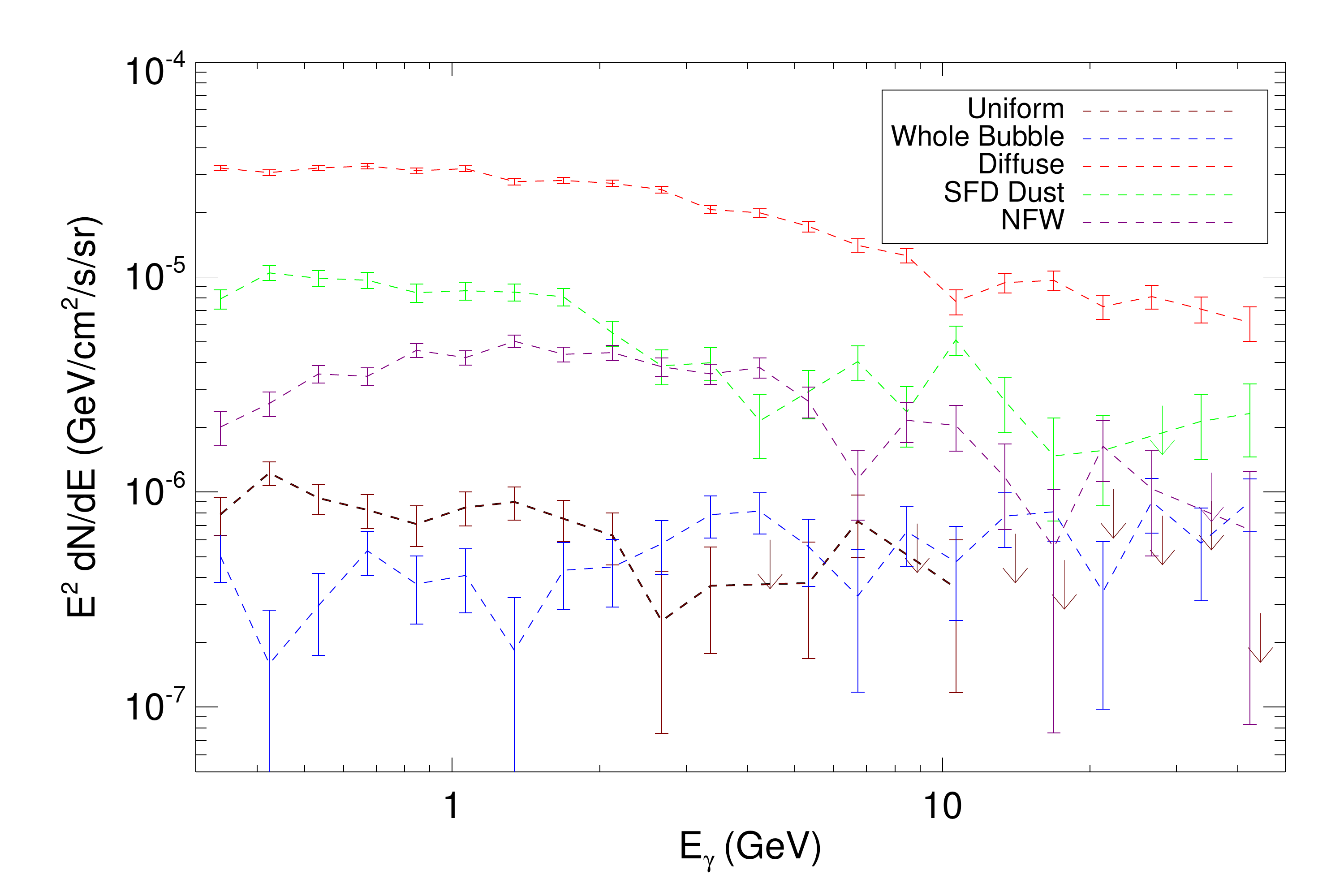}
\caption{In the left frame, we show the spectra correlated with the various templates, from a fit with the usual backgrounds as well as the Schlegel-Finkbeiner-Davis (SFD) dust map, with the standard ROI. The right frame shows the results of the same fit, but also including a dark matter template with $\gamma=1.2$. The spectra for the dust map and diffuse model represent the average flux correlated with those templates outside the $|b| < 1^\circ$ mask and within $5^\circ$ of the Galactic Center.}
\label{fig:coeffsextraSFD}
\end{figure*}

\begin{figure*}
\includegraphics[width=3.5in]{{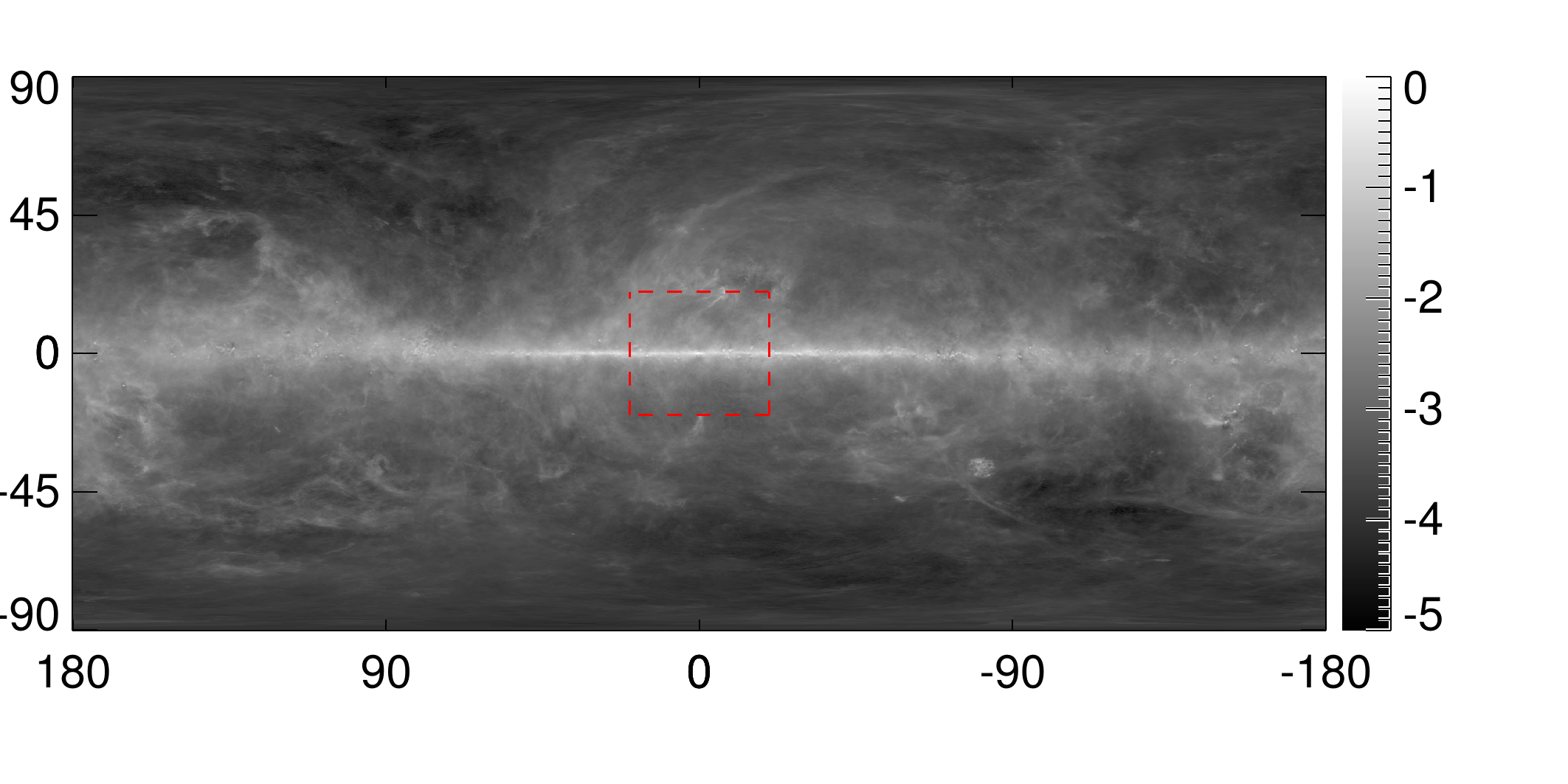}}
\includegraphics[width=3.5in] {{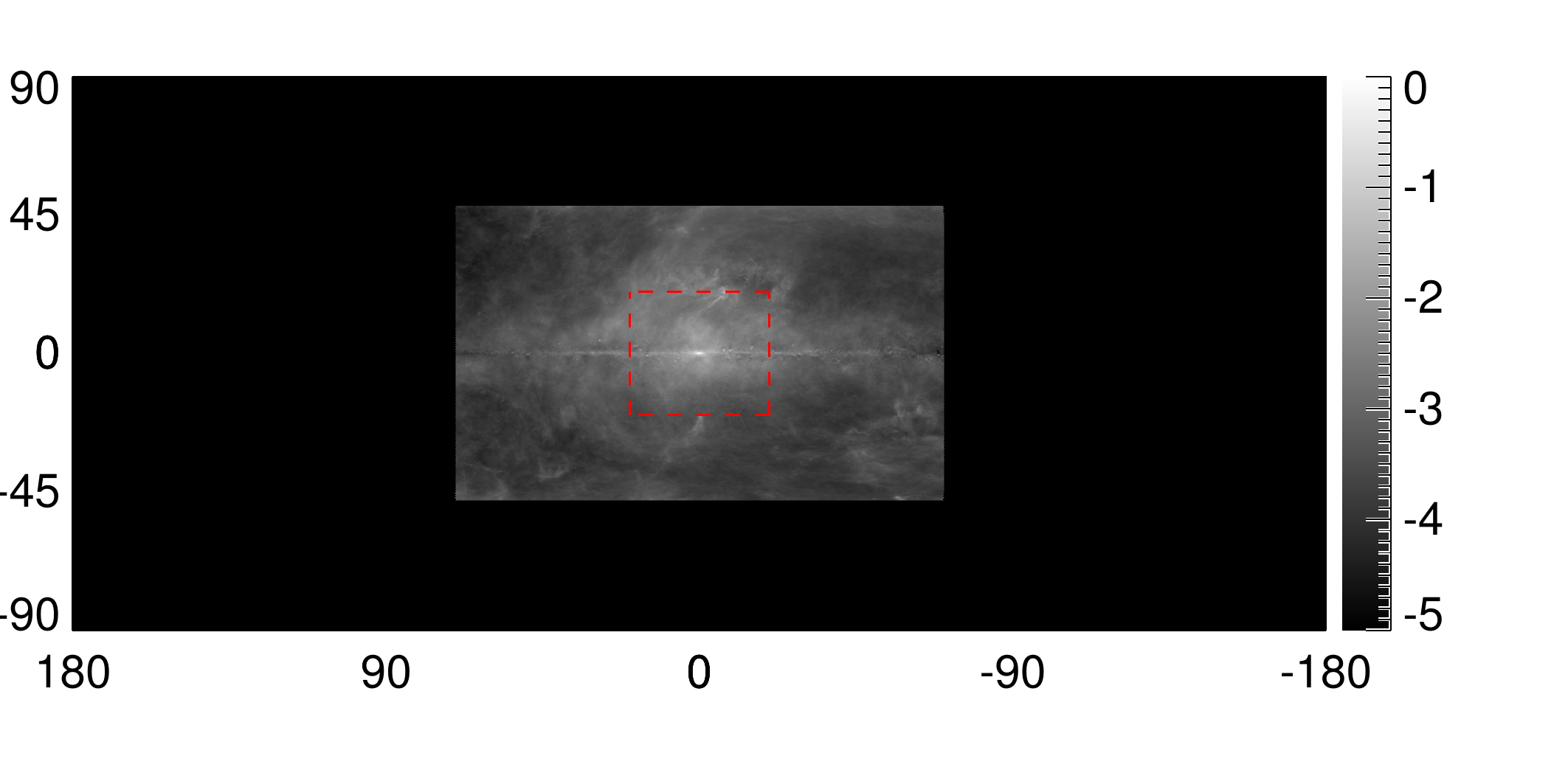}}
\caption{Left frame: The Schlegel-Finkbeiner-Davis dust map, used as a tracer for emission from proton-gas interactions. Right frame: An example of a dust-modulation template, created by multiplying the dust map by $f(r)/g(r)$, in the case where $f(r)$ is a projected squared NFW with $\gamma=0.8$. Red dashed lines indicate the boundaries of our standard ROI. All maps are given in logarithmic (base 10) units, normalized to the brightest point in each map. The modulated-dust template is artificially set to zero for $|b| > 45^{\circ}$ and $|l| > 70^{\circ}$, to avoid errors due to the denominator factor becoming small; as these bounds lie outside our ROI, they will not affect our results. See text for details.}
\label{fig:healcartdeltad}
\end{figure*}

\subsection{Variation in the $\pi^0$ Contribution to the Galactic Diffuse Emission}
\label{app:sfddust}

Although the spectrum of the observed excess does not appear to be consistent with gamma rays produced by interactions of proton cosmic rays with gas, one might wonder whether the \emph{difference} between the true spectrum and the model might give rise to an artificially peaked spectrum. While we fit the spectrum of emission correlated with the \emph{Fermi} diffuse model from the data, the model contains at least two principal emission components with quite different spectra (the gamma rays from the inverse Compton scattering of cosmic-ray electrons, and those from the interactions between cosmic-ray protons and gas), and their ratio is essentially fixed by our choice to use a single template for the diffuse Galactic emission (although we do allow for an arbitrary isotropic offset). Mismodeling of the cosmic-ray spectrum or density in the inner Galaxy could also give rise to residual differences between the data and model.

As a first step in exploring such issues, we consider relaxing the constraints on the background model by adding the Schlegel-Finkbeiner-Davis (SFD) map of interstellar dust \cite{Schlegel:1997yv} as an additional template. This dust map has previously been used effectively as a template for the gas-correlated gamma-ray emission~\cite{Dobler:2009xz,Su:2010qj}. By allowing its spectrum to vary independently of the \emph{Fermi} diffuse model, we hope to absorb systematic differences between the model and the data correlated with the gas. While the approximately spherical nature of the observed excess (see Sec.~\ref{morphology}) makes the dust template unlikely to absorb the majority of this signal, if the spectrum of the excess were to change drastically as a result of this new component, that could indicate a systematic uncertainty associated with the background modeling. 

In Fig.~\ref{fig:coeffsextraSFD}, we show the results of a template fit using the three background templates described in Sec.~\ref{inner}, as well as the SFD dust map. The additional template improves the fit markedly, and absorbs significant emission across a broad range of energies. However, when the dark matter template is added, the fit still strongly prefers its presence and recovers the familiar spectrum with power peaked at $\sim$1-3 GeV.


\begin{figure*}
\includegraphics[width=3.0in]{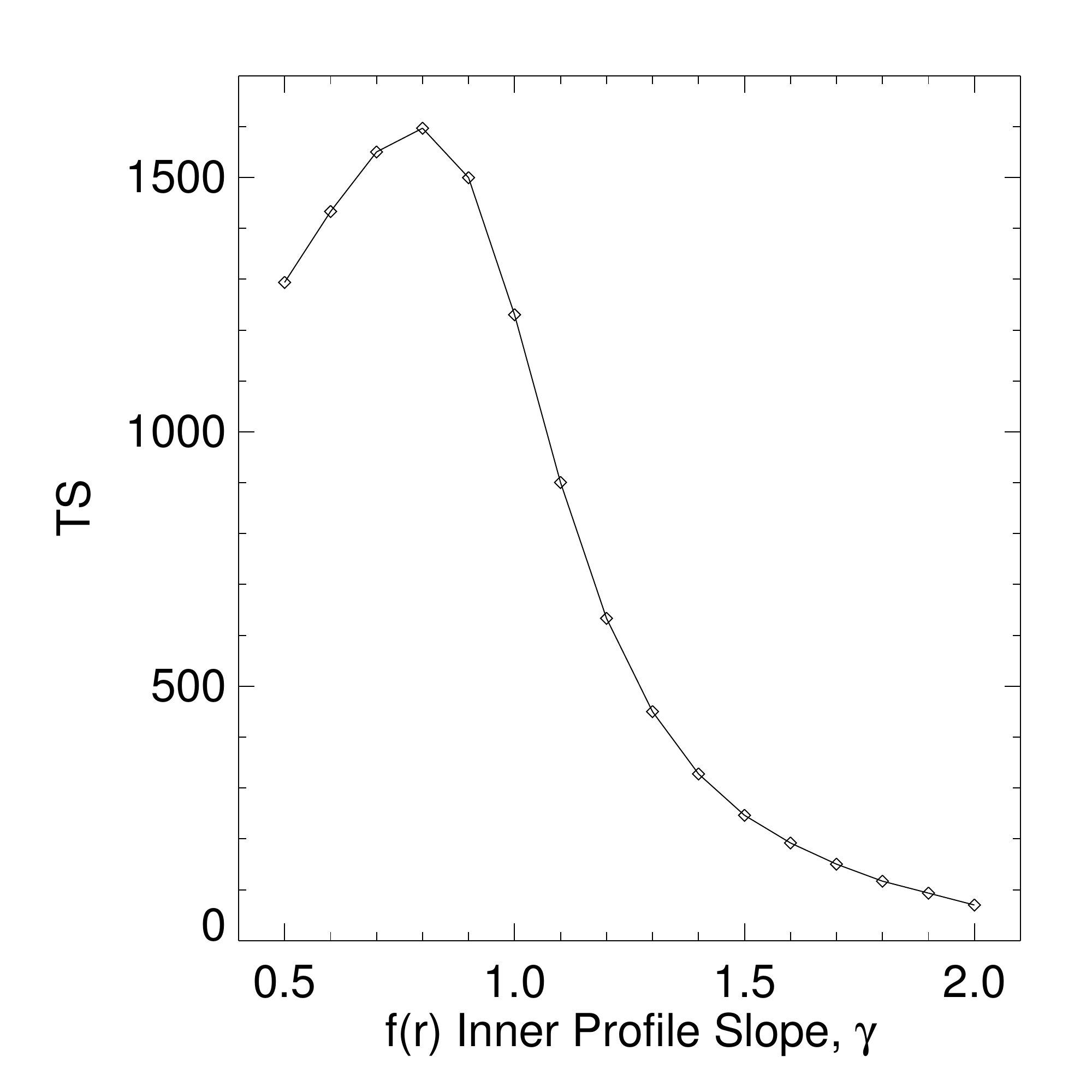}
\includegraphics[width=3.0in]{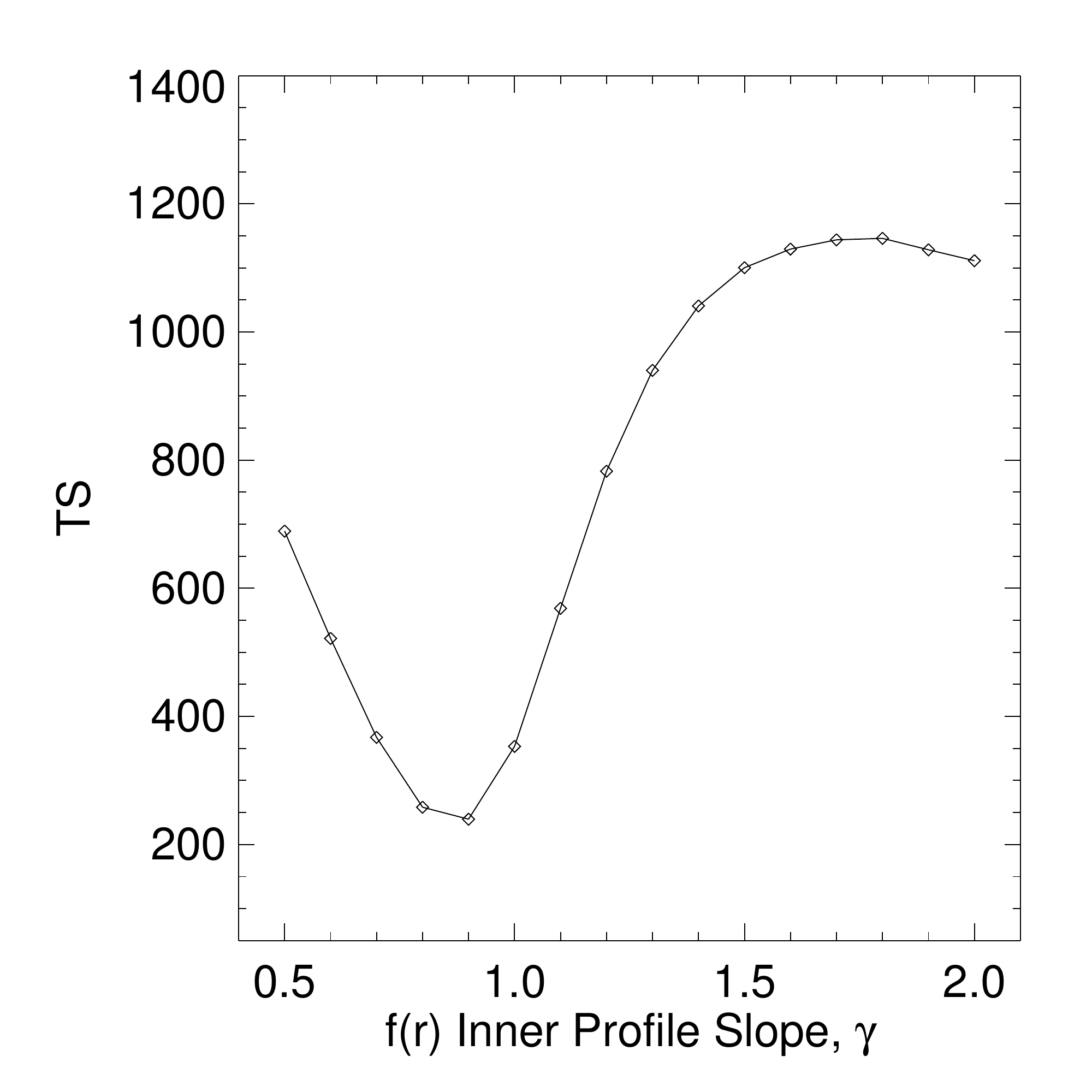}
\caption{In the left frame, we plot the improvement in TS between the template fit performed using known backgrounds and a modulated Schlegel-Finkbeiner-Davis dust map, and the fit using only the known backgrounds, as a function of the inner profile slope $\gamma$ of the $f(r)$ template used in constructing the modulation. In the right frame, we show the improvement in TS when a $\gamma=1.18$ dark matter template is added to the previous fit, as a function of the inner profile slope $\gamma$ of the $f(r)$ template.}
\label{fig:chisqdeltad}
\end{figure*}

\begin{figure*}
\includegraphics[width=0.40\textwidth ]{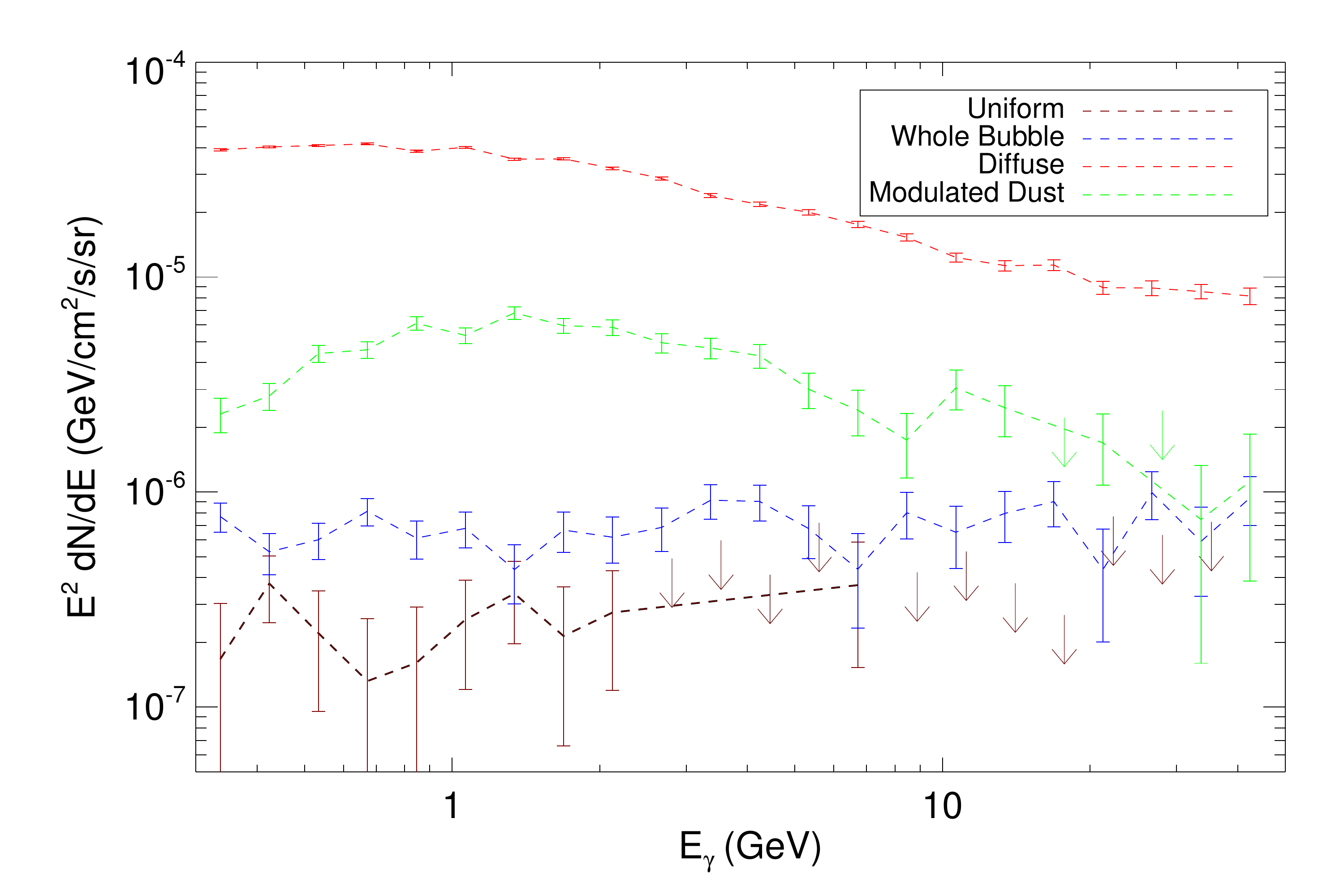}
\includegraphics[width=0.40\textwidth ]{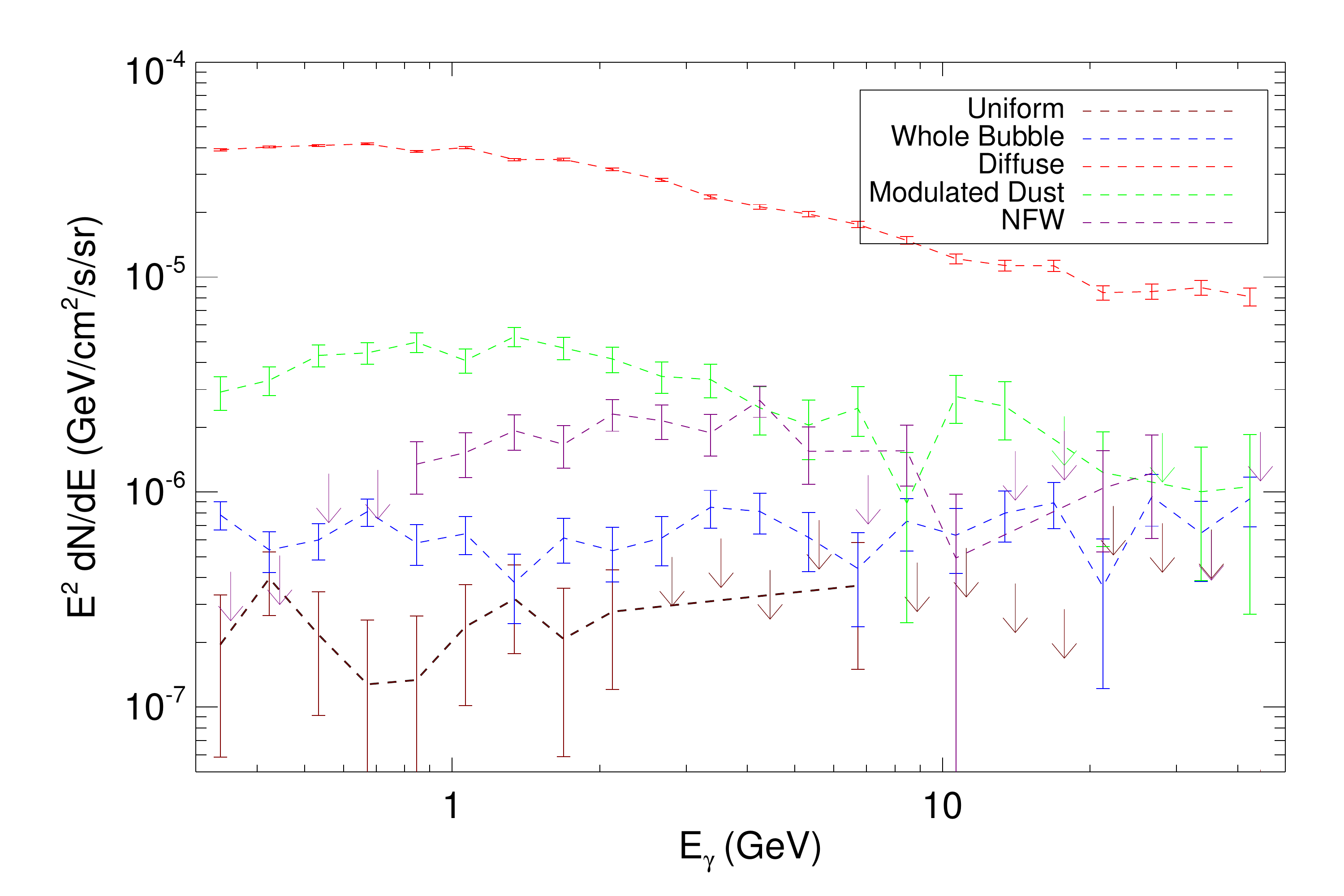}
\caption{The left frame shows the spectra obtained from a template fit employing the standard backgrounds and a modulated dust template, choosing $f(r)$ with $\gamma=0.8$ (see text), in the standard ROI. In the right frame, we plot the coefficients from the same template fit, but with an additional $\gamma=1.18$ dark matter template included. The normalization of the spectrum for the modulated dust template is described in the text; the normalization of the diffuse model spectrum is as in Fig. \ref{fig:coeffsextraSFD}. Due to the large variation in the amplitudes of the different spectra, we use a log scale; where the central values are negative, we instead plot the $3\sigma$ upper limit in that bin.}
\label{fig:coeffsdeltad}
\end{figure*}


\subsection{Modulating the $\pi^0$ Contribution}

The use of the SFD dust map as a tracer for the emission from cosmic-ray proton interactions with gas (producing neutral pions) is predicated on the assumption that the distribution of cosmic-ray protons is approximately spatially uniform. In this appendix, we demonstrate the robustness of the observed signal against the relaxation of this assumption. Specifically we consider an otherwise unmotivated modulation of the gas-correlated emission that seems most likely to be capable of mimicking the signal: the proton density at energies of a few tens of GeV increasing toward the Galactic Center in such a way as to produce the spatially concentrated spectral feature found in the data. Since the gas density is strongly correlated with the Galactic Disk while the signal appears to be quite spherically symmetric (see Sec.~\ref{morphology}), this would require the modulation from varying the cosmic-ray proton density to be aligned perpendicular to the Galactic Plane.

To this end we created additional templates of the form:
\begin{equation} {\rm Modulation} = ({\rm SFD~dust~map})\times \frac{f(r)}{g(r)}, \end{equation}
where $f(r)$ is a projected squared NFW template and $g(r)$ is a simple data-driven characterization of how the SFD dust map falls off with increasing galactic latitude and longitude. In this sense we have factored out how the dust map itself increases towards the Galactic Center and replaced this with a slope that matches a generalized NFW profile. Different modulations were generated by varying $f(r)$, which was done by choosing various values of the NFW inner slope, $\gamma$, from 0.5 to 2.0 in 0.1 increments. In order to determine $g(r)$, the dust map was binned in longitude and latitude and a rough functional form was chosen for each. For longitude, we analyzed the region with $|l|<70^{\circ}$, and fit the profile of the dust map with a Gaussian. For latitude, we considered $|b|<45^{\circ}$ and determined a best-fit using a combination of an exponential and linear function. These two best-fits were then multiplied to give $g(r)$. Each of the new templates were normalized such that the average value of all pixels with an angle between 4.9 and 5.1 degrees from the Galactic Center was set to unity. This was done in order to aid a comparison with the projected squared NFW template, which is normalized similarly. An example of the final template is shown in Fig.~\ref{fig:healcartdeltad}, which was created using an $f(r)$ with $\gamma=0.8$. 

Note that there is no particular physics motivation behind this choice of modulating function; we are attempting to create a dust-correlated map that mimics the observed signal as closely as possible, even if it is not physically reasonable. Since the dust map is integrated along the line of sight, the modulation we have performed is also not precisely equivalent to the effect of changing the cosmic ray density in the inner Galaxy -- this analysis serves as a test of correlation with the gas, but the modulation should \emph{not} be interpreted as a cosmic-ray density map.

Each of the modulated-dust templates was combined with the three background templates described in Sec.~\ref{inner} and run through the maximum likelihood analysis. The results can be seen in the left frame of Fig.~\ref{fig:chisqdeltad}. Generically, the modulated-dust template acquires an appreciable coefficient in a similar energy range to the observed excess. (This should not be surprising, as the modulated-dust templates have been designed to absorb the excess to the greatest degree possible.) The spectrum associated with the template fit using an $f(r)$ of $\gamma=0.8$, near where the $\chi^2$ was improved most, is shown in the left frame of Fig.~\ref{fig:coeffsdeltad}. Nevertheless, when a dark matter template was added to the analysis, there was always a substantial improvement in quality of the fit, as shown in the right frame of Fig.~\ref{fig:chisqdeltad} for a dark matter template with an inner slope of $\gamma=1.18$.

When the dark matter template and modulated dust map are added to the fit together, both acquire non-negligible coefficients, as shown in the right frame of Fig.~\ref{fig:coeffsdeltad}. The modulated dust map is correlated with a soft spectrum, similar to that of the diffuse model, while the dark matter template acquires power in the $\sim 1-5$ GeV range around the peak of the excess. The presence of the modulated dust map \emph{does} in this case substantially bias the extracted spectrum for the dark matter template -- this is not greatly surprising, as by construction the two templates are very similar in shape.

The observant reader may note that the TS of the best-fit modulated dust map is actually \emph{greater} than the TS for the dark matter template. However, it appears this may be due to the modulated dust map doing a better job of picking up unmodeled emission correlated with the \emph{dust}, rather than with the few-GeV excess. If the SFD dust map is added to the fit to provide an additional degree of freedom to the diffuse model, as described in App. \ref{app:sfddust}, the TS for the best fit dark matter template becomes 1748, compared to 1302 for the best fit modulated dust.

\begin{figure*}
\includegraphics[width=0.45\textwidth]{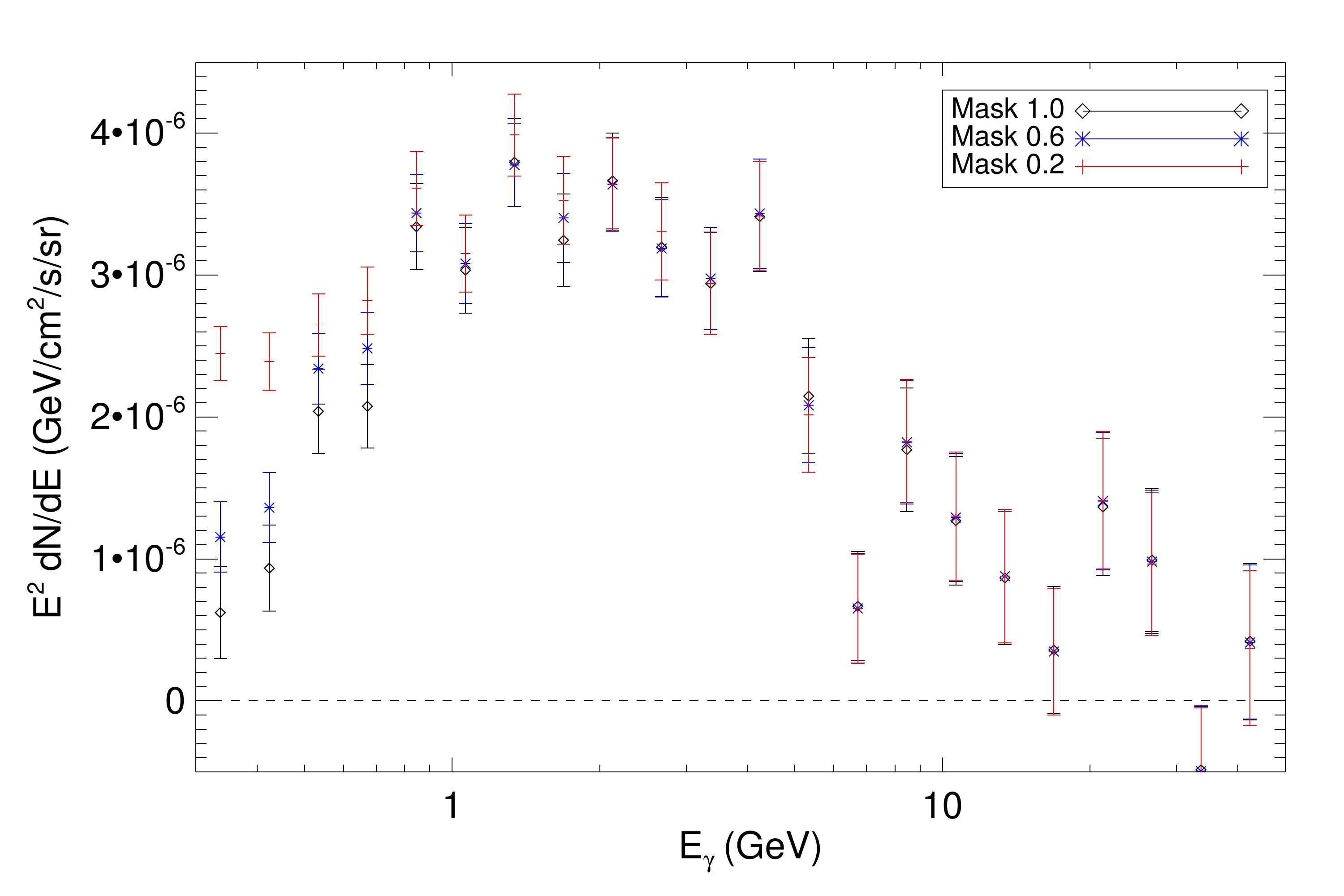}\\
\includegraphics[width=0.45\textwidth]{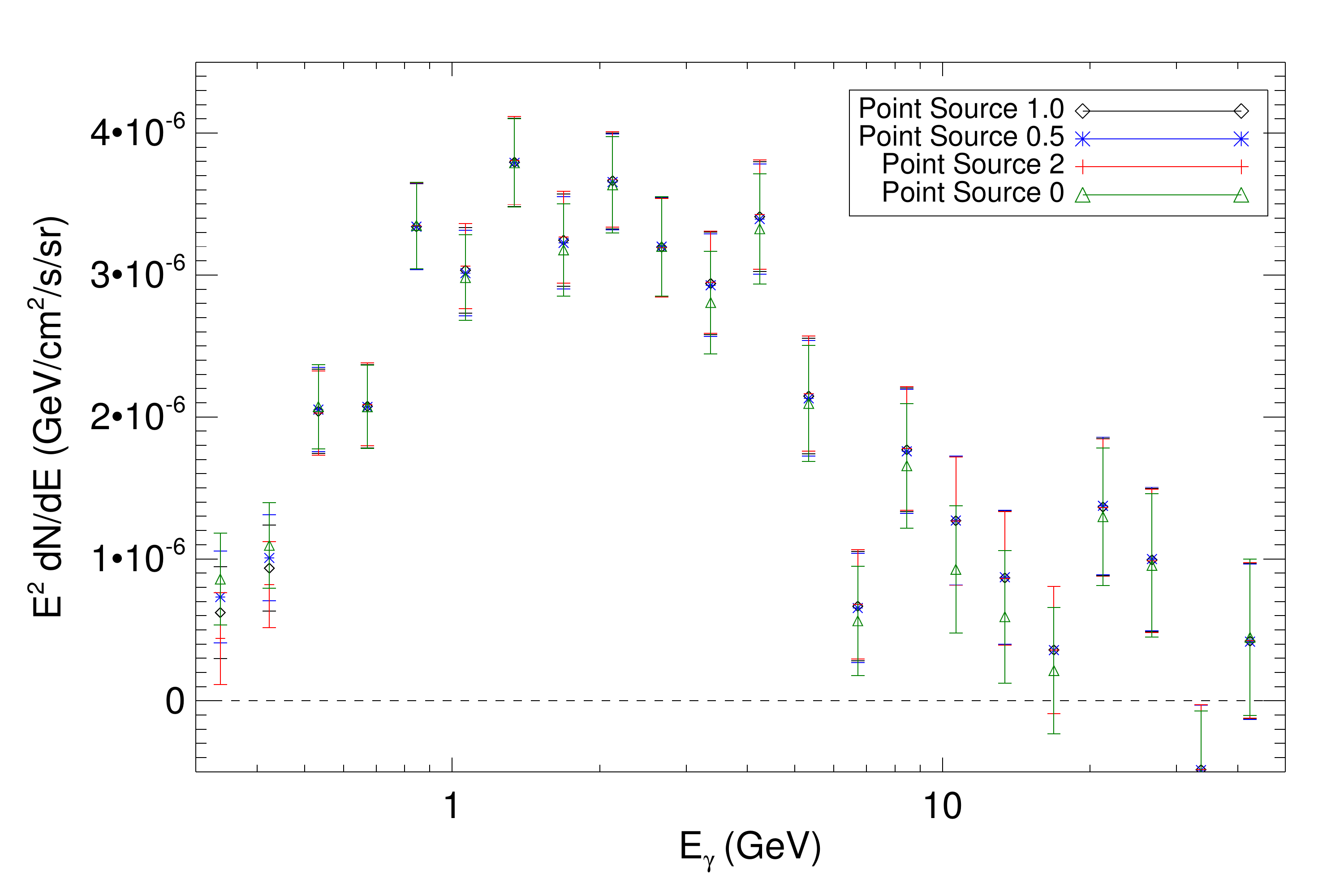}
\includegraphics[width=0.45\textwidth]{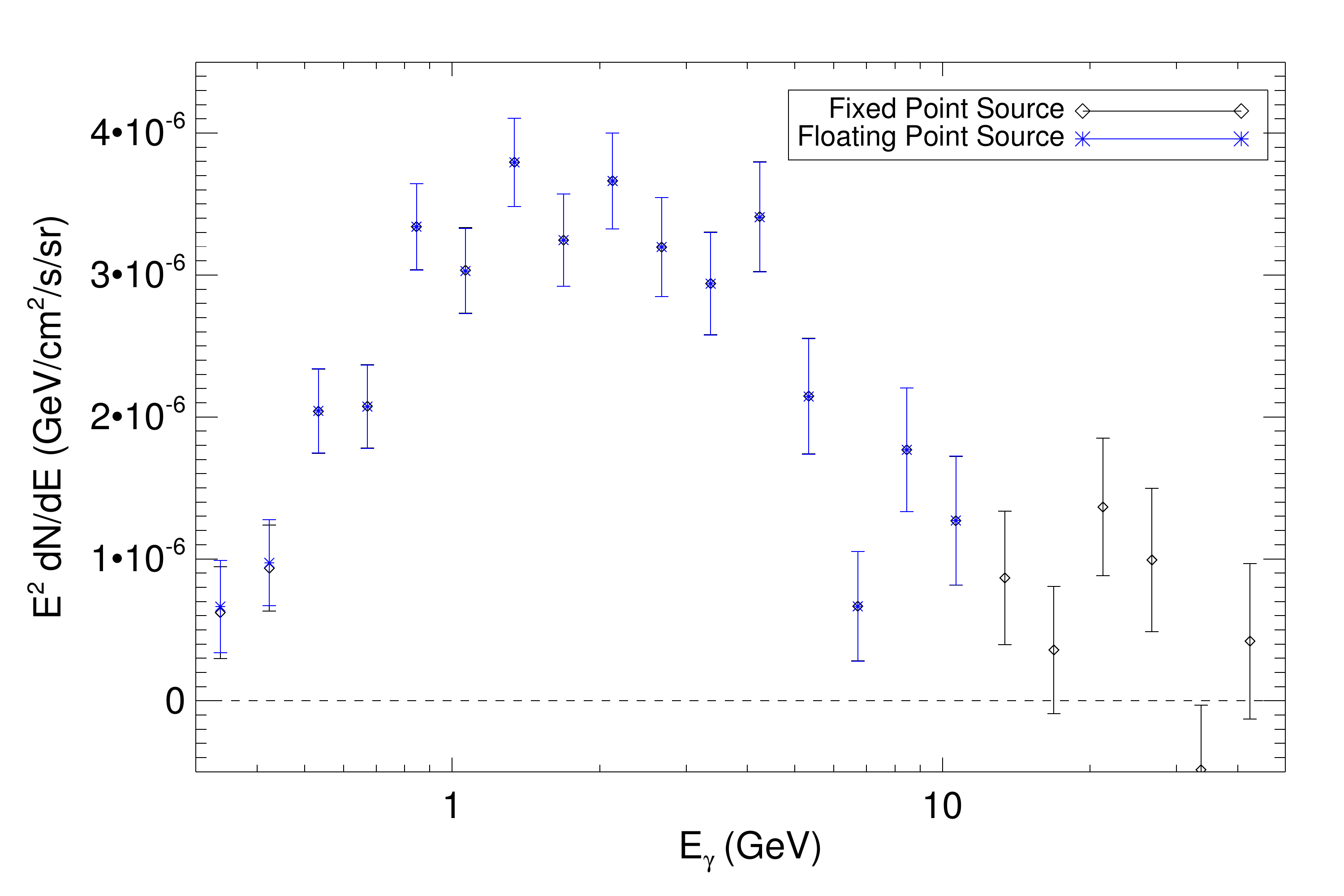}
\caption{Upper panel: Here we show the impact of changing the point source mask radius, shrinking its original size from 1.0 to 0.6 and 0.2. We see that only at the lower energies is there any impact. Lower panel, left frame: We show the result of subtracting the point sources multiplied by a several values: 0, 0.5, 1 and 2. Lower panel, right frame: We show the difference of allowing the point source model to float at each energy as opposed to keeping it fixed. In the floating case we only perform the fit up to 10 GeV; beyond this point it becomes numerically unstable. Our NFW template has $\gamma=1.2$ for all fits.}
\label{fig:psmask}
\end{figure*}

\begin{figure}
\includegraphics[width=0.46\textwidth ]{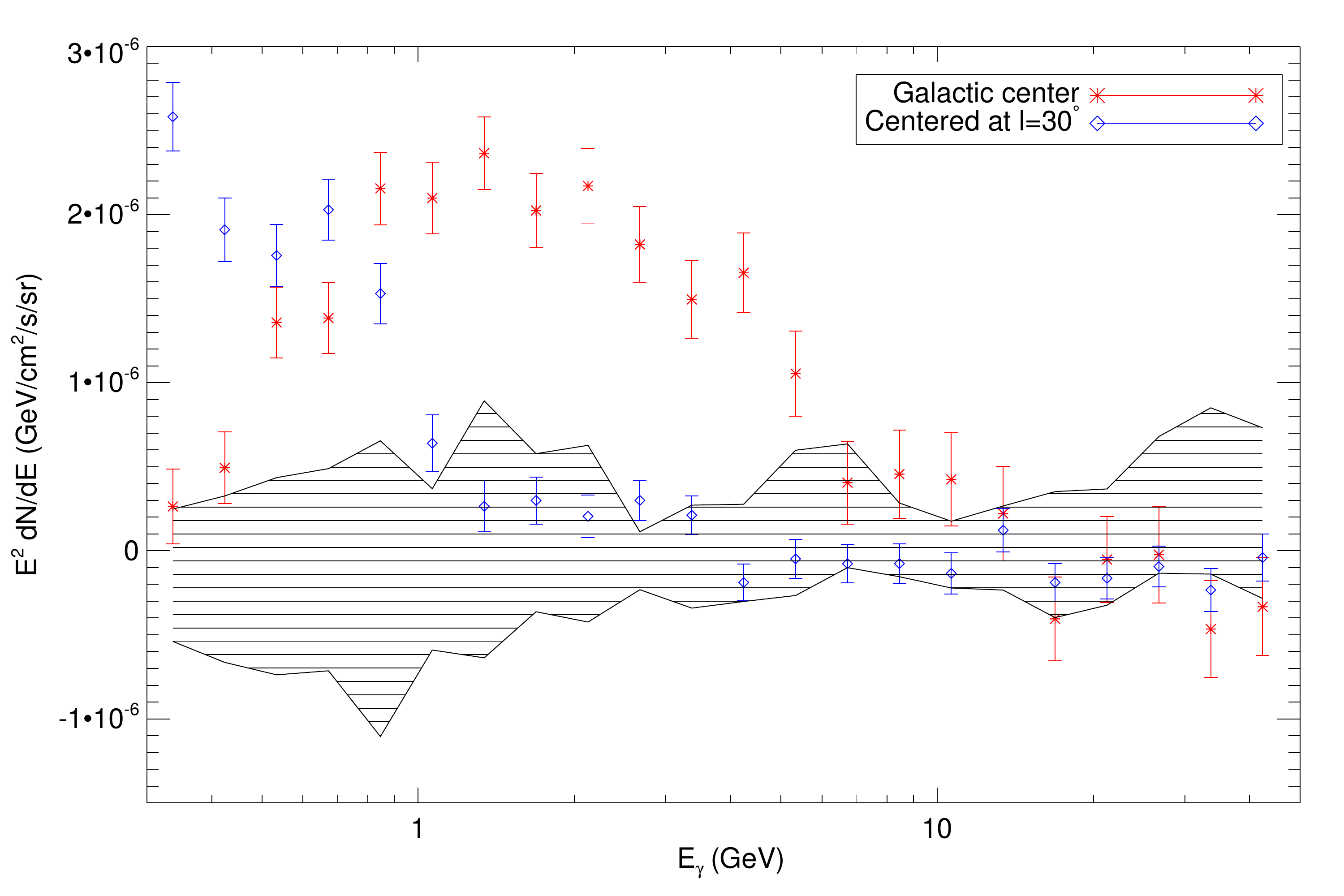}
\caption{Red stars indicate the Galactic Center spectrum, whereas blue diamonds indicate the spectrum correlated with a DM-annihilation-like template (corresponding to an NFW profile with an inner slope $\gamma = 1.3$) centered at $b=0^\circ, l=30^\circ$, instead of at the Galactic Center. The band of horizontal lines indicates the spread of the best-fit spectra correlated with DM-annihilation-like templates shifted in $30^\circ$ increments along the Galactic plane: for the ten other cases sampled ($l=60^\circ, 90^\circ, ..., 330^\circ$), the emission correlated with the DM-annihilation-like template was nearly an order of magnitude below the Galactic Center excess at its peak, with no evidence of spectral similarity. In this case we perform the fit over the full-sky ROI (with an appropriate best-fit $\gamma$), rather than our standard ROI, to ensure stability of the fit and keep the fitted normalizations of the background templates similar over the different runs.}
\label{fig:planescan}
\end{figure}

\begin{figure}
\includegraphics[width=0.45\textwidth]{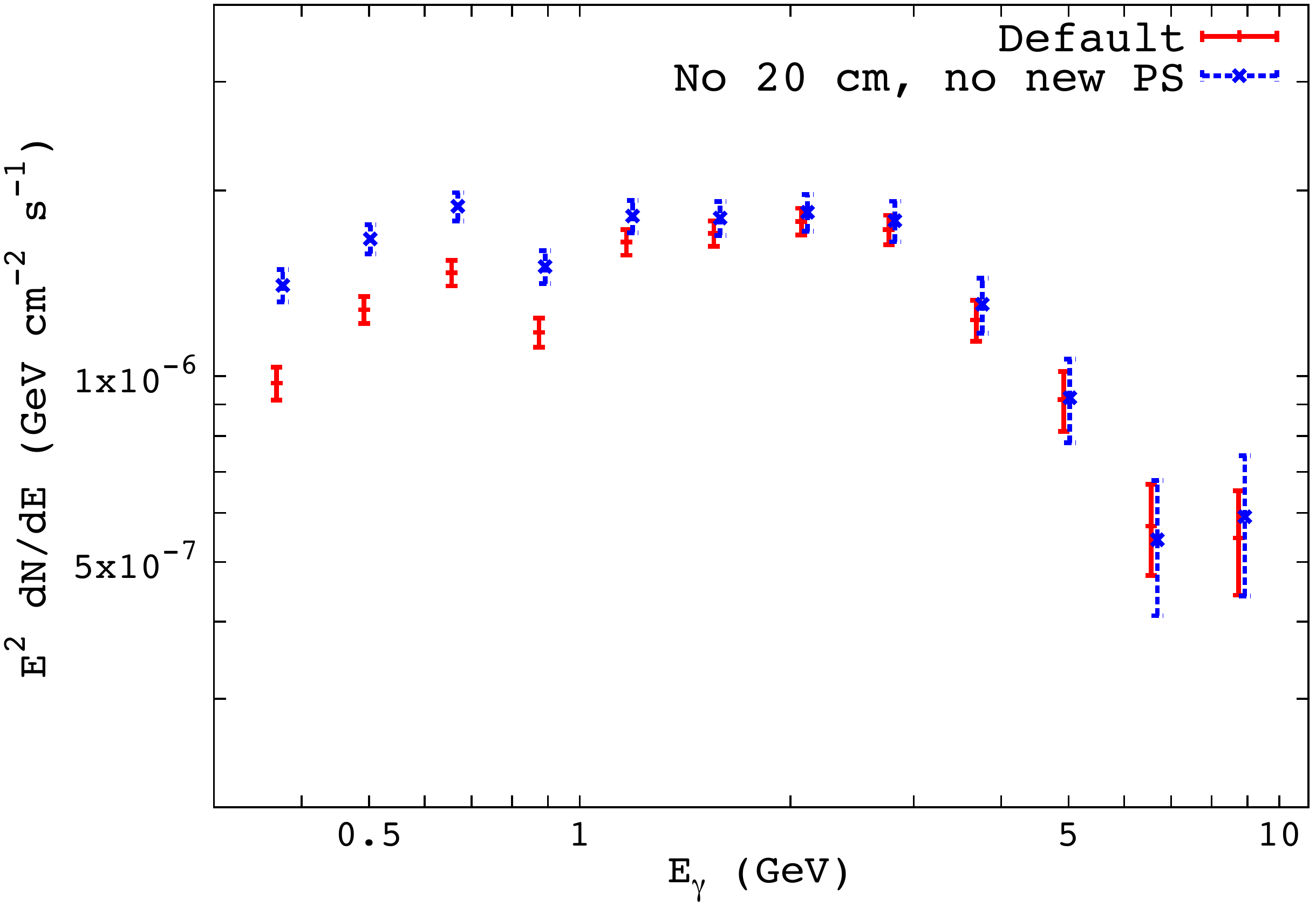}
\hspace{0.2in}
\includegraphics[width=0.45\textwidth]{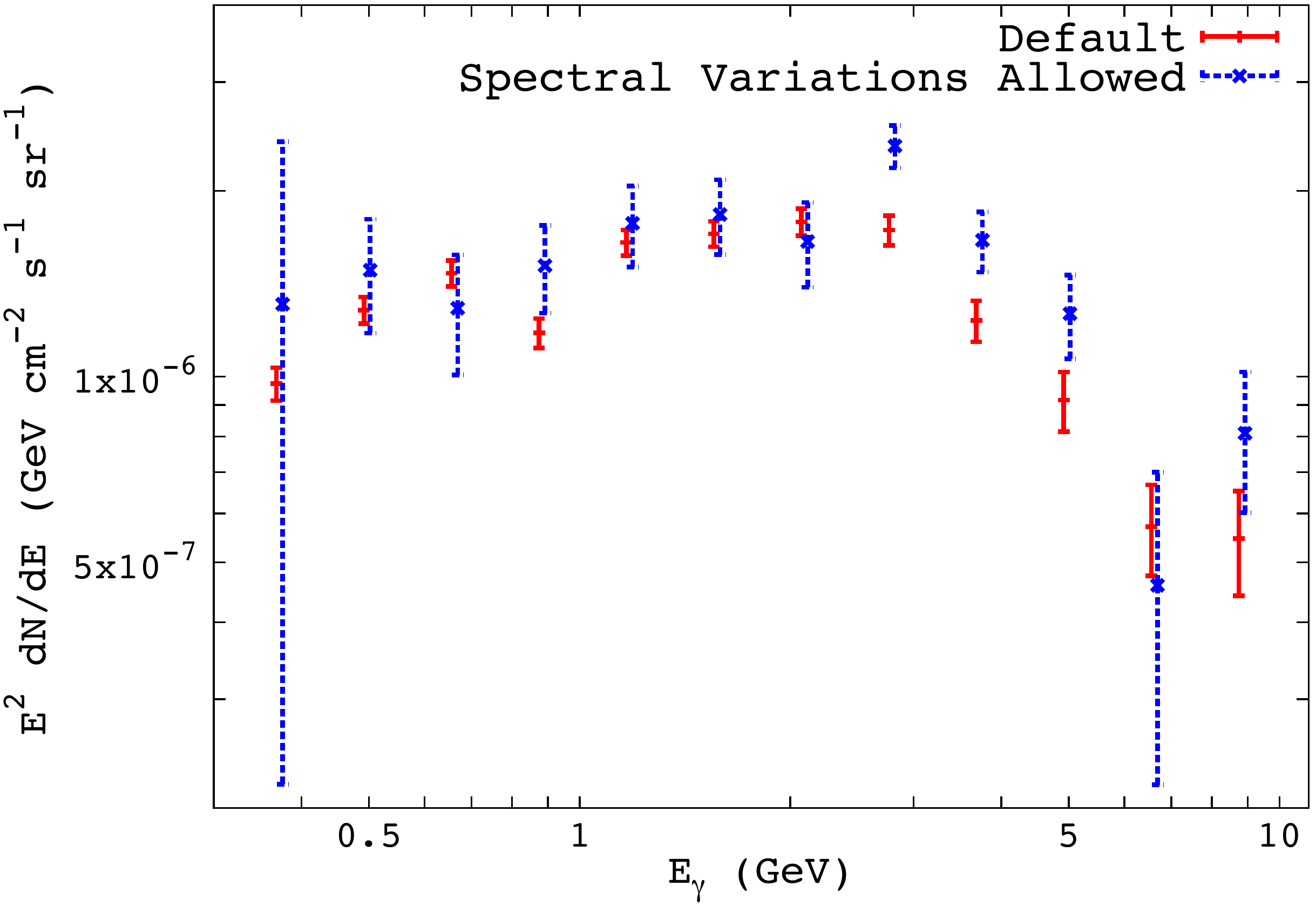}
\caption{Top frame:  A comparison of the spectrum of the dark matter template found in our default Galactic Center analysis to that found when the 20 cm template and two additional point sources are not included in the fit (for $\gamma=1.3)$. The exclusion of these additional components from the fit leads to a softer spectrum at energies below $\sim$1 GeV, but does not influence the spectrum or intensity of the dark matter residual at higher energies.  Bottom frame: The spectrum of the dark matter template found in our Galactic Center analysis under our default assumptions, and when the flux in each energy bin is allowed to float freely for each of the isotropic, 20 cm, and Galactic diffuse components. Although the error bars become larger when this additional freedom is allowed, the residual excess remains and is robust across a wide range of energies. See text for details.}
\label{timappendix}
\end{figure}

The above conclusions were checked to be robust against the choice of ROI and diffuse model; very similar results were found when the analysis was repeated using the full sky or the \texttt{p7v6} model.

It thus appears that a spatial modulation of the gas-correlated emission with coincidental similarities to a dark matter signal could significantly bias the extracted spectrum, but it is difficult (at least within the tests we have performed) to absorb the excess completely. The Galactic Center analysis also finds no evidence for correlation between the excess and known gas structures. Thus even if the $\pi^0$ background has been modeled incorrectly, this deficiency seems unlikely to provide an explanation for the observed signal.

\section{Modifications to the Point Source Modeling and Masking for the Inner Galaxy}
\label{app:pointsource}

As the point sources are concentrated along the Galactic disk and toward the Galactic Center, mismodeling of point sources might plausibly affect the extraction of the signal. To study the potential impact of mismodeling, and check the validity of our point source model, in the Inner Galaxy analysis, we perform the following independent tests:

\begin{itemize}
\item We allow the overall normalization of the point source model to float independently in each energy bin (the relative normalizations of different sources at the same energy are held fixed).
\item We halve or double the flux of all sources in the point source model, relative to the values given in the \texttt{2FGL} catalog.
\item We omit the point source model from the fit entirely.
\item We furthermore investigate the impact of our (fairly arbitrary) choice of mask radius, which is set at the $95\%$ containment radius of the (energy-dependent) PSF by default.
\end{itemize}

Plots showing the results of these various checks are found in Fig.~\ref{fig:psmask}. We find that the impact on the spectrum of even quite severe errors in the point source modeling (such as omitting it entirely or multiplying all source fluxes by a factor of two) is negligible, with the standard mask. Reducing the mask to a very small value has a greater effect, but is still only substantial at the lowest energies; we attribute the extra emission here to leakage from unmasked and poorly-subtracted bright sources.


\section{Shifting the Dark Matter Contribution Along the Plane}
\label{app:shift}

The maps of Fig. 6 show residual bright structure along the Galactic plane. The presence of other bright excesses with the same spectrum along the disk, not simply in the Galactic Center, could favor astrophysical explanations for the signal. To test this possibility, we shift the DM-annihilation-like spatial template along the Galactic plane in 30$^\circ$ increments; as usual, the other templates in the fit are the \emph{Fermi} Bubbles, the diffuse model and an isotropic offset. For numerical stability and consistency of the background modeling, we perform these fits over the full sky rather than the standard ROI. All templates are normalized so that their spectra reflect the flux five degrees from their centers. For ten of the twelve points sampled, the emission correlated with this template is very small; the cross-hatched band in Fig. \ref{fig:planescan} shows the full range of the central values for these ten cases. For the point centered at $l=30^\circ$, there is substantial emission correlated with the template at energies below 1 GeV, but its spectrum is very soft, resembling the Galactic plane more than the excess at the Galactic Center. The last point is the Galactic Center.

We have performed the same test shifting the center of the DM-annihilation-like template in $5^\circ$ increments from $l=-30^\circ$ to $l=30^\circ$. The templates centered at $l = \pm 5^\circ$ absorb emission associated with the Galactic Center excess, albeit with lower amplitude; none of the other cases detect any excess of comparable size with a similar spectrum.

\newpage

\section{Variations to the Galactic Center Analysis}
\label{app:gc}

In the default set of templates used in our Galactic Center analysis, we have employed
astrophysical emission models which include several additional
components that are not included within the official \textit{Fermi} diffuse models or source catalogs. These
include the two point sources described in Ref.~\cite{YusefZadeh:2012nh} and a model tracing the
20 cm synchrotron emission. In models without a dark matter contribution, these structures are extremely
significant; the addition of the 20 cm template is preferred with TS=130, and the inclusion of the additional two point sources is favored with TS=15.9 and 59.3, respectively.

Upon including the dark matter template in the fit, however, the significance of
these additional components is lessened substantially. In this fit, the addition of
the 20 cm template and the two new point sources is preferred at only TS=12.2, 21.8, and 14.6, respectively. Additionally, our best-fit models attribute extremely soft spectra to each of these sources. The 20 cm component has a hard spectrum at
low energies but breaks to a spectral index of -3.3 above 0.6 GeV.  The
spectral indices of the two point sources are -3.1 and -2.8, respectively.
The total improvement in TS for the addition of these combined sources is 47.6.

In the upper frame of Fig.~\ref{timappendix}, we compare the spectrum of the dark matter template found in our default analysis to that found when the 20 cm template and two additional point sources are not included (for $\gamma=1.3)$. The exclusion of these additional components from the fit leads to a softer spectrum at energies below $\sim$1 GeV, but
does not influence the spectrum or intensity of the dark matter
residual at higher energies.

In our default Galactic Center analysis, the isotropic emission is taken to follow a power-law form, while the emission associated with the 20 cm template is allowed to follow a broken power-law, and the Galactic diffuse model adopts a spectrum as given by the model provided by the \textit{Fermi} Collaboration. As a test of the robustness of our results to these assumptions, we perform our fit once again, allowing the flux in each energy bin to float freely for each of the isotropic, 20 cm, and Galactic diffuse components.  In the lower frame of Fig.~\ref{timappendix}, we compare the spectrum extracted in this exercise to that found using our default assumptions. Although the error bars become larger, the residual excess is found to be robust across a wide range of energies.

Lastly, to explore the possibility that the gas distribution as implicitly described by the diffuse model has a systematically biased radial distribution, we performed our fits after distorting the morphology of the diffuse model template such that it becomes brighter at a higher or lower rate as one approaches the center of the Galaxy. However, we found this variation to yield no significant improvement in our fits.

\end{appendix}

\end{document}